\documentclass[twocolumn,twocolappendix]{aastex631}
\hypersetup{linkcolor=blue,citecolor=blue,filecolor=blue,urlcolor=blue}
\received{September 9, 2022}
\revised{December 12, 2022}
\accepted{February 11, 2023}

\usepackage{graphicx}	% Including figure files
\usepackage{amsmath}	% Advanced maths commands
\usepackage{diagbox}
\usepackage{multirow}
\usepackage{array}
\usepackage{verbatim}
\newcolumntype{M}[1]{>{\centering\arraybackslash}m{#1}}
\submitjournal{ApJS}

\shorttitle{The cluster pressure profile from tSZ stacks}
\shortauthors{Tramonte et al.}
\graphicspath{{./}{figures/}}
\begin{document}

\title{Exploring the mass and redshift dependence of the cluster pressure profile with stacks on thermal SZ maps}
\author{Denis Tramonte}\thanks{D. Tramonte, Denis.Tramonte@xjtlu.edu.cn}
\affiliation{Purple Mountain Observatory, No. 8 Yuanhua Road, Qixia District, Nanjing 210034, China}
\affiliation{Department of Physics, Xi'an Jiaotong-Liverpool University, 111 Ren'ai Road, \\    \quad Suzhou Dushu Lake Science and Education Innovation District, Suzhou Industrial Park, Suzhou 215123, P.R. China}
\affiliation{NAOC-UKZN Computational Astrophysics Center (NUCAC), University of KwaZulu-Natal, Durban, 4000, South Africa}
\affiliation{School of Chemistry and Physics, University of KwaZulu-Natal, Westville Campus, Private Bag X54001, Durban, South Africa}

\author{Yin-Zhe Ma}\thanks{Y.-Z. Ma, ma@ukzn.ac.za}
\affiliation{NAOC-UKZN Computational Astrophysics Center (NUCAC), University of KwaZulu-Natal, Durban, 4000, South Africa}
\affiliation{School of Chemistry and Physics, University of KwaZulu-Natal, Westville Campus, Private Bag X54001, Durban, South Africa}
\affiliation{Purple Mountain Observatory, No. 8 Yuanhua Road, Qixia District, Nanjing 210034, China}
\affiliation{National Institute for Theoretical and Computational Sciences (NITheCS), South Africa}
%%\thanks{$^{\dagger}$\url{ma@ukzn.ac.za1}}

\author{Ziang Yan}
\affiliation{Ruhr University Bochum, Faculty of Physics and Astronomy, Astronomical Institute (AIRUB), German Centre for Cosmological Lensing, 44780 Bochum, Germany}
\affiliation{Department of Physics and Astronomy, University of British Columbia, 6224 Agricultural Road, Vancouver, BC, V6T 1Z1, Canada}

\author{Matteo Maturi}
\affiliation{Center for Astronomy - University of Heidelberg, Albert-Ueberle-Stra{\ss}e 2, 69120 Heidelberg, Germany}
\affiliation{Institute of Theoretical Physics - University of Heidelberg, Albert-Ueberle-Stra{\ss}e 2, 69120 Heidelberg, Germany}

\author{Gianluca Castignani}
\affiliation{Department of Physics and Astronomy "A. Righi" - Alma Mater Studiorum University of Bologna, Via Piero Gobetti 93/2, 40129 Bologna, Italy}
\affiliation{INAF - Astrophysics and Space Science Observatory of Bologna, Via Piero Gobetti 93/3, 40129 Bologna, Italy}

\author{Mauro Sereno}
\affiliation{INAF - Astrophysics and Space Science Observatory of Bologna, Via Piero Gobetti 93/3, 40129 Bologna, Italy}
\affiliation{INFN - Bologna Section, Viale Berti Pichat 6/2, 40127 Bologna, Italy}

\author{Sandro Bardelli}
\affiliation{INAF - Astrophysics and Space Science Observatory of Bologna, Via Piero Gobetti 93/3, 40129 Bologna, Italy}

\author{Carlo Giocoli}
\affiliation{INAF - Astrophysics and Space Science Observatory of Bologna, Via Piero Gobetti 93/3, 40129 Bologna, Italy}
\affiliation{Department of Physics and Astronomy "A. Righi" - Alma Mater Studiorum University of Bologna, Via Piero Gobetti 93/2, 40129 Bologna, Italy}
\affiliation{INFN - Bologna Section, Viale Berti Pichat 6/2, 40127 Bologna, Italy}

\author{Federico Marulli}
\affiliation{Department of Physics and Astronomy "A. Righi" - Alma Mater Studiorum University of Bologna, Via Piero Gobetti 93/2, 40129 Bologna, Italy}
\affiliation{INAF - Astrophysics and Space Science Observatory of Bologna, Via Piero Gobetti 93/3, 40129 Bologna, Italy}
\affiliation{INFN - Bologna Section, Viale Berti Pichat 6/2, 40127 Bologna, Italy}

\author{Lauro Moscardini}
\affiliation{Department of Physics and Astronomy "A. Righi" - Alma Mater Studiorum University of Bologna, Via Piero Gobetti 93/2, 40129 Bologna, Italy}
\affiliation{INAF - Astrophysics and Space Science Observatory of Bologna, Via Piero Gobetti 93/3, 40129 Bologna, Italy}
\affiliation{INFN - Bologna Section, Viale Berti Pichat 6/2, 40127 Bologna, Italy}

\author{Emanuella Puddu}
\affiliation{INAF - Osservatorio Astronomico di Capodimonte, Salita Moiariello 16, I-80131, Napoli, Italy}

\author{Mario Radovich}
\affiliation{INAF - Padua Astronomical Observatory, Vicolo dell’Osservatorio, 5, 35122 Padova, Italy}

\author{Ludovic Van Waerbeke}
\affiliation{Department of Physics and Astronomy, University of British Columbia, 6224 Agricultural Road, Vancouver, BC, V6T 1Z1, Canada}

\author{Angus H. Wright}
\affiliation{Ruhr University Bochum, Faculty of Physics and Astronomy, Astronomical Institute (AIRUB), German Centre for Cosmological Lensing, 44780 Bochum, Germany}

%\correspondingauthor{Denis Tramonte}
%\email{tramonte@pmo.ac.cn}

%\correspondingauthor{Yin-Zhe Ma}
%\email{ma@ukzn.ac.za}

\begin{abstract}
We provide novel constraints on the parameters defining the universal pressure profile (UPP) 
within clusters of galaxies, and explore their dependence on the cluster mass and redshift, 
from measurements of Sunyaev-Zel'dovich Compton-$y$ profiles. We employ both the 
\textit{Planck} 2015 MILCA and the ACT-DR4 $y$ maps over the common 
$\sim 2,100\,\text{deg}^2$ footprint. We combine existing cluster catalogs based on KiDS, SDSS and DESI 
observations, for a total of 23,820 clusters spanning the mass range 
	$10^{14.0}\,\text{M}_{\odot}<M_{500}<10^{15.1}\,\text{M}_{\odot}$ and the redshift range $0.02<z<0.98$. 
We split the clusters into three independent bins in mass and redshift; for each combination
 we detect the stacked SZ cluster signal and extract the mean $y$ angular profile. 
 The latter is predicted theoretically adopting a halo model framework, and MCMCs are employed to
estimate the UPP parameters, the hydrostatic mass bias $b_{\rm h}$ and possible cluster miscentering effects. 
We constrain  $[P_0,c_{500},\alpha,\beta]$ to $[5.9,2.0,1.8,4.9]$ with \textit{Planck} and to $[3.8,1.3,1.0,4.4]$ with ACT using the full cluster sample, in agreement with previous findings. We do not find any compelling evidence for a residual mass or redshift dependence, thus expanding the validity of the cluster pressure profile over much larger $M_{500}$ and $z$ ranges; this is the first time the model has been tested on such a large (complete and representative) cluster sample. Finally, we obtain loose constraints on the hydrostatic mass bias in the range 0.2-0.3, again in broad agreement with previous works.
\end{abstract}


\defcitealias{gong19}{G19}

%=================================================================================================
%=================================================================================================

%=================================================================================================
%===================================  INTRODUCTION ===============================================	 
%=================================================================================================

\section{Introduction}
\label{sec:introduction}

Galaxy clusters are an invaluable cosmological probe, providing information on the geometry of the 
Universe, on the growth of cosmic structures, and, at lower scales, on the processes of galaxy formation 
and evolution~\citep{voit05, allen11}. The majority of cluster baryonic matter (up to 90\%) is found as 
a diffuse component referred to as the intra-cluster medium (ICM), which is shock-heated and ionized in 
the strong cluster gravitational field, up to temperatures of 5-10\,keV. A proper characterization of 
the ICM physical properties is of great interest, not only for allowing an indirect calibration of the mass 
proxies based on ICM observations, but also for providing useful insights in the processes of galaxy 
evolution and feedback.  

\begin{table*} 
\centering
\caption{Summary of previous estimates on the parameters entering the UPP expression in equation~\eqref{eq:upp}. For each work we report the reference in the literature, the number of clusters used in the study (when applicable), the used physical observables and the best-fit values for the UPP parameters. Boldface values were kept fixed in the corresponding fit. For more details we redirect to Appendix~\ref{app:upp_summary}.}
\label{tab:upp_summary}
\renewcommand{\arraystretch}{0.9}
\begin{tabular}{M{3.4cm}|M{1.8cm}M{2.3cm}|M{1.8cm}M{1.4cm}M{1.4cm}M{1.5cm}M{1cm}}
\hline
\multirow{4}{*}{\textbf{Reference}} & \multicolumn{2}{c|}{\multirow{2}{*}{\textbf{Data set}}} & \multicolumn{5}{c}{\multirow{2}{*}{\textbf{UPP parameters}}} \\
 & & & & & & & \\
\cline{2-8}
 & \multirow{2}{*}{\textbf{Objects}} & \multirow{2}{*}{\textbf{Observables}} & \multirow{2}{*}{$\boldsymbol{P_0}$} & \multirow{2}{*}{$\boldsymbol{c_{500}}$} & \multirow{2}{*}{$\boldsymbol{\alpha}$} & \multirow{2}{*}{$\boldsymbol{\beta}$} & \multirow{2}{*}{$\boldsymbol{\gamma}$}\\
 & & & & & & & \\
\hline
\multirow{2}{3.4cm}{\centering \citet{nagai07}} & \multirow{2}{1.8cm}{\centering 16 clusters} & \multirow{2}{2.3cm}{\centering X-ray, simulations} & \multirow{2}{1.8cm}{\centering 3.3} & \multirow{2}{1.4cm}{\centering 1.8} & \multirow{2}{1.4cm}{\centering 1.3} & \multirow{2}{1.5cm}{\centering 4.3} & \multirow{2}{1cm}{\centering 0.7}\\
 & & & & & & & \\
\multirow{2}{3.4cm}{\centering \citet{arnaud10}} & \multirow{2}{1.8cm}{\centering 33 clusters} & \multirow{2}{2.3cm}{\centering X-ray, simulations} & \multirow{2}{1.8cm}{\centering 8.403$\,h_{70}^{-3/2}$} & \multirow{2}{1.4cm}{\centering 1.177} & \multirow{2}{1.4cm}{\centering 1.0510} & \multirow{2}{1.5cm}{\centering 5.4905} & \multirow{2}{1cm}{\centering 0.3081}\\
 & & & & & & & \\
\multirow{2}{3.4cm}{\centering \citet{planck_ir_v}} & \multirow{2}{1.8cm}{\centering 62 clusters} & \multirow{2}{2.3cm}{\centering SZ, X-ray} & \multirow{2}{1.8cm}{\centering 6.41} & \multirow{2}{1.4cm}{\centering 1.81} & \multirow{2}{1.4cm}{\centering 1.33} & \multirow{2}{1.5cm}{\centering 4.13} & \multirow{2}{1cm}{\centering $\mathbf{0.31}$}\\[1ex]
 & & & & & & & \\
\citet{sayers16} & 47 clusters & SZ, X-ray & $9.13\pm2.98$ & $\mathbf{1.18}$ & $\mathbf{1.0510}$ & $6.13\pm0.76$ & $\mathbf{0.3081}$\\[1ex]
\citet{gong19}  & $\sim10^5$ LRGs & SZ & $2.18^{+9.02}_{-1.98}$ & $1.05^{+1.27}_{-0.47}$ & $1.52^{+1.47}_{-0.58}$ & $3.91^{+0.87}_{-0.44}$ & $\mathbf{0.31}$\\[1ex]
\multirow{2}{3.4cm}{\centering \citet{ma21}} & \multirow{2}{1.8cm}{\centering -} & \multirow{2}{2.3cm}{\centering SZ, WL (convergence)} & \multirow{2}{1.8cm}{\centering $9.68^{+10.02}_{-7.11}$} & \multirow{2}{1.4cm}{\centering $2.71^{+0.92}_{-0.93}$} & \multirow{2}{1.4cm}{\centering $5.97^{+1.81}_{-4.73}$} & \multirow{2}{1.5cm}{\centering $3.47^{+1.39}_{-0.60}$} & \multirow{2}{1cm}{\centering $\mathbf{0.31}$}\\[1ex]
 & & & & & & & \\
\citet{ma21} & - & SZ, WL (shear)  & $6.62^{+2.06}_{-1.65}$ & $1.91^{+1.07}_{-0.65}$ & $1.65^{+0.74}_{-0.50}$ & $4.88^{+1.18}_{-2.46}$ & $\mathbf{0.31}$\\[1ex]
\multirow{2}{3.4cm}{\centering \citet{pointecouteau21}} & \multirow{2}{1.8cm}{\centering 31 clusters} & \multirow{2}{2.3cm}{\centering SZ} & \multirow{2}{1.8cm}{\centering $3.36^{+0.90}_{-0.71}$} & \multirow{2}{1.4cm}{\centering $\mathbf{1.18}$} & \multirow{2}{1.4cm}{\centering $1.08^{+0.13}_{-0.11}$} & \multirow{2}{1.5cm}{\centering $4.30\pm0.12$} & \multirow{2}{1cm}{\centering $\mathbf{0.31}$}\\
 & & & & & & & \\
\multirow{2}{3.4cm}{\centering \citet{he21}} & \multirow{2}{1.8cm}{\centering 33 clusters} & \multirow{2}{2.3cm}{\centering X-ray, simulations} & \multirow{2}{1.8cm}{\centering 5.048} & \multirow{2}{1.4cm}{\centering 1.217} & \multirow{2}{1.4cm}{\centering 1.192} & \multirow{2}{1.5cm}{\centering $\mathbf{5.490}$} & \multirow{2}{1cm}{\centering 0.433}\\
 & & & & & & & \\[0.3ex]
\hline
\end{tabular}
\end{table*}

The high temperature of the ICM plasma has made it a traditional target for X-ray observations~\citep{sarazin88}, 
a property that has been exploited by different generations of satellite missions to build X-ray cluster 
catalogs~\citep{voges99, hicks08, mehrtens12,klein21}. A complementary probe is the observation of the thermal Sunyaev-Zel'dovich (tSZ) effect~\citep{sunyaev72}, a secondary anisotropy of the cosmic microwave background 
(CMB) radiation which is produced when CMB photons interact with a population of high-energy electrons 
via inverse Compton scattering. The resulting temperature fluctuations with respect to the CMB temperature 
$T_{\rm CMB}$ can be expressed as:
\begin{equation}
	\frac{\Delta T}{T_{\rm CMB}} = f(\xi)y,
\end{equation}
where the dependence on the scaled frequency $\xi \equiv h\nu/(k_{\rm B}T_{\rm CMB})$, $h$ and $k_{\rm B}$ being 
the Planck and Boltzmann constants, is encoded in the function $f(\xi)=\xi\coth{(\xi/2)}-4$; the result is a 
decrease (increase) of the CMB temperature at frequencies below (above) 217 GHz. The magnitude of the effect is quantified by the Compton parameter $y$ which is proportional to the electron pressure 
integrated along the line of sight (LoS):
\begin{equation}
	\label{eq:y}
	y = \frac{\sigma_{\rm T}}{m_{\rm e}c^2}\int_{\rm LoS} \text{d}l \,P_{\rm e}(l),
\end{equation}
where $\sigma_{\rm T}$ is the Thomson cross-section, $m_{\rm e}c^2$ the electron rest energy and $P_{\rm e}(l)$ the 
electron pressure at a physical LoS separation $l$. 

The high energy electrons found in the ICM make the tSZ effect an ideal probe to detect and study galaxy 
clusters~\citep{birkinshaw99,carlstrom02}. Unlike the X-ray brightness, which is proportional to the 
squared electron density $n_{\rm e}^2$, the Compton parameter is proportional to $n_{\rm e}$, which 
implies that it has a higher sensitivity to low mass densities and can be used to trace the ICM out to larger 
separations from the cluster core. In addition, it is independent of the cluster redshift\footnote{Strictly 
speaking, tSZ observations detect the cluster signal integrated over the beam solid angle $Y_{\rm SZ}$, 
which is proportional to the temperature-weighted mass of the cluster $M\,\langle T_{\rm e}\rangle$ and 
to the inverse square of the angular diameter distance at the cluster redshift, 
$Y_{\rm SZ}\propto M\,\langle T_{\rm e}\rangle\,D_{\rm A}(z)^{-2}$~\citep{carlstrom02}. This introduces 
a marginal redshift dependence. In fact, tSZ detections of clusters at $z>1$ are not common, possibly 
due to a lower ICM temperature or the contamination from radio loud AGNs.}, and tSZ observations can be 
conveniently carried out at radio and microwave frequencies from ground-based observatories. The advances 
in tSZ observational techniques during the past two decades have yielded dedicated tSZ-detected cluster 
catalogs and the reconstruction of the Compton parameter signal over extended areas of the sky, using 
both satellite missions like \textit{Planck}~\citep{planck_15_xxvii} and ground-based facilities like 
the Atacama Cosmology Telescope~\citep[ACT,][]{hilton21} and the South Pole Telescope~\citep[SPT,][]{bleem20}.  

Both in the case of X-ray or tSZ data, a proper characterization of the ICM eventually translates into 
the modeling of the local electron pressure. The analysis presented in~\citet{nagai07} first proposed 
a generalized Navarro-Frenk-White~\citep[NFW,][]{navarro97} parametrization as a universal model for the electron pressure 
profile, in the form:
\begin{eqnarray}
	\label{eq:upp}
	\mathbb{P}(x) &\equiv & \frac{P_{\rm e}(r)}{P_{500}} \nonumber \\
	&=& \dfrac{P_0}{(c_{500}x)^{\gamma}[1+(c_{500}x)^{\alpha}]^{(\beta-\gamma)/\alpha}},
\end{eqnarray}
where $P_{\rm e}(r)$ is the electron pressure at a physical separation $r$ from the cluster center, 
$P_{500}$ is the characteristic pressure expected in a self-similar model (Section~\ref{ssec:clusterpress}), 
carrying the dependence on the cluster mass and redshift, and the profile is expressed as a function of the 
scaled radial separation\footnote{Unless explicitly stated, it is understood that overdensity masses and radii 
are referred to the Universe critical density $\rho_{\rm c}(z)$ at the considered redshift; in formulae,
$M_{\Delta}=4\pi\Delta\rho_{\rm c}(z)R_{\Delta}^3/3$, with $\Delta$ the overdensity value.} $x\equiv r/R_{500}$. 
This universal pressure profile (hereafter UPP) is parametrized in terms of the concentration\footnote{In some work, 
the UPP is expressed as a function of $x'=r/r_{\rm s}$, where the scale radius $r_{\rm s}$ is related to 
the overdensity radius $R_{500}$ via the concentration parameter, $c_{500}=R_{500}/r_{\rm s}$.} $c_{500}$, an 
overall normalization factor $P_0$, and the parameters $\gamma$, $\alpha$ and $\beta$ which are, respectively, 
the profile slopes at small ($x\ll 1/c_{500}$), intermediate ($x\sim 1/c_{500}$) and large ($x\gg 1/c_{500}$) 
separations from the cluster center. \citet{nagai07} provided best-fit estimates for the parameters based on 
profiles reconstructed from \textit{Chandra} data and results from hydrodynamical simulations.  

In the past decade, several other works have contributed to constraining the UPP parameters using different 
observables and techniques. A summary of the results on the fitted UPP parameters from these works is 
reported in Table~\ref{tab:upp_summary}, while for a more comprehensive summary we redirect to 
Appendix~\ref{app:upp_summary}. Most of these studies employed a limited set of well-characterized and high-significance clusters to measure the cluster pressure profile~\citep{arnaud10,planck_ir_v,sayers16,pointecouteau21,he21}. 
A substantially different approach was instead adopted in~\citet{gong19}, hereafter G19, where the UPP 
parameters were fitted over the stack of a large number ($\sim10^5$) of regions surrounding 
luminous red galaxies (LRGs) at 
$z\lesssim0.5$, assuming the latter are good tracers of massive dark matter halos. This type of work foregoes the 
profile reconstruction for individual objects and focuses instead on the mean ICM properties of an extended 
sample. The results were in agreement with previous studies, thus proving the feasibility of this approach 
for characterizing the UPP. Besides, the large number statistics provided by the sample in~\citetalias{gong19} 
allowed the authors to split the LRG sample into three redshift bins, finding hints of a redshift evolution of 
the UPP parameters; the inclusion of an explicit redshift dependence in the normalization pressure $P_{500}$ 
proved effective in improving the reduced $\chi^2$ when fitting the combination of the three redshift bins. Finally, other, more 
indirect estimates of the UPP parameters were obtained from the cross-correlation between tSZ and weak lensing 
(WL) data~\citep{hojjati15,hojjati17,ma21}.

\begin{figure*} % +++++++++++++++++++++++++++++++++++++++++++++++++++++++++++++++
\includegraphics[trim= 0mm 0mm 0mm 0mm, scale=0.29]{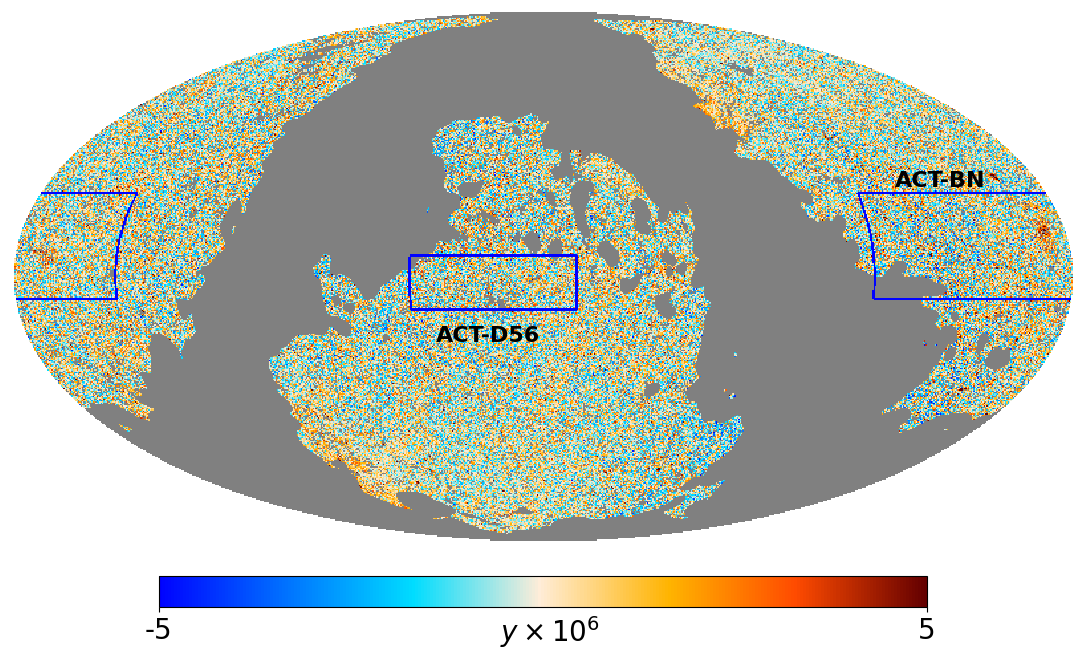}\qquad
\includegraphics[trim= 20mm 0mm 0mm 0mm, scale=0.23]{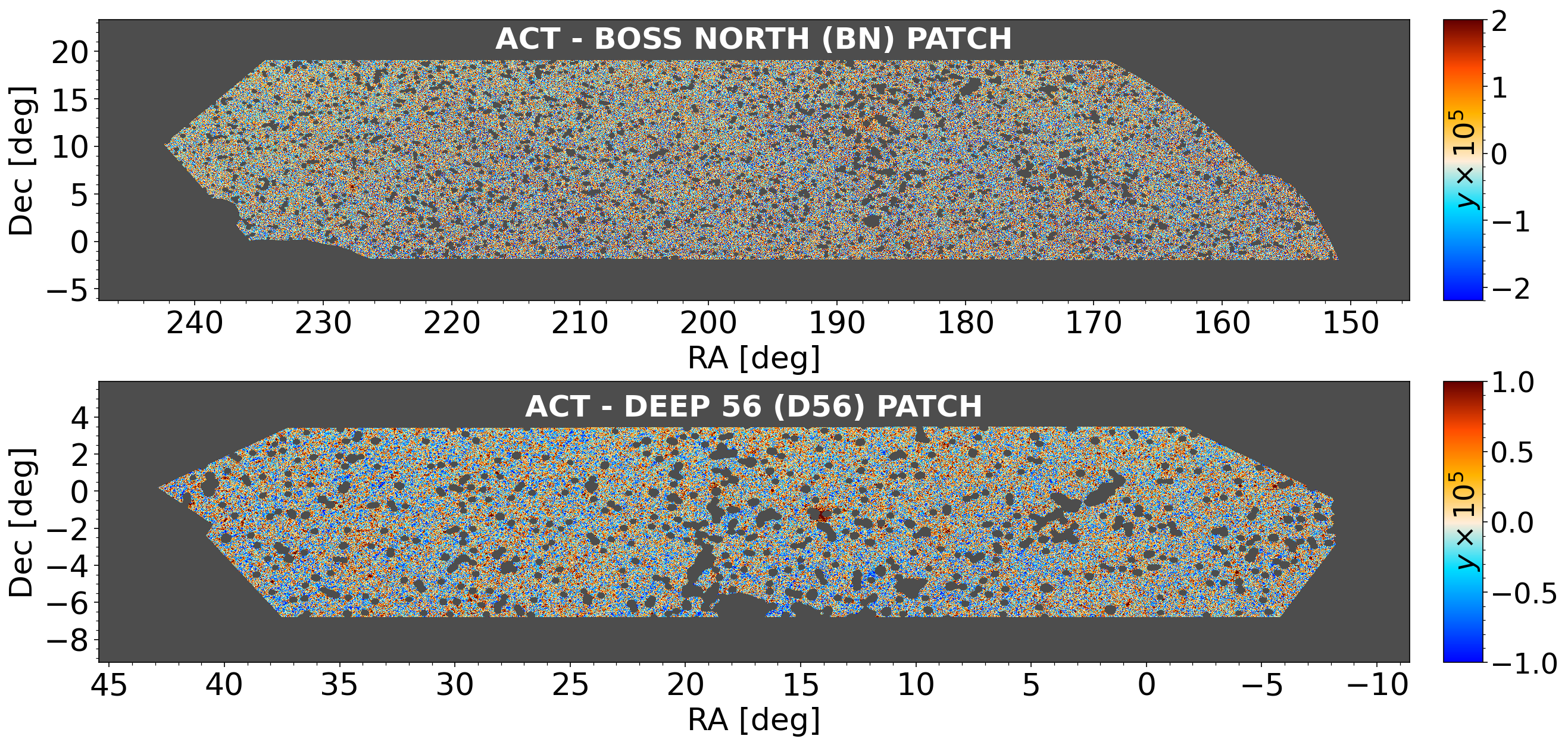}
	\caption{\textit{Left}: the all-sky \textit{Planck} 2015 MILCA Compton $y$-map, plotted in 
	equatorial coordinates with the Galactic plane 40\% mask and the point source mask 
	overlapped; the position of the two ACT patches is also marked with blue boxes. \textit{Right}: the 
	two patches of the ACT-DR4 $y$-map, combined with the local projection of the mask 
	adopted for \textit{Planck}. The color scale units are different in \textit{Planck} and ACT to 
	better show the features in the corresponding maps; in both cases, masked regions are shown in 
	gray color.\label{fig:maps}}	
\end{figure*}  % +++++++++++++++++++++++++++++++++++++++++++++++++++++++++++++++

The present study aims at extending this line of work by providing novel estimates of the UPP parameters. Our 
analysis is based on cluster stacks on $y$ maps; the UPP parameters are then fitted on the reconstructed mean 
$ y$ profiles. The main difference with~\citetalias{gong19} is that we will use a real cluster catalog instead 
of reverting to LRGs as positional tracers. Besides, we will be using a {\it complete} cluster sample; this is a major 
difference compared to the cluster-based studies listed in Table~\ref{tab:upp_summary}, whose high-significance 
cluster samples were non-complete and in some cases even non-representative (see Appendix~\ref{app:upp_summary} for 
a more extensive discussion). Our sample is obtained by merging 
existing cluster catalogs, yielding a total of $\sim2.3\times10^4$ clusters spanning the $M_{500}$ mass range 
$[10^{14},10^{15.1}]\,\text{M}_{\odot}$ and the photometric redshift range $[0.1,0.8]$. This is another important
difference compared with the analysis in~\citetalias{gong19}, which targeted mostly the mass range of rich 
groups ($M_{500}\lesssim 10^{14}\,\text{M}_{\odot}$). Furthermore, our large sample allows 
us to split the data set not only in different redshift bins but also in different mass bins, thus exploring in more detail possible deviations from the universality of the pressure profile. Finally, we will obtain independent 
results from both \textit{Planck} and ACT Compton maps for the same cluster sample.

This paper is organized as follows. We begin by describing the two data sets employed in our analysis, namely the 
Compton $y$-maps (Section~\ref{sec:ymaps}) and the cluster catalogs (Section~\ref{sec:catalogs}). 
Section~\ref{sec:methodology} presents the methodology we adopt for generating our reference sample, stacking the 
clusters, and extracting the associated angular $y$ profiles with their uncertainties. The formalism we employ to 
model the cluster $y$ signal is detailed in Section~\ref{sec:theory}, whereas the parameter estimation analysis is 
presented in Section~\ref{sec:estimation}. Finally, Section~\ref{sec:conclusions} reports the conclusions. Throughout 
this paper we adopt a spatially-flat $\Lambda$CDM cosmological model with parameter values $h=0.674$, $\Omega_{\rm m}=0.315$, 
$\Omega_{\rm b}=0.0493$, $\sigma_8=0.811$ and $n_{\rm s}=0.965$~\citep{planck_18_vi}.

%=================================================================================================
%=================================================================================================

%=================================================================================================
%================================    COMPTON PARAMETER MAPS 	 ================================= 
%=================================================================================================

\section{Compton parameter maps}
\label{sec:ymaps}

In this work we conduct the stacking analysis on two different Compton parameter maps, obtained by \textit{Planck}
and ACT. We describe each in details in the following.

\subsection{\textit{Planck} data} %================================================
\label{ssec:planck}

We employ the all-sky Compton parameter map delivered by the \textit{Planck} Collaboration and described 
in~\citet{planck_15_xxii}. The map is publicly available at the \textit{Planck} Legacy 
Archive\footnote{\url{https://pla.esac.esa.int/\#maps}.}, and can be downloaded in {\tt HEALPix} format~\citep{gorski05} 
with a pixelization set by the resolution parameter $N_{\rm side}=2048$ (corresponding to a pixel size of $\sim 1.8\,\text{arcmin}$). The map was generated via a tailored 
linear combination of \textit{Planck} individual frequency maps. Two versions of the map are available, obtained 
from two different implementations of the Internal Linear Combination (ILC) algorithm, namely the Modified Internal 
Linear Combination Algorithm~\citep[MILCA, ][]{hurier13} and the Needlet Independent Linear Combination~\citep[NILC,][]{remazeilles11} 
method.  In the following we will adopt the MILCA map only, as we verified that the use of the NILC map yields 
 results on the Compton parameter profiles compatible within the final error bars. Prior to their linear combination, 
\textit{Planck} channel maps were first degraded to a common resolution of 10 arcmin, which is the reference full-width 
at half maximum (FWHM) value for the final Compton map. 

We adopt a suitable mask to avoid contamination from residual Galactic foregrounds and strong extragalactic radio 
sources. The \textit{Planck} Legacy Archive provides a point source mask and different Galactic masks tailored to the 
tSZ analysis. We combine the 40\% Galactic plane mask and the point source mask, excluding a total 50.6\% of the sky. 
The \textit{Planck} MILCA Compton map, combined with our adopted mask, is shown in the first panel of Fig.~\ref{fig:maps}.

\begin{figure*}  % +++++++++++++++++++++++++++++++++++++++++++++++++++++++++++++++
\includegraphics[trim= 0mm 0mm 0mm 0mm, scale=0.2]{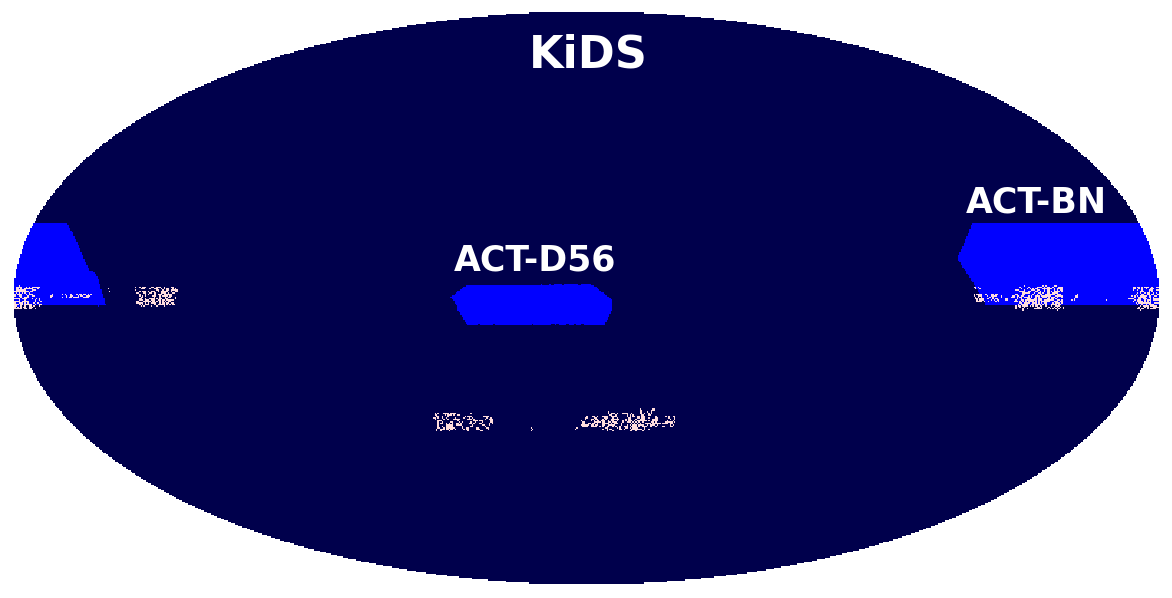}
\includegraphics[trim= 0mm 0mm 0mm 0mm, scale=0.2]{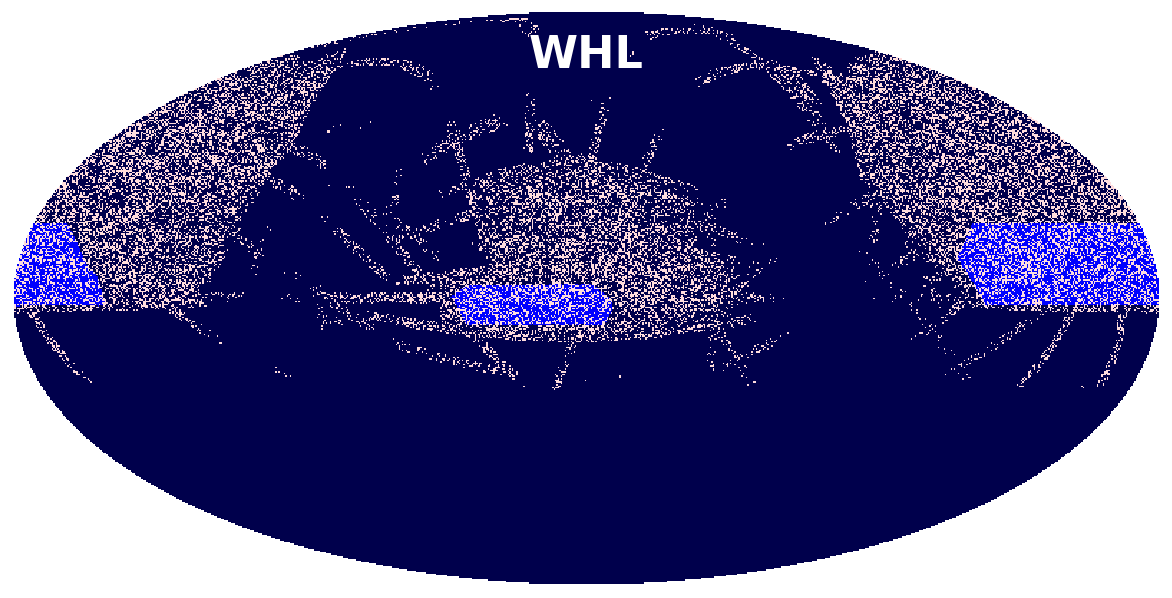}
\includegraphics[trim= 0mm 0mm 0mm 0mm, scale=0.2]{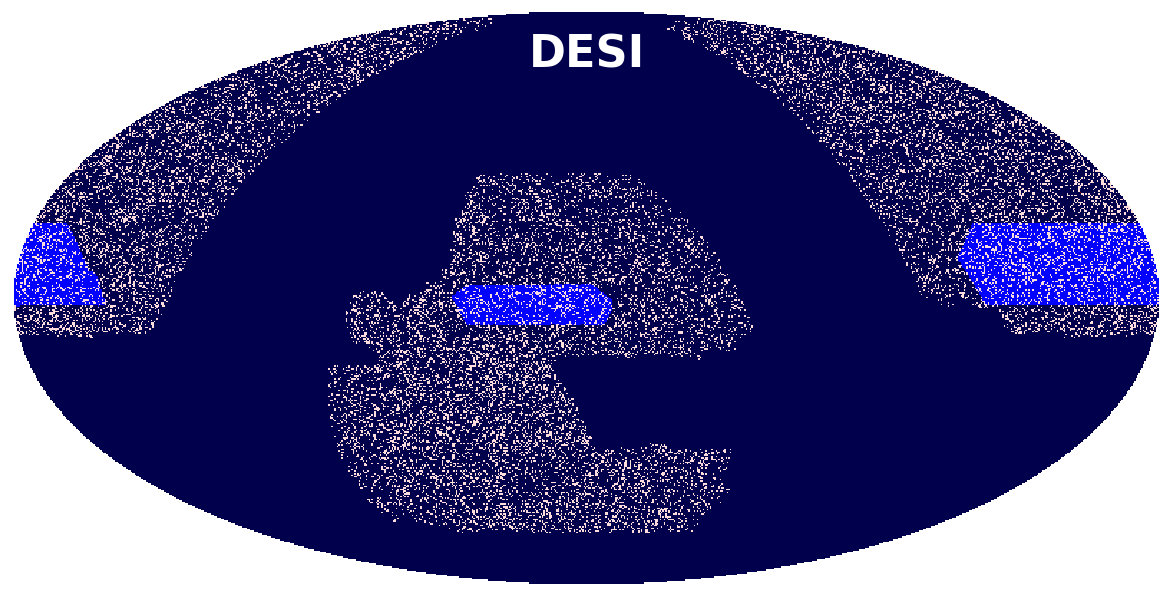}
	\caption{Footprint of the two ACT patches (lighter blue) adopting the same equatorial frame 
	as in the first panel of Fig.~\ref{fig:maps}, compared with the angular distribution of 
	our chosen cluster catalogs (white color), namely KiDS (left), WHL (middle) and DESI 
	(right).\label{fig:footprints}}
\end{figure*}  % +++++++++++++++++++++++++++++++++++++++++++++++++++++++++++++++

\subsection{ACT data} %================================================
\label{ssec:act}

We employ the ACT Compton $y$-map described in~\citet{madhavacheril20} and publicly accessible at the LAMBDA 
website\footnote{\url{https://lambda.gsfc.nasa.gov/product/act/act_dr4_derived_maps_get.cfm}.}. This map was also built 
with an ILC approach using \textit{Planck} individual frequency maps\footnote{Hence, \textit{Planck} and ACT tSZ maps are 
not completely independent. This, however, is not an issue in our study, as the choice of using both data sets has been made 
to better exploit all the available tSZ data. In the end, the higher resolution of ACT is still the main factor making a 
difference in the resulting angular Compton profiles.} up to $545\,\text{GHz}$ and ACT maps at 98 and $150\,\text{GHz}$. 
The two experiments are complementary in terms of angular sensitivity: whereas ACT has higher resolution than \textit{Planck} 
and provides superior quality data at small scales, \textit{Planck} large-scale data are free from the atmospheric noise 
affecting ACT maps. The limited available sky area of $\sim 2,100\,\text{deg}^2$ allowed the authors to apply an ILC implementation 
tailored to a projected 2-dimensional analysis, which is a novel approach compared to previous spherical harmonics-based 
approaches to analyze all-sky data. The final map covers two disjoint patches in the sky, labelled as BOSS North (BN, covering 
$1,633\,\text{deg}^2$) and Deep 56 (D56, covering $456\,\text{deg}^2$), where deep observational ACT data from 2014 and 2015 
were available. Unlike \textit{Planck} maps, ACT maps for these patches are provided as two-dimensional arrays in a \textit{plate carr\'ee}  
(equirectangular) projection, with a pixel size of $\sim0.5\,\text{arcmin}$. The resolution of these ACT maps is $\text{FWHM}=1.6\,\text{arcmin}$.

Although individual frequency maps underwent a process of source subtraction prior to their combination, possible residuals 
are still present in the final product. As no specific point-source mask is available for ACT tSZ maps, we revert to using the 
same combined Galactic and point source \textit{Planck} mask also for these maps, by projecting it on the plane areas covered by 
ACT\footnote{Although this mask may potentially miss some point sources entering the ACT footprint, it can be considered conservative
as the larger \textit{Planck} beam size would result in a larger masked area around each compact source. We also explicitly tested that 
the use of no mask at all produces negligible variations in the ACT profiles presented in Section~\ref{fig:profiles}, thus proving that the 
specific masking strategy is not a critical issue in a stacking analysis like the one presented in this paper.}.
The resulting masked maps are plotted in the two right panels of Fig.~\ref{fig:maps}.

%=================================================================================================
%================================    CLUSTER CATALOGUES     ====================================== 
%=================================================================================================

\section{Cluster catalogs}
\label{sec:catalogs}
We consider a composite cluster sample obtained by merging three independent catalogs, described individually in 
the following. The methodology we adopt for their combination is addressed in Sections~\ref{ssec:mcal} and~\ref{ssec:merging}. 
Notice that in the end we will only consider clusters with mass $M_{500}>10^{14}\,\text{M}_{\odot}$, as lower 
masses would not yield a significant detection in the stacks described in Section~\ref{ssec:stacks}.

\subsection{AMICO-KiDS DR3 Catalog} %================================================
\label{ssec:kids}
We employ the cluster catalog described in~\citet{maturi19}, which was obtained by running the Adaptive Matched 
Identifier of Clustered Objects (AMICO) algorithm~\citep{bellagamba18} on the third data release of the Kilo Degree 
Survey~\citep[KiDS-DR3,][]{dejong17}. KiDS-DR3 data provide photometric redshifts for $\sim48.7$ million sources 
over $447\,\text{deg}^2$. AMICO is based on a linear optimal matched filter which (in this particular application) 
exploits information on galaxy positions, $r$-band magnitude, and redshift distribution to build a three-dimensional 
map of the amplitude $A$ over the volume spanned by the galaxy catalog. The quantity $A$, which is evaluated with its 
associated variance $\sigma_A$, is related to the likelihood of finding a galaxy cluster; the location with the 
highest likelihood is then identified as the first cluster candidate. The signal associated with the latter is 
subsequently removed from the $A$ map, before re-evaluating the likelihood and searching for the second cluster candidate. 
The process is repeated until reaching a low-limit signal-to-noise ($S/N$) ratio $A/\sigma_A=3$; the final output catalog 
contains 12,939 clusters identified with $S/N>3$, over the redshift range $0.078<z<0.754$. We restrict our analysis to 
the 7,957 clusters with $S/N>3.5$; the resulting sample has a typical $\gtrsim95\%$ purity over the whole redshift 
range and a completeness $\gtrsim 90\%$ for $M_{500}>10^{14}\,\text{M}_{\odot}$ at $z<0.6$~\citep{maturi19}. 

The measured amplitude $A$ served as the primary mass proxy for each detection; cluster masses were assigned on the 
basis of an $A-M_{200}$ scaling relation, where the baseline mean values for the overdensity mass $M_{200}$ were computed 
from KiDS lensing data~\citep{bellagamba19}. As it is customary to employ $M_{500}$ as the mass definition in tSZ studies, 
we convert the AMICO masses into $M_{500}$ assuming a NFW profile and using the concentration model from~\citet{ishiyama21}. For each cluster we also 
generate a population of 200 random values of masses, normally distributed around the $M_{200}$ value and with a dispersion 
set by the available uncertainty $\sigma_{M_{200}}$; we convert each of these values to $M_{500}$ and adopt their root mean square (\textit{rms}) as our 
estimate for the uncertainty $\sigma_{M_{500}}$. The resulting, mean uncertainty on $\log_{10}{(M_{500}/M_{\odot})}$ for KiDS 
clusters is 0.194\,dex. When queried to match the ACT footprint, the KiDS catalog contributes with 3,318 clusters (806 
clusters with $M_{500}>10^{14}\,\text{M}_{\odot}$), all located in the ACT-BN patch (Fig.~\ref{fig:footprints}, left panel). 

\subsection{SDSS-DR12 WHL Catalog} %================================================
\label{ssec:whl}
The catalog described in~\citet{wen12} identified 132,684 galaxy clusters based on SDSS-DR8~\citep{aihara11} 
photometric data in the redshift range $0.05\leq z<0.80$. The catalog has a $\gtrsim94\%$ purity over the whole 
sample and a $\gtrsim95\%$ completeness for clusters with $M_{200}>10^{14} M_{\odot}$ at $z<0.42$; cluster masses 
were estimated based on their richness and optical luminosity. A more recent update of the catalog, based on 
SDSS-DR12~\citep{alam15}, is presented in~\citet{wen15}; hereafter we shall label such catalog as 
WHL\footnote{This acronym contains the initials of the authors of the original publication~\citep{wen12}. The catalog 
is publicly available at \url{https://vizier.u-strasbg.fr/viz-bin/VizieR?-source=J/ApJ/807/178}.}. The update 
not only took advantage of the improved SDSS data quality, which provided additional spectroscopic redshift information 
for a total of $\sim2.3$ million galaxies and allowed to improve cluster detection at high redshifts, but also exploited a 
better-defined mass proxy. The new catalog includes 25,419 additional clusters detected around bright galaxies at high 
redshift and provides updated redshift and richness estimates for the previously identified objects. Each cluster position 
is defined by the coordinates of its brightest cluster galaxy (BCG), and its redshift is estimated as the mean of the spectroscopic redshifts of 
member galaxies (when available). As for the mass estimation, the analysis firstly defined a calibration sample by 
merging existing cluster samples with mass proxies based on X-ray data~\citep{vikhlinin09,mantz10,piffaretti11,takey14} 
or tSZ-data~\citep{hasselfield13,planck_15_xxvii}. Common clusters across different catalogs were used to yield a 
homogeneous mass definition throughout the composite sample, by scaling it to the definition adopted in~\citet{vikhlinin09}. 
The final calibration sample consisted of 1191 clusters overlapping with available SDSS data, with 
mass\footnote{Unlike~\citet{wen12}, where masses are quoted as $M_{200}$, the work in~\citet{wen15} adopts the $M_{500}$ definition.} 
$M_{500}>0.3\times10^{14}\,\text{M}_{\odot}$ and redshift $0.05<z<0.75$. For these clusters, the total $r$-band luminosity 
within $R_{500}$, corrected by a redshift-dependent factor, was found to be well correlated with the cluster mass $M_{500}$ 
with a scatter of 0.17 dex. The associated scaling relation can be applied to estimate the mass of all clusters in the 
updated WHL catalog.

For our analysis, we employ the updated WHL catalog with a total of 158,103 clusters spanning the mass range 
$M_{500}\in[10^{12.3},10^{15.5}]\,\text{M}_{\odot}$ and the redshift range $[0.03,0.80]$, about $77\%$ of which with 
spectroscopic redshift information. As there is no mass error estimate availabe for individual clusters, we evaluate a mean 
uncertainty on $M_{500}$ as follows. We consider clusters in the calibration sample and convert their measured richness 
into mass $M_{500}^{\rm (scal)}$, adopting the same scaling relation that was employed in~\citet{wen15} to compute the 
mass estimates for the updated cluster catalog. For the calibration sample clusters, the independent mass estimate 
$M_{500}^{\rm (lit)}$ from the literature is also available\footnote{These mass estimates $M_{500}^{\rm (lit)}$
are the ones obtained from the 
aforementioned list of X-ray and tSZ studies, after the rescaling performed by~\citet{wen15} to homogenize the mass definition
in the calibration sample.}; we consider then the scatter values 
$M_{500}^{\rm (scal)}-M_{500}^{\rm (lit)}$ for the calibration clusters and take their {\it rms} as the common mass uncertainty 
for the WHL catalog we use in our analysis. The result is an uncertainty of 0.187\,dex in $\log_{10}{(M_{500}/M_{\odot})}$, 
which is slightly more conservative than the value quoted for the scatter of the scaling relation in~\citet{wen15}. The query 
for matching the ACT footprints yields 27,367 clusters, of which 20,967 in the BN patch and 6,400 in the D56 patch 
(Fig.~\ref{fig:footprints}, central panel); after applying the mass cut $M_{500}>10^{14}\,\text{M}_{\odot}$, the numbers 
are 18,597, 14,186 and 4,411, respectively. 

\subsection{DESI-DR8 Catalog} %================================================
\label{ssec:desi}
This cluster catalog was obtained directly from galaxy samples in the DESI Legacy Imaging 
Surveys~\citep{dey19} Data Release 8 (DESI-DR8 hereafter). The catalog production is described 
in~\citet{yang21} and is based on the updated version of the halo-based group/cluster finder presented 
in~\citet{yang05} and later employed in~\citet{yang07}. In this case, the cluster mass is computed 
based on the measured cluster luminosity. The group finder follows an iterative approach: at each stage, 
the cumulative group luminosity distribution is computed from the known luminosity of member galaxies; 
abundance matching with the cumulative halo mass function allows assigning to each group a tentative mass, 
which in turn allows the update of the membership information. The process starts by assuming that each galaxy is 
a group candidate, and continues until convergence in the galaxy membership information and in the derived 
mass-to-luminosity ratios. In this case, the estimated mass is defined for an overdensity $\Delta=180$ 
with respect to the mean matter density of the Universe at the cluster redshift ($M_{180,\text{m}}$), 
and the cluster position is assigned to its geometrical, luminosity-weighted center. 
The authors first tested this halo finder on a mock galaxy catalog generated from the ELUCID 
simulation~\citep{wang16}, comparing the results with the output of a traditional Friend-of-Friends 
algorithm~\citep{davis85}: in $\sim90\%$ of groups with mass $M_{180,\text{m}}\gtrsim10^{12.5}h^{-1}\text{M}_{\odot}$, 
the halo finder correctly identified more than 60\% of the member galaxies, with a quoted mean mass 
uncertainty of 0.2 dex for masses $M_{180,\text{m}}\gtrsim10^{13.5}h^{-1}\text{M}_{\odot}$; the purity 
resulted $>90\%$ for groups with mass $M_{180,\text{m}}\gtrsim10^{12}h^{-1}\text{M}_{\odot}$ and reached 
$\sim100\%$ for $M_{180,\text{m}}\gtrsim10^{14.5}h^{-1}\text{M}_{\odot}$. 

The authors subsequently applied the group finder to DESI-DR8 to yield positions, redshifts and masses 
for $\sim92$ million objects, the majority of which being low mass groups with less than 3 member galaxies. 
For the purpose of the present analysis we clearly restrict the sample to the most massive objects; 
after converting the mass to the $M_{500}$ definition with the concentration model from~\citet{ishiyama21}, 
we apply the mass cut $M_{500}>10^{14}\,\text{M}_{\odot}$, resulting in a total 
of 110,908 objects spanning the redshift range $z\in[0.02,0.97]$. Given our lower mass cut, and the lack 
of error estimates on individual cluster masses, we can adopt the quoted value 0.2\,dex as the common uncertainty 
on $\log_{10}{(M_{500}/M_{\odot})}$. We find 13,018 clusters overlapping with ACT maps, 
10,253 in the BN patch and 2,765 in the D56 patch (Fig.~\ref{fig:footprints}, right panel).

\begin{figure*} % +++++++++++++++++++++++++++++++++++++++++++++++++++++++++++++++
\includegraphics[trim= 0mm 0mm 0mm 0mm, scale=0.31]{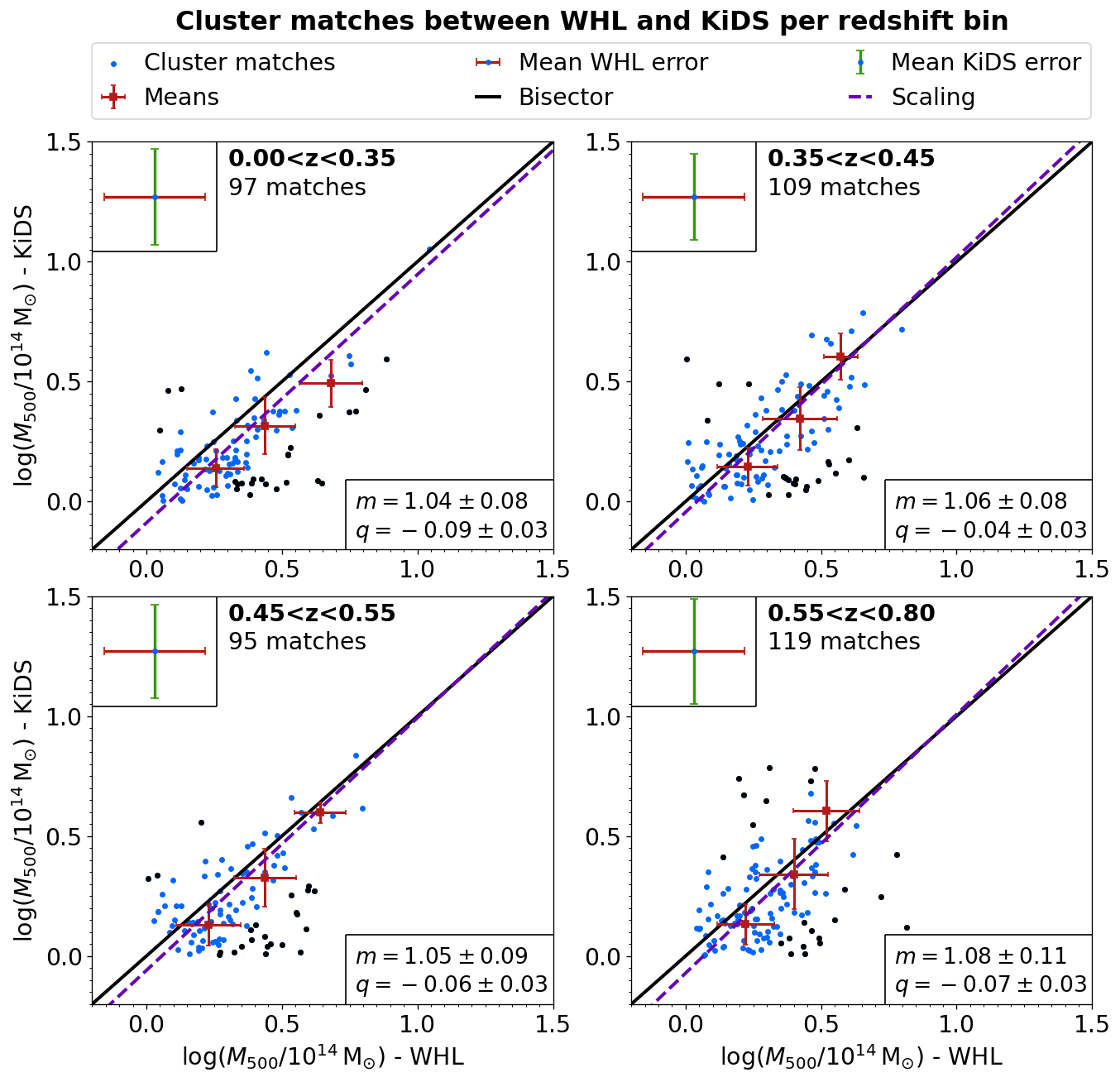}
\includegraphics[trim= 0mm 0mm 0mm 0mm, scale=0.31]{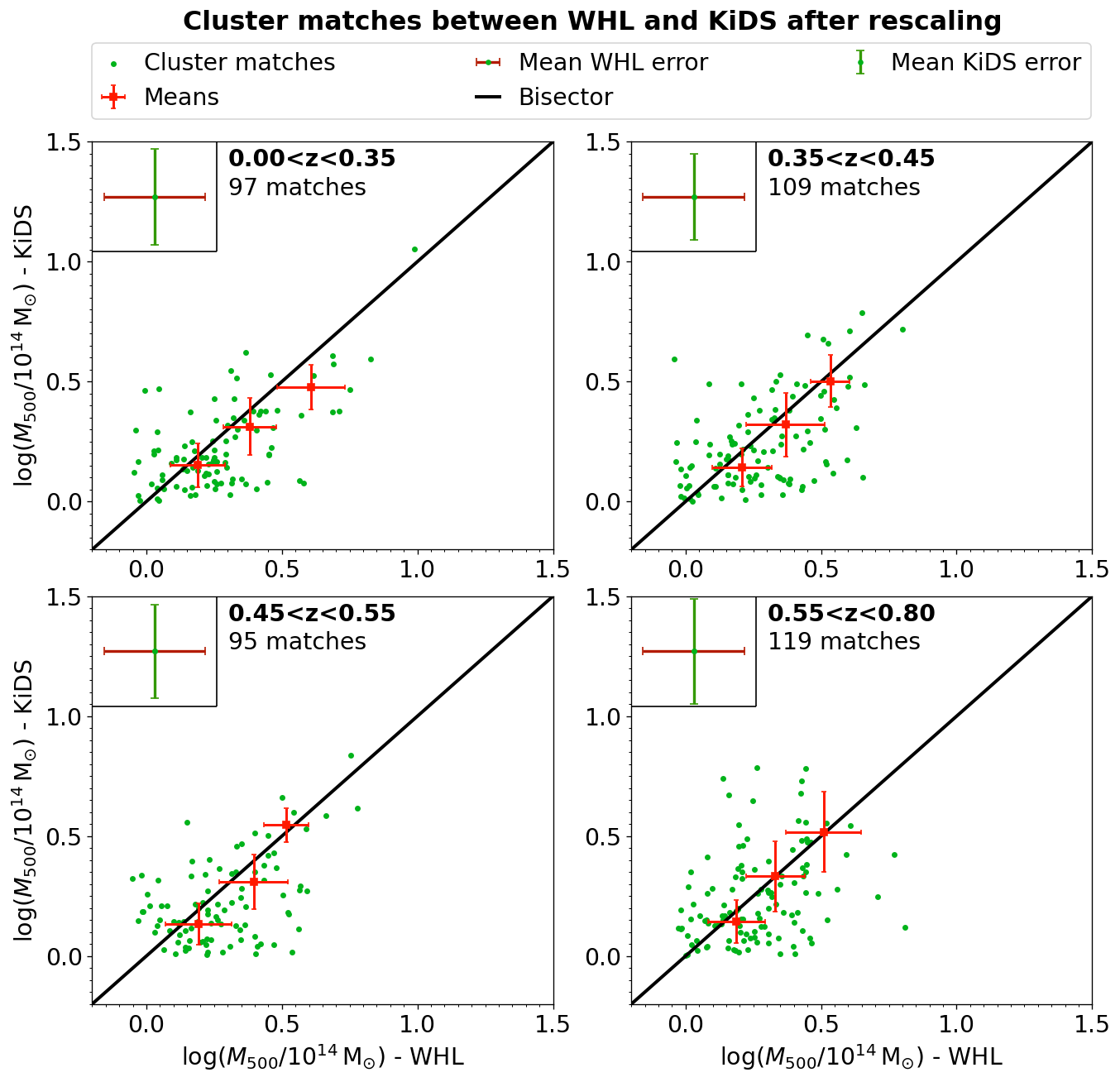}
	\caption{Summary of the mass calibration analysis described in Section~\ref{ssec:mcal}, for the 
	case of KiDS-WHL comparison. \textit{Left panels}: for each redshift bin, we plot a comparison 
	between masses from the two catalogs for the retrieved cluster matches using dots; the bisector
	$y=x$ is shown for comparison as a solid black line. The mean error bars for the mass estimates of 
	each match are shown in a separate box in the top-left corner. The red squares and their error bars 
	represent the mean and standard deviation for the masses of matches binned along the bisector, 
	and serve to better visualize the trend of the points 
	compared to the bisector. The purple dashed line represents the linear regression adopted to scale 
	WHL masses to KiDS masses according to Eq.~\eqref{eq:linfit}, with the scaling parameters 
	quoted in the low-right corner of each plot; matches plotted in black were considered outliers and 
	not included in the linear fit. \textit{Right panels}: same as for the left panels, but after 
	applying the rescaling to WHL masses; the scaling overall improves the agreement between the 
	masses of matched clusters.\label{fig:matches_kw}}
\end{figure*} % +++++++++++++++++++++++++++++++++++++++++++++++++++++++++++++++

\subsection{Mass calibration} %================================================
\label{ssec:mcal}
With the aim of employing the largest possible statistics in our study of the cluster pressure 
profile, we merge together the three catalogs described above. In order to avoid potential 
biases in the subsequent analysis, care needs to be taken in this operation to ensure that the 
cluster mass definition is consistent across all catalogs we will combine. Although in 
every case we adopt the $M_{500}$ definition, masses from different catalogs are based on different 
observables, scaling relations and methodologies.

We then search for cluster matches between pairs of catalogs, and compare the associated 
mass values. Our matching criterion is purely positional: two clusters in different catalogs are 
considered to be the same object if their projected linear separation on the sky $\Delta r$ and their redshift 
separation $\Delta z$ satisfy the conditions $\Delta r <0.5\,\text{Mpc}$ and $\Delta z < 0.05\,(1+\bar{z})$, 
where $\bar{z}$ is the mean redshift of the two clusters (this expression for the limit in $\Delta z$ takes 
into account the possible higher errors on $z$ measurements at higher redshifts). This query is applied to the 
full catalogs (not restricted to the ACT footprint), but keeping the constraint 
$M_{500}>10^{14}\,\text{M}_{\odot}$. Besides, the search for matches is conducted in four disjoint
redshift intervals, namely $[0.00,0.35]$, $[0.35,0.45]$, $[0.45,0.55]$, $[0.55,0.80]$; the different 
interval size takes into account the non-uniform redshift distributions of our cluster catalogs, 
and allows to yield a comparable number of matches in each bin. The reason why we perform this analysis
in different redshift bins is to allow for a possible $z$ evolution in the agreement of the mass 
estimates from different catalogs. The number of chosen redshift bins is rather arbitrary; in our case, 4 bins 
are enough to show any redshift trend while retaining a number of matches per bin which is large enough 
for the subsequent analysis. 

\begin{figure*} % +++++++++++++++++++++++++++++++++++++++++++++++++++++++++++++++
\includegraphics[trim= 0mm 0mm 0mm 0mm, scale=0.31]{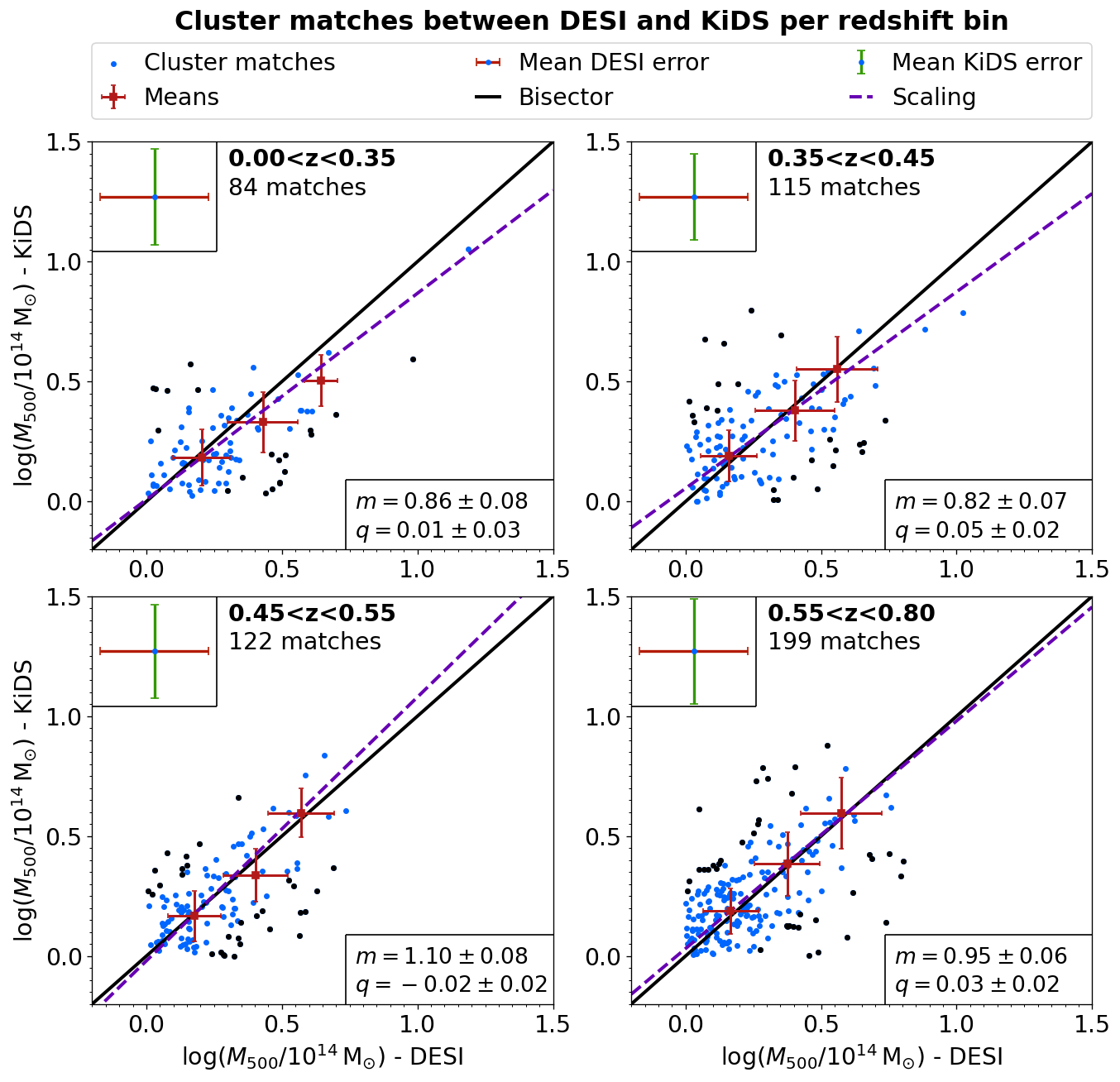}
\includegraphics[trim= 0mm 0mm 0mm 0mm, scale=0.31]{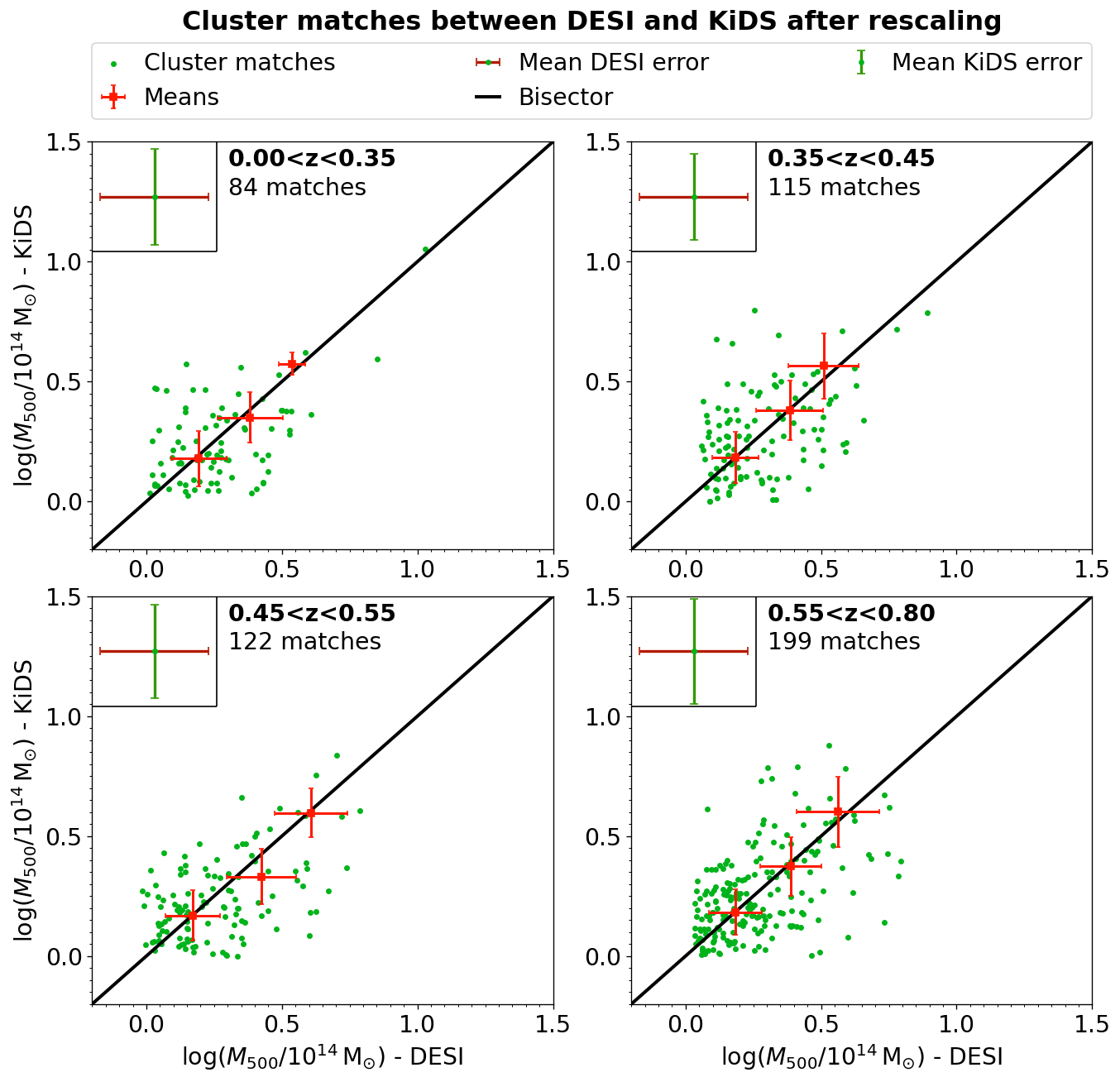}
	\caption{Same as in Fig.~\ref{fig:matches_kw}, but for the comparison between KiDS and DESI.\label{fig:matches_kd}}
\end{figure*}

\begin{figure*} % +++++++++++++++++++++++++++++++++++++++++++++++++++++++++++++++
\includegraphics[trim= 0mm 0mm 0mm 0mm, scale=0.31]{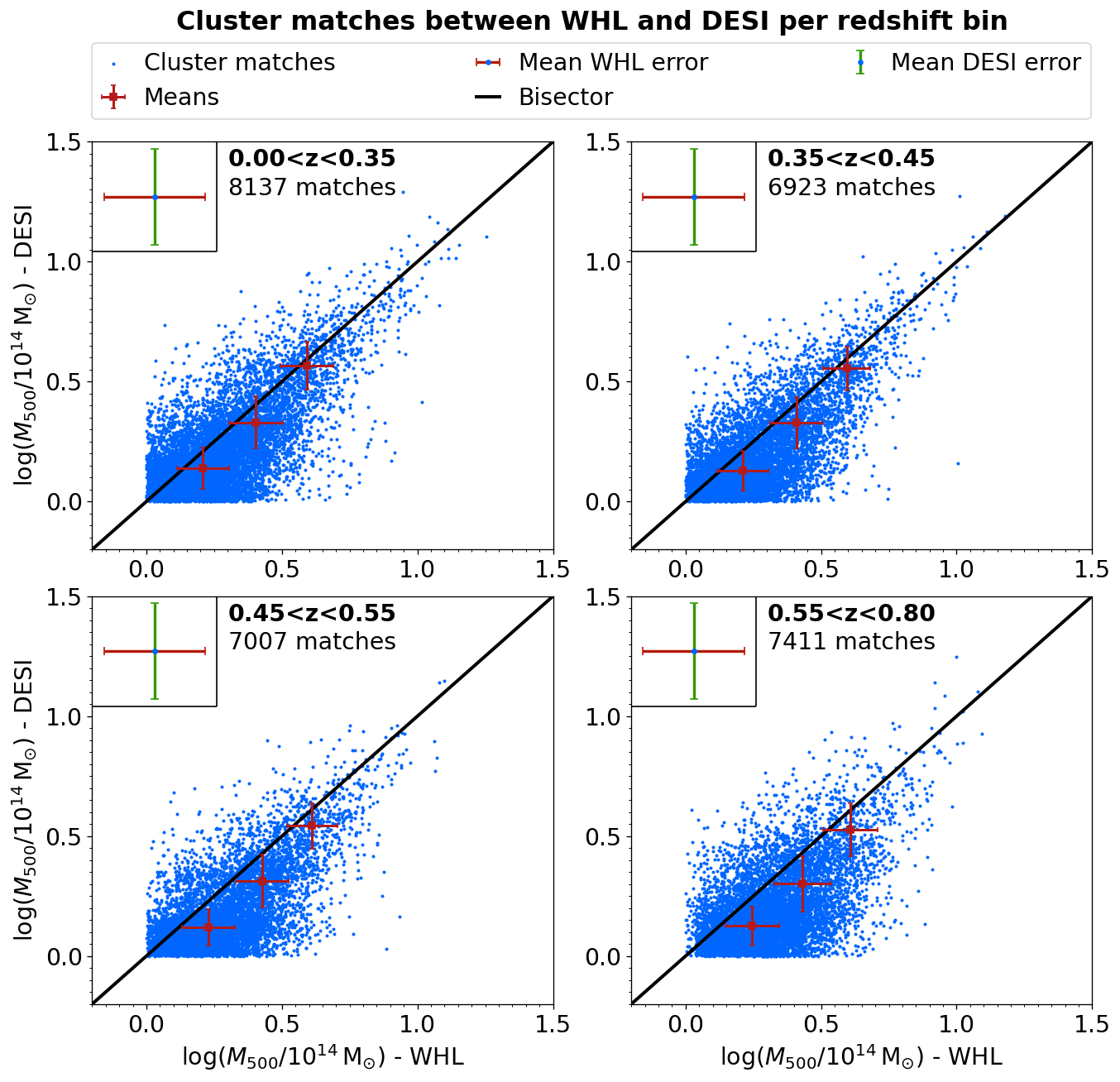}
\includegraphics[trim= 0mm 0mm 0mm 0mm, scale=0.31]{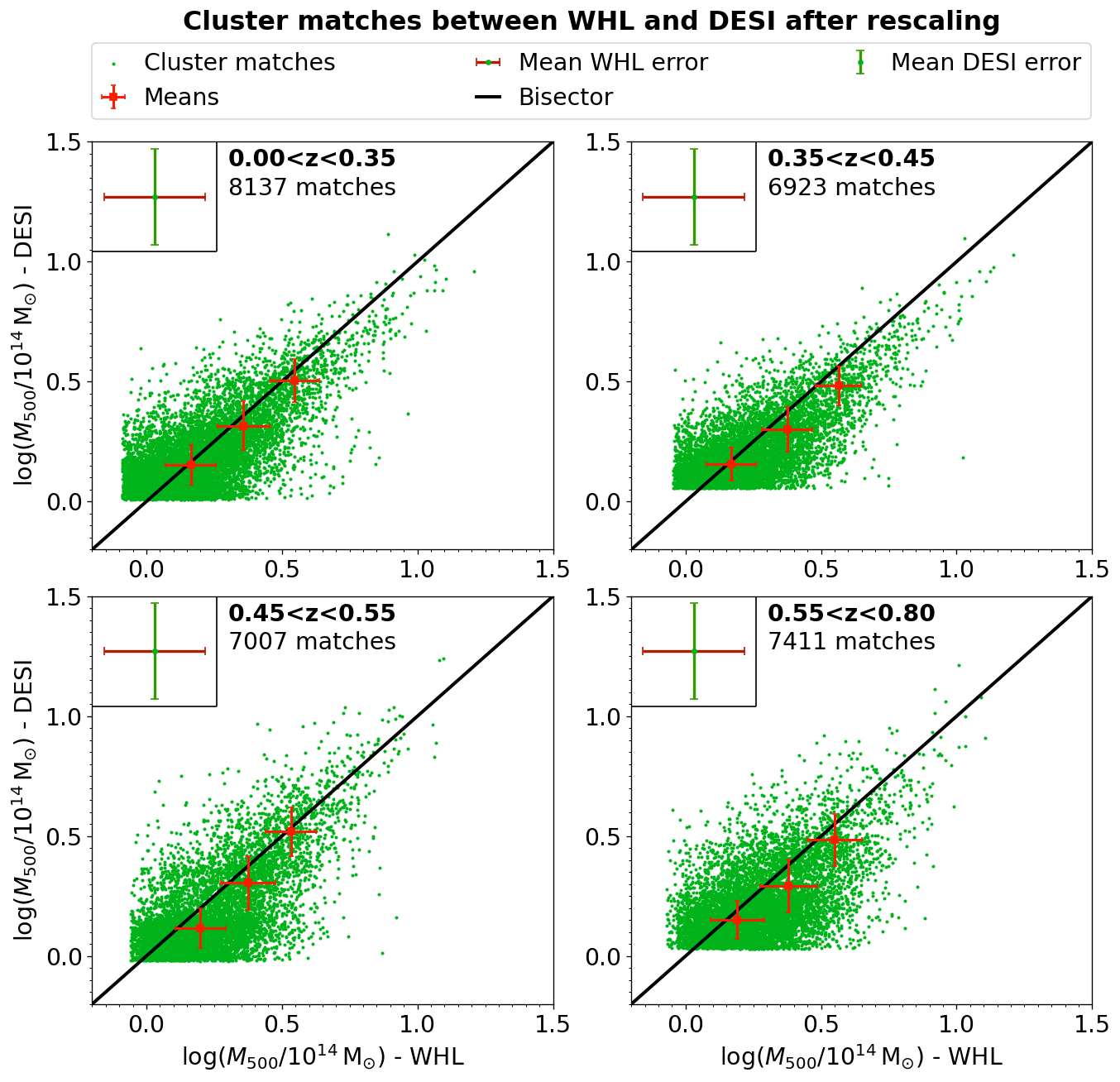}
	\caption{Similar to Fig.~\ref{fig:matches_kw} and~\ref{fig:matches_kd}, but for the comparison
	between DESI and WHL. In this case no direct linear regression was employed to scale WHL masses  
	to match DESI values or vice-versa; hence, the left panels do not include any fitting results. The right panels show the comparison
	between DESI and WHL after being both independently scaled to match the mass values from KiDS. Even in 
	this case, the scaling yields a better agreement between the masses of matched clusters from the two catalogs.\label{fig:matches_wd}}
\end{figure*}

The results of the query for cluster matches are shown in the left panels of Figs.~\ref{fig:matches_kw},~\ref{fig:matches_kd} 
and~\ref{fig:matches_wd} for the combinations KiDS-WHL, KiDS-DESI and WHL-DESI respectively. 
Each plot refers to one redshift interval and shows the comparison between the masses of the matched 
clusters as obtained from the corresponding catalogs, together with the bisector $y=x$ that represents 
the ideal case of equality. In order to avoid excessive clutter in these plots, the mean error bars 
associated with the mass estimates are shown in a separate box in each top-left corner. 
We typically find around 100 cluster matches per redshift bin when considering the KiDS catalog, and 
typically more than 7000 matches for the combination of the larger catalogs WHL and DESI. In all 
plots the masses are quoted in logarithmic units of $10^{14}\,\text{M}_{\odot}$, which is a convenient 
choice as it sets our low mass threshold at the origin.

As expected, the intrinsic scatter of the scaling relations adopted to estimate each cluster mass, 
combined with the uncertainty in the measurement of the associated mass proxy, determines a visible 
scatter of the points around the bisector. However, if the distribution of points also shows any 
clear trend deviating from the bisector, the mass estimates we are adopting could be systematically 
biased. This effect is better visualized by splitting the points into different mass bins and 
just considering the associated mean masses computed for each catalog. In principle, the bins 
could be defined along each of the two coordinate axes, by choosing the corresponding catalog as 
a reference. In this case, however, the mass estimates from all catalogs have significant error 
bars, which makes this strategy for splitting the points unreliable. In order not to make a preferential 
choice for any catalog, we group instead the points in bins of equal separation from the line $y=-x$, 
or in other words, we bin the points along the bisector, with the boundaries for each bin being the 
lines $y=-x+\Delta$ (with $\Delta\simeq0.2$). In Figs.~\ref{fig:matches_kw} to~\ref{fig:matches_wd} 
the mean masses of the 
points in each bin are shown as red squares, while their standard deviations are quantified by the 
associated error bars. 

 % +++++++++++++++++++++++++++++++++++++++++++++++++++++++++++++++

In general, considering the mass uncertainties shown in the top left corner of each plot, individual 
matches are compatible with the bisector within $1\sigma$; the mean points, however, reveal in 
some cases a trend or an offset that deviates from the bisector. It is then meaningful to correct for 
this effect following the procedure adopted in~\citet{wen15}, where the same issue was encountered 
when building the calibration cluster sample by merging pre-existing independent catalogs. We then choose a 
reference catalog for the mass definition and re-scale the masses of the other two catalogs to match the 
values from the reference one. Masses for the KiDS clusters are obtained via a richness-mass scaling relation 
calibrated on weak lensing measurements, which directly probe the cluster mass without any assumption 
on their physical state; hence, they can be considered more reliable than the masses quoted in WHL, 
which rely on the tSZ and X-ray masses 
defined in the calibration sample, or in DESI, which relies on the abundance matching between halo 
luminosity and mass function. We shall therefore take KiDS as the reference catalog, and scale the 
masses in the other catalogs accordingly. For each redshift bin the scaling has the form:
\begin{equation}
	\label{eq:linfit}
	\log_{10}{\tilde{M}_{\rm KiDS}} =m_{\text{X},i}\,\log_{10}{\tilde{M}_{\rm X}} + q_{\text{X},i},
\end{equation}
where $\tilde{M} \equiv M/(10^{14}\,\text{M}_{\odot})$, $X$ is either WHL or DESI, $i$ selects the redshift bin, 
and the parameters $m$ and $q$ are obtained via linear regression on the masses of the associated matches. 

In this context results from the linear regression can easily be biased by outliers. The latter could be a result of 
spurious matches, or of the combination of the intrinsic scatters in the scaling relations adopted to derive the 
cluster mass in the two catalogs (combined with the uncertainties in the measurements of the mass proxies themselves).
In order for our fits not to be biased by these outliers, we identify and remove them according to the 
following procedure. In each of 
the mass bins bounded by the $y=-x+\Delta$ edge lines, we compute the orthogonal distance $d$ of each 
point to the bisector, and evaluate the associated standard deviation $\sigma_d$; for each bin, we then 
discard all points for which $d>2\,\sigma_d$ from the linear regression analysis. These outliers are 
shown in black color in the left panels of Figs.~\ref{fig:matches_kw},~\ref{fig:matches_kd} and~\ref{fig:matches_wd}. 
For the remaining points, due to the large error bars in both axes, an ordinary least square fit would 
not be suitable in our case; we adopt instead an orthogonal distance regression method using the \texttt{SciPy ODR} 
package\footnote{\url{https://docs.scipy.org/doc/scipy/reference/odr.html}.}. The resulting linear scalings 
are plotted as dashed purple lines in Figs.~\ref{fig:matches_kw} and~\ref{fig:matches_kd}, where a 
box in each low-right corner reports the best-fit parameters. 
We notice that in all cases the intercept satisfies $|q|<0.1$; this result already suggests a broad 
consistency between mass estimates from different catalogs, as in the ideal case of $m=1$ the intercept 
quantifies the mean offset between the two mass definitions. In the case of the KiDS-WHL comparison, the 
slope is always compatible with 1, while we find larger deviations for the KiDS-DESI case. However, we 
stress that the mean shift in mass resulting from applying the scaling to the cluster samples we use in the 
subsequent analysis (not just the sub-samples of matched clusters) is equal to 0.06 dex for WHL 
and 0.02 dex for DESI, with maximum shifts of 0.09 dex and 0.16 dex respectively. Hence, our mass 
correction is always below the initial mass uncertainty for individual clusters, and on average much smaller. 

 % +++++++++++++++++++++++++++++++++++++++++++++++++++++++++++++++

We can now use Eq.~\eqref{eq:linfit} with the associated best-fit parameters to 
rescale the masses of WHL and DESI in each redshift bin. As DESI is the only catalog that extends at 
$z>0.8$, its highest redshift clusters have no matches with KiDS; hence, these cluster masses are not rescaled. 
The right panels of Figs.~\ref{fig:matches_kw} to~\ref{fig:matches_wd} show once more the comparison of the 
masses for cluster matches, but this time after applying the mass rescaling; again, we overplot to the points 
the bisector and the mean masses with standard deviations for the $y=-x+\Delta$ bins. As already mentioned, 
our mass correction is overall small, so these plots resemble the ones shown in the left-hand panels. Still, 
it is possible to appreciate how the agreement between the mean-mass points and the bisector has marginally 
improved. It is interesting to consider the comparison between WHL and DESI in Fig.~\ref{fig:matches_wd}; we 
notice that, in this case, no fitting function is overplotted to the plots in the left panel, as we did not 
consider a direct rescaling of WHL masses to match DESI values, or vice-versa. Instead, both catalogs were 
re-scaled independently to match KiDS mass values in their associated matches. The comparison between the 
rescaled WHL and DESI masses is shown in the right panel of Fig.~\ref{fig:matches_wd}. Again, the mean points 
over the $y=-x+\Delta$ bins show a better agreement with the bisector, thus corroborating the consistency of 
our approach for mass recalibration.  

In Appendix~\ref{app:systematics} we quantify what impact this mass rescaling has on the final results of our
study. We will repeat the stacking analysis and the parameter estimation on a different version of our cluster 
sample, obtained by merging the three catalogs without any explicit mass rescaling. We will show that the 
differences in the final results are indeed much smaller than the statistical uncertainties we quote on our parameter
estimates.

\begin{figure*} % +++++++++++++++++++++++++++++++++++++++++++++++++++++++++++++++
\includegraphics[trim= 0mm 0mm 0mm 0mm, scale=0.25]{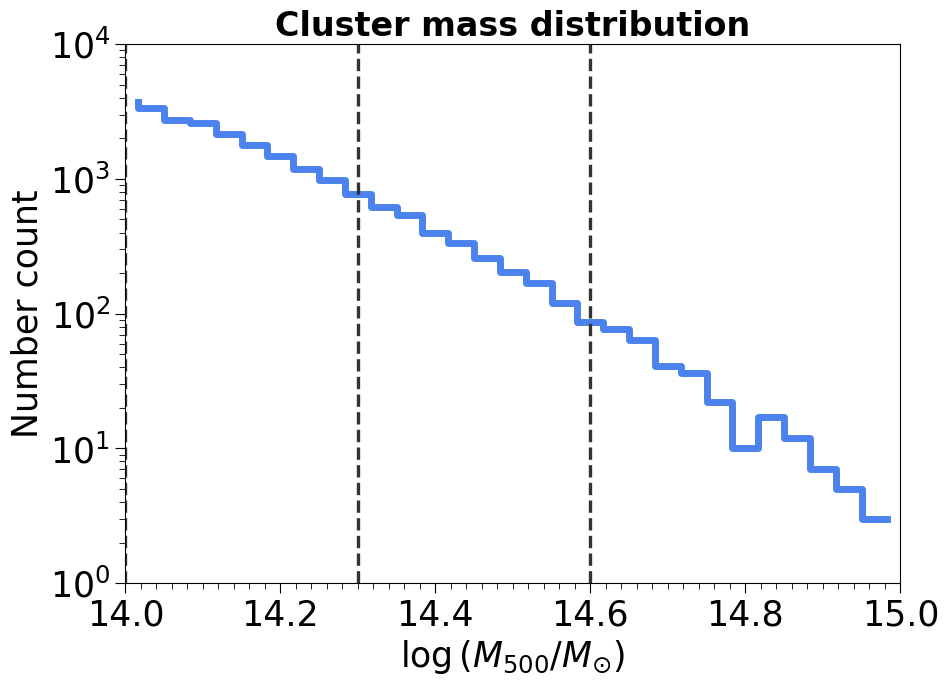}
\includegraphics[trim= 5mm 0mm 0mm 0mm, scale=0.25]{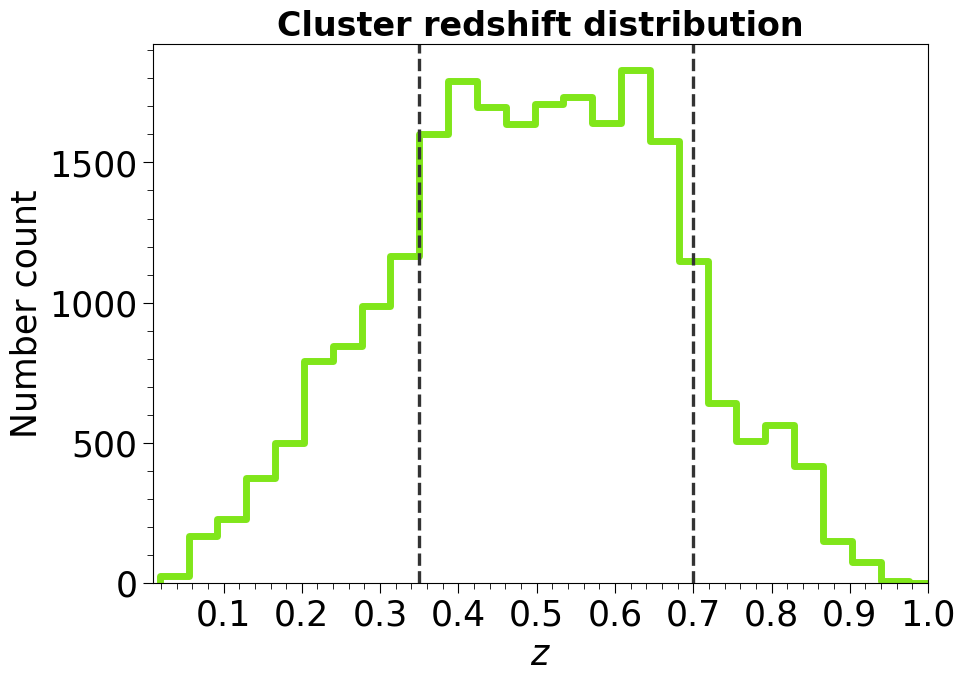}
\includegraphics[trim= 0mm 0mm 0mm 0mm, scale=0.25]{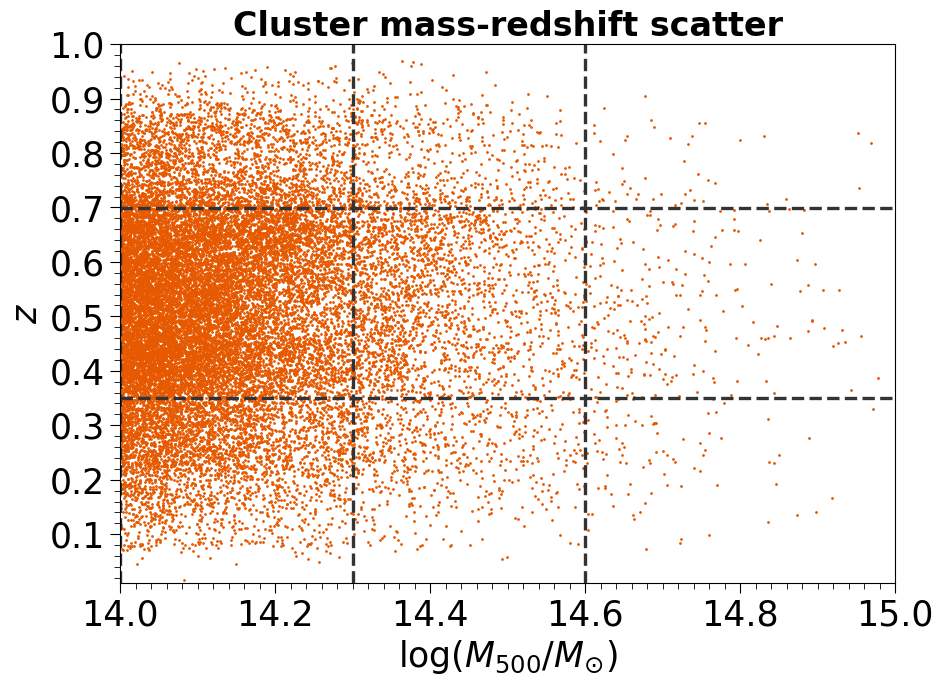}
\caption{Statistics of the full cluster sample employed in this study. Left and middle panels show the cluster 
	number distributions in mass and redshift, respectively, while the right panel shows the joint distribution 
	of the two variables. In each panel, the dashed lines mark the boundaries of the chosen mass and/or redshift 
	bins, as detailed in Section~\ref{ssec:binning}.\label{fig:distrib}}
\end{figure*} % +++++++++++++++++++++++++++++++++++++++++++++++++++++++++++++++

\subsection{The joint cluster catalog} %================================================
\label{ssec:merging}

The mass rescaling described in the previous section is applied to the WHL and DESI catalogs, including 
also clusters with mass lower than $10^{14}\,\text{M}_{\odot}$; the cut $M_{500}>10^{14}\,\text{M}_{\odot}$
is then applied subsequently to the mass-rescaled catalogs. This allows clusters with initial mass 
below our threshold to be included in the chosen sample if the rescaling increases their $M_{500}$ above 
$10^{14}\,\text{M}_{\odot}$. All catalogs are then queried to match the footprint of the two ACT 
patches, which results in 806, 15,330 and 14,477 clusters in KiDS, WHL and DESI respectively.  
We then remove cluster repetitions; the latter are identified adopting the same positional 
matching criteria detailed in Section~\ref{ssec:mcal}. From WHL and DESI we remove all clusters that 
match KiDS positions while keeping the corresponding entries in the KiDS catalog; as the latter has 
substantially lower statistics compared to WHL and DESI, this choice ensures no KiDS cluster is discarded.
For each remaining repetition between WHL and DESI, we keep the entry from one of these catalogs
based on a random choice with equal probability. Overall, we remove 3,437 clusters from WHL and 3,356 
clusters from DESI. 

These cleared catalogs are then merged together to obtain the final sample we employ for our 
study, which contains 23,820 clusters in total. The sample spans the mass range 
$M_{500}\in [10^{14.0},10^{15.1}]\,\text{M}_{\odot}$ and the redshift range $z\in[0.02,0.97]$;
a summary of the number of clusters for different source catalogs and ACT patches is reported in 
Table~\ref{tab:dataset}. In the end, WHL 
and DESI contribute with a comparable number of clusters in each ACT patch. The mass 
and redshift distributions of the final cluster sample, combining objects from both ACT 
patches, are shown in Fig.~\ref{fig:distrib}.

\begin{table}
\centering
\caption{Summary of the contributions from the KiDS, WHL and DESI catalogs to the final cluster sample employed in this analysis. For each catalog we report the number $N$ of clusters overlapping with the ACT footprint, the numbers $N_{\rm BN}$ and $N_{\rm D56}$ of clusters located in the BN and D56 patches respectively, and their redshift spans. These values are referred to the WHL and DESI samples obtained after mass recalibration (Section~\ref{ssec:mcal}) and the removal of overlapping objects (Section~\ref{ssec:merging}).}
\label{tab:dataset}
\renewcommand{\arraystretch}{1.3}
\begin{tabular}{c|cccc}
\hline
Catalog & $N$ & $N_{\rm BN}$ & $N_{\rm D56}$ & $\left[z_{\text{min}},z_{\text{max}}\right]$ \\
\hline
KiDS  & 806 & 806 & 0 & [0.08,0.74] \\
WHL   & 11893 & 8937 & 2956 & [0.04,0.78] \\
DESI  & 11121 & 8735 & 2386 & [0.02,0.97] \\
\hline
Total & 23820 & 18478 & 5342 & [0.02,0.97] \\
\hline
\end{tabular}
\end{table}
 %----------------------------------------------------------------

%=================================================================================================
%=================================================================================================

%=================================================================================================
% =================================   METHODOLOGY  ===============================================	
%=================================================================================================

\section{The mean cluster Compton profiles}
\label{sec:methodology}
In this section we describe the methodology we adopt to measure the mean angular $y$ profiles 
from our cluster samples, and discuss the results. 

\subsection{Cluster binning}
\label{ssec:binning}

The main goal of this study is to explore the dependence of the cluster pressure profile 
on mass and redshift. To this aim, we define a set of three bins in $\log_{10}(M_{500})$, 
bounded by the values $[14.0,14.3,14.6,15.1]$, and three bins in $z$, bounded by the values 
$[0.00,0.35,0.70,1.00]$. We then split our cluster catalog into a set of nine sub-samples, 
each belonging to the combination of a mass bin and a redshift bin. The last panel in 
Fig.~\ref{fig:distrib} is a scatter plot showing the distribution of clusters in the 
$(\log_{10}{(M_{500})},z)$ plane; this plot shows that there is no strong correlation between 
the two variables, most likely as a result of the catalogs being nearly complete above the 
chosen mass cut, so that the choice of a redshift bin is independent of the choice of 
a mass bin and it is meaningful to adopt the same set of $M_{500}$ boundaries across the 
whole $z$ range. We also consider the marginalized samples obtained by joining all 
redshift values for each mass bin, and vice-versa. Finally, we include the full cluster 
sample as the marginalized case over both variables. We obtain this way a total of 16 
different cluster samples (the full sample, 9 disjoint sub-samples and 6 marginalized 
cases), which we employ separately to measure the mean Compton parameter 
profile. Hereafter, we shall refer to them as $M$-$z$ bins. 

Our choice for the number of bins in $M_{500}$ and $z$ ensures we have enough statistics for 
the stacking analysis in each $M$-$z$ bin, while at the same time allowing to investigate possible 
dependences in mass and redshift. We follow the same choice adopted in~\citetalias{gong19} and 
consider three bins in redshift, almost equally spaced over the range spanned by the cluster 
sample. We also adopt three bins for the 
mass; however, given the steep decrease in the number of objects for increasing masses, in this 
case the top $M_{500}$ bin is chosen wider in order to retain sufficient statistics for the 
stacking analysis described in Section~\ref{ssec:stacks}. Clearly, in the end we still have a 
different number of clusters in each $M$-$z$ bin; this, however, is not an issue in our analysis. 
The formalism we adopt to model our measurements, which is described in 
Section~\ref{ssec:application}, naturally accounts for the number of objects and the mass 
span of each case. The final number of clusters included in each $M$-$z$ bin is 
reported in Table~\ref{tab:stats}.

\begin{figure*} % +++++++++++++++++++++++++++++++++++++++++++++++++++++++++++++++
\includegraphics[trim= 0mm 0mm 0mm 0mm, scale=0.53]{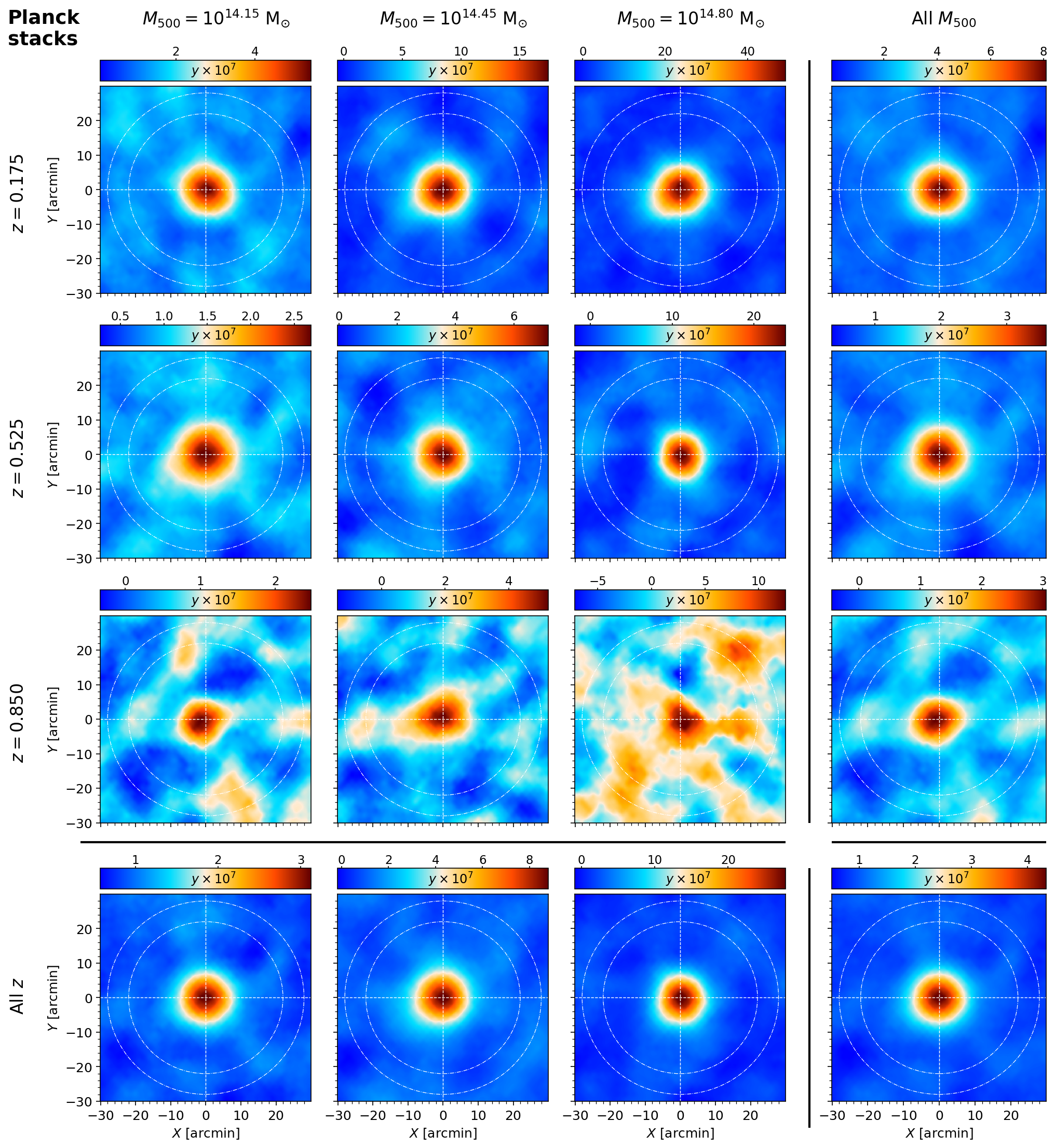}
\caption{Stacking results for the \textit{Planck} $y$-map. Each row (column) represents a 
	selected redshift (mass) bin, with the last one showing the marginalized case over 
	the full redshift (mass) range. Notice that the color scale reported above each panel 
	is not the same and is chosen to saturate each stack. In each panel, the nominal center 
	of the stack is marked by dashed white lines, while dot-dashed lines mark the inner 
	and outer boundaries of the annulus employed to estimate the mean background 
	level.\label{fig:planck_stacks}}
\end{figure*}  % +++++++++++++++++++++++++++++++++++++++++++++++++++++++++++++++

\begin{figure*} % +++++++++++++++++++++++++++++++++++++++++++++++++++++++++++++++
\includegraphics[trim= 0mm 0mm 0mm 0mm, scale=0.53]{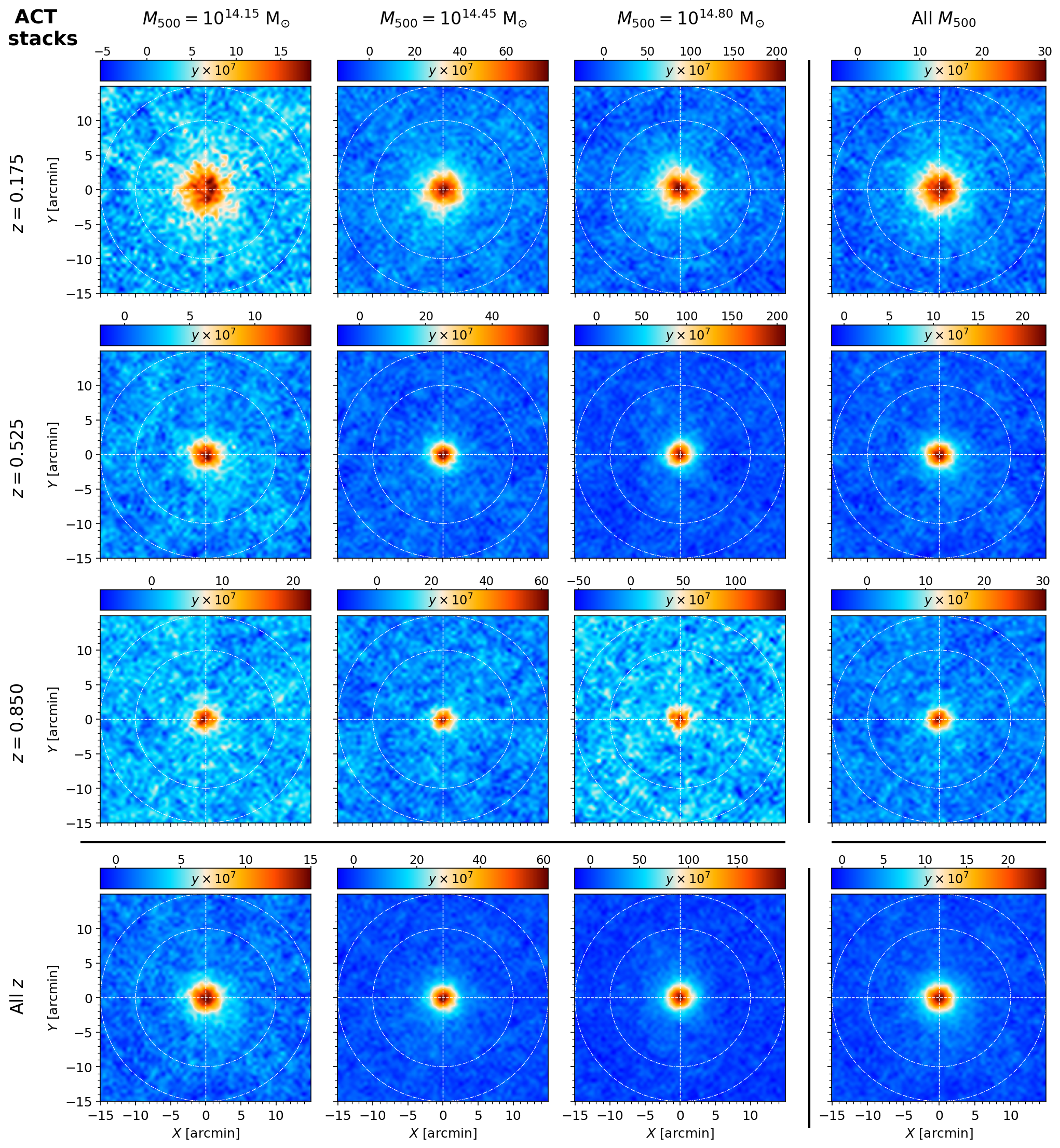}
\caption{Same as in Fig.~\ref{fig:planck_stacks} but for the ACT $y$-map.\label{fig:act_stacks}}
\end{figure*} % +++++++++++++++++++++++++++++++++++++++++++++++++++++++++++++++

\begin{figure*}  % +++++++++++++++++++++++++++++++++++++++++++++++++++++++++++++++
\includegraphics[trim= 0mm 0mm 0mm 0mm, scale=0.28]{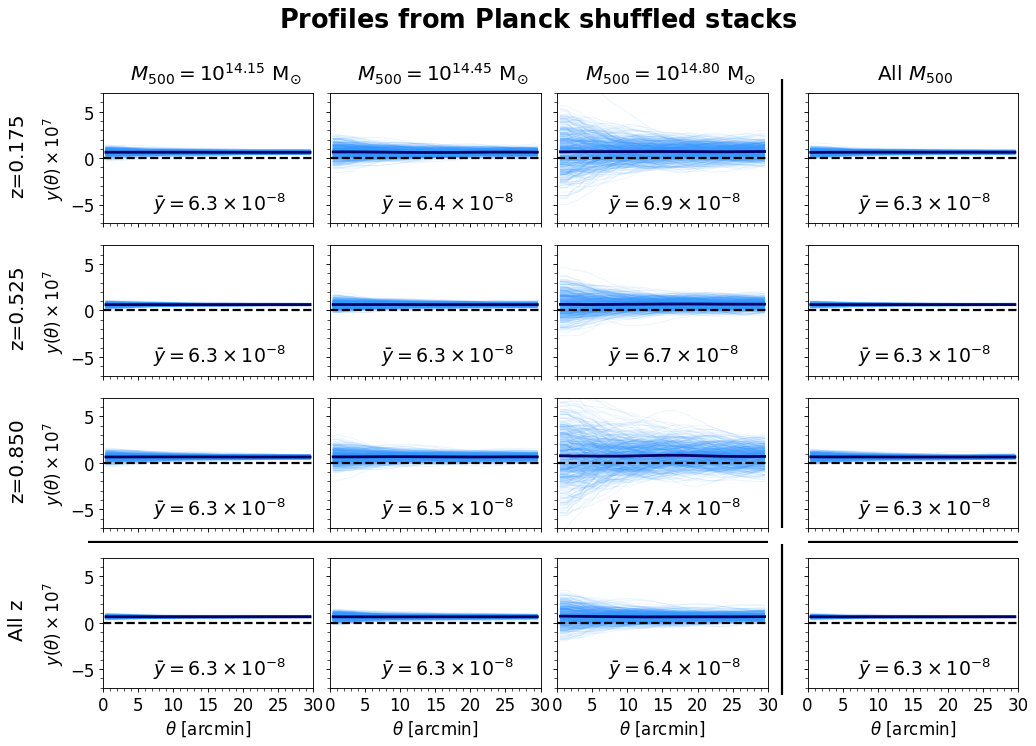} 
\includegraphics[trim= 0mm 0mm 0mm 0mm, scale=0.28]{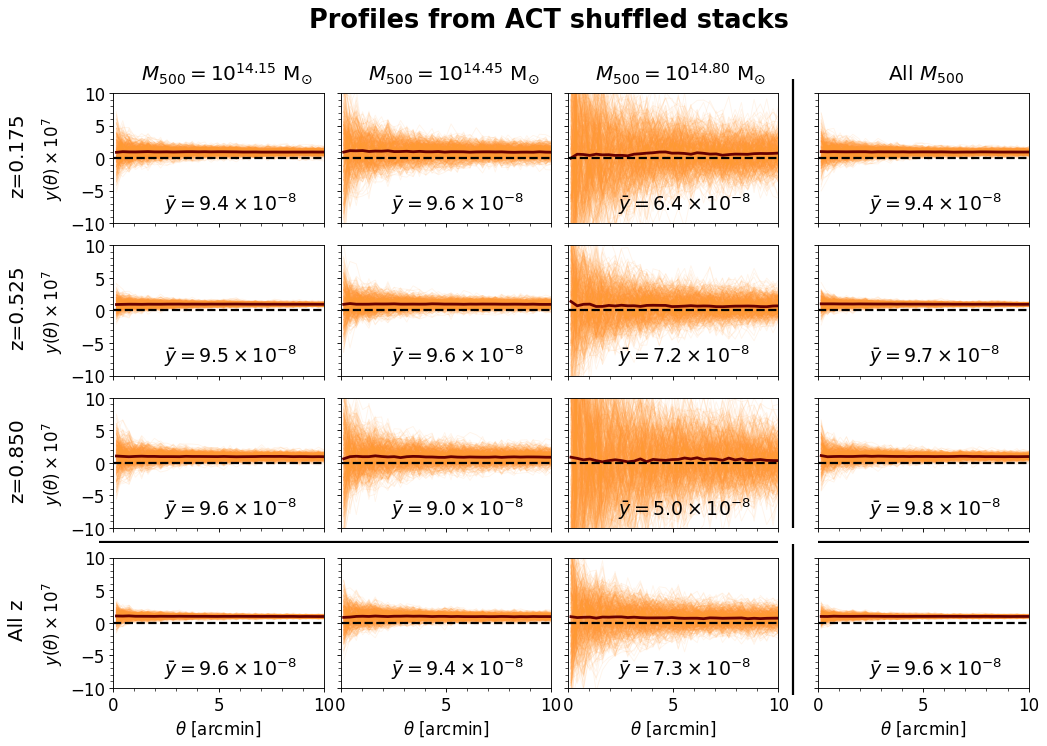} 
	\caption{Estimation of the mean background value for each $M$-$z$ bin for 
	both \textit{Planck} (\textit{left}) and ACT (\textit{right}) maps. For each panel
	we show the profiles obtained by stacking 500 replicas of the corresponding cluster 
	sample, each obtained by randomly shuffling the cluster positions. The thicker solid 
	line shows instead the average profile; the mean amplitude $\bar{y}$ of the latter over 
	the considered angular range is quoted explicitly in each panel. The zero level is 
	marked by a dashed line.\label{fig:shuffled}}
\end{figure*}  % +++++++++++++++++++++++++++++++++++++++++++++++++++++++++++++++

\begin{figure*}  % +++++++++++++++++++++++++++++++++++++++++++++++++++++++++++++++
\includegraphics[trim= -4mm 0mm 0mm 0mm, scale=0.34]{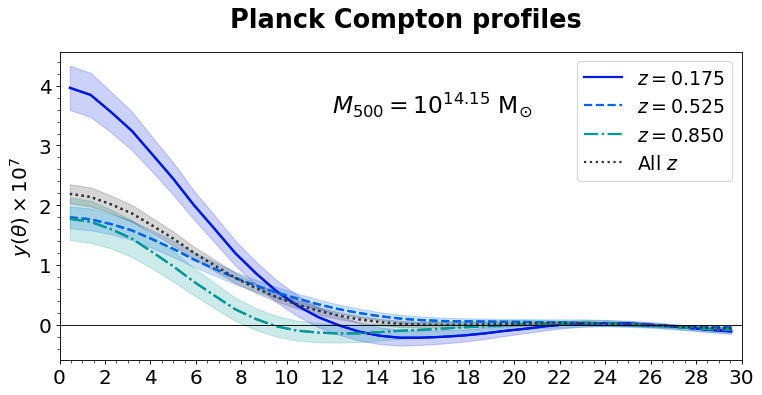} 
\includegraphics[trim= -4mm 0mm 0mm 0mm, scale=0.34]{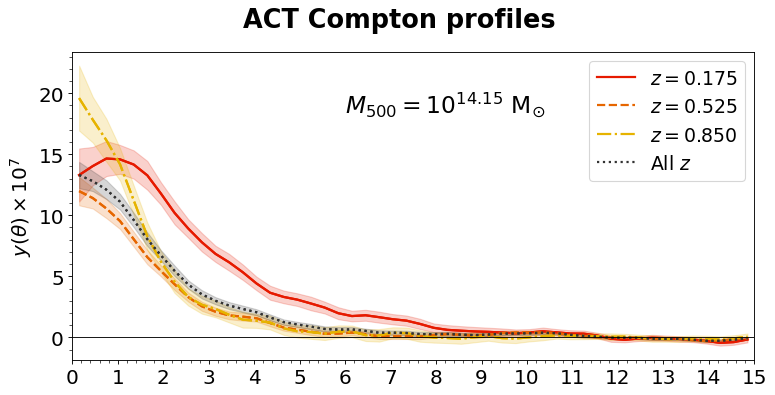} \\
\includegraphics[trim= 0mm 0mm 0mm 0mm, scale=0.34]{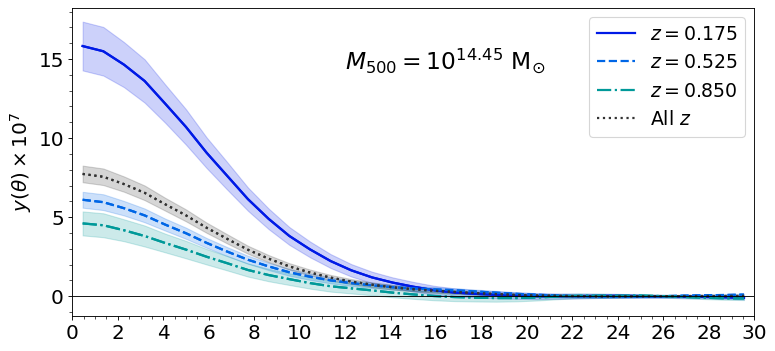}  
\includegraphics[trim= -4mm 0mm 0mm 0mm, scale=0.34]{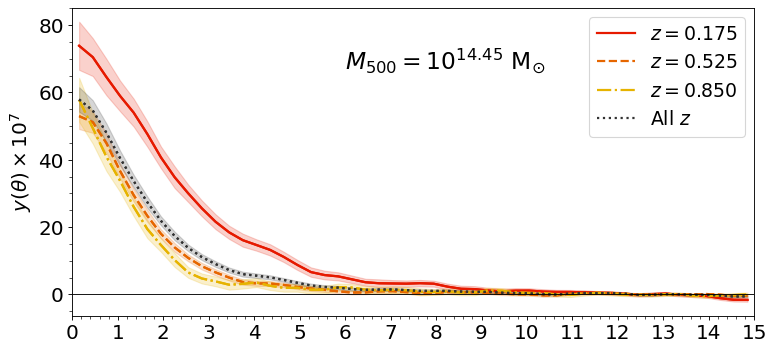} \\
\includegraphics[trim= 0mm 0mm 0mm 0mm, scale=0.34]{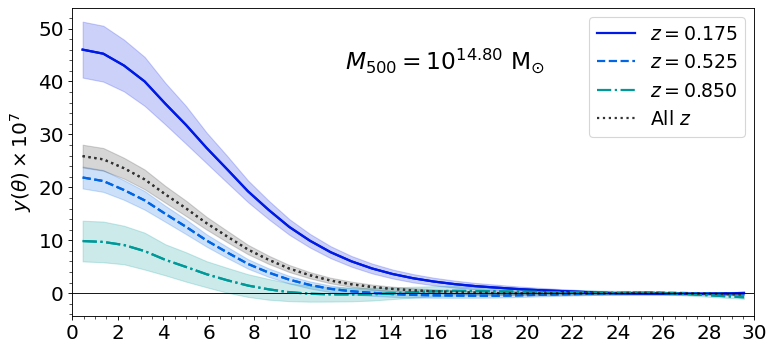} 
\includegraphics[trim= 0mm 0mm 0mm 0mm, scale=0.34]{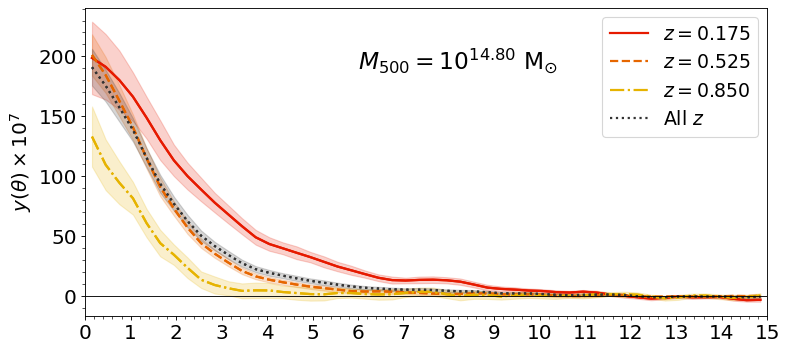} \\
\includegraphics[trim= -4mm 0mm 0mm 0mm, scale=0.34]{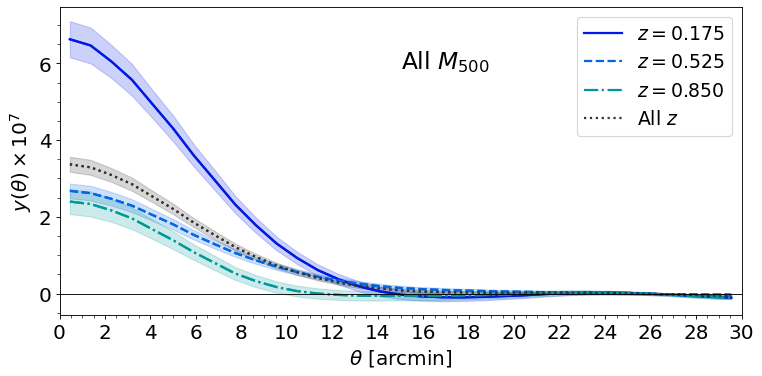} 
\includegraphics[trim= -4mm 0mm 0mm 0mm, scale=0.34]{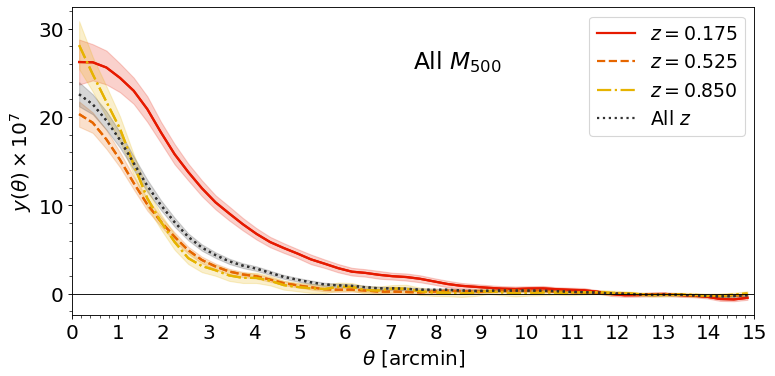} \\
\caption{Radial profiles obtained from the stacks shown in Figs.~\ref{fig:planck_stacks} 
	and~\ref{fig:act_stacks}. The left column shows the results for \textit{Planck}, 
	the right column the results for ACT. Each row corresponds to a chosen $M_{500}$ bin, 
	with the bottom row showing the marginalized case over all masses. Each panel shows the 
	results from the three redshift bins and for the marginalized case over all redshifts; 
	the shaded area around each profile quantifies the associated 1-$\sigma$ uncertainty. \label{fig:profiles}}
\end{figure*}  % +++++++++++++++++++++++++++++++++++++++++++++++++++++++++++++++

\begin{figure*} % +++++++++++++++++++++++++++++++++++++++++++++++++++++++++++++++
\includegraphics[trim= 0mm 0mm 0mm 0mm, scale=0.24]{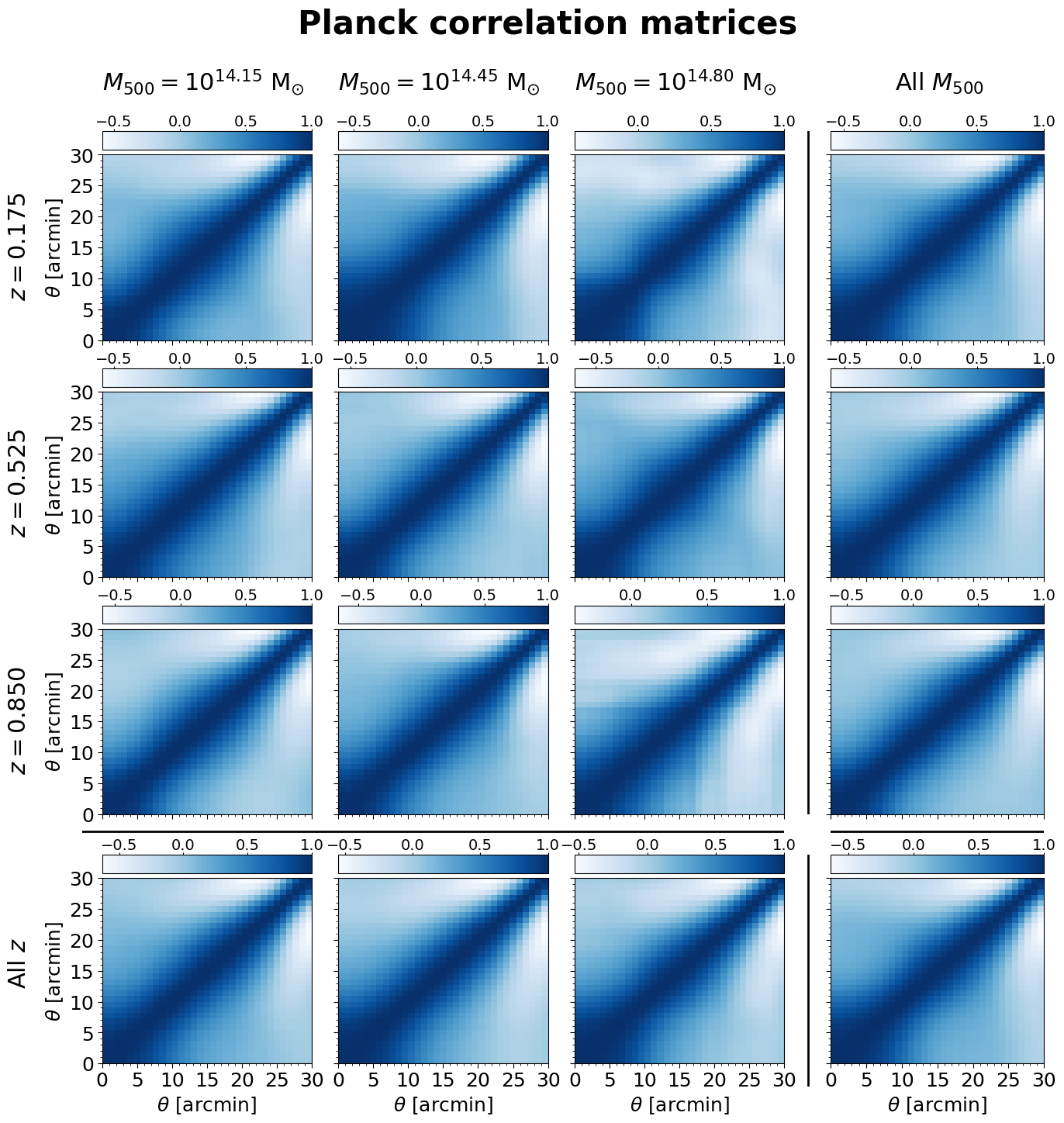}\quad\quad
\includegraphics[trim= 0mm 0mm 0mm 0mm, scale=0.24]{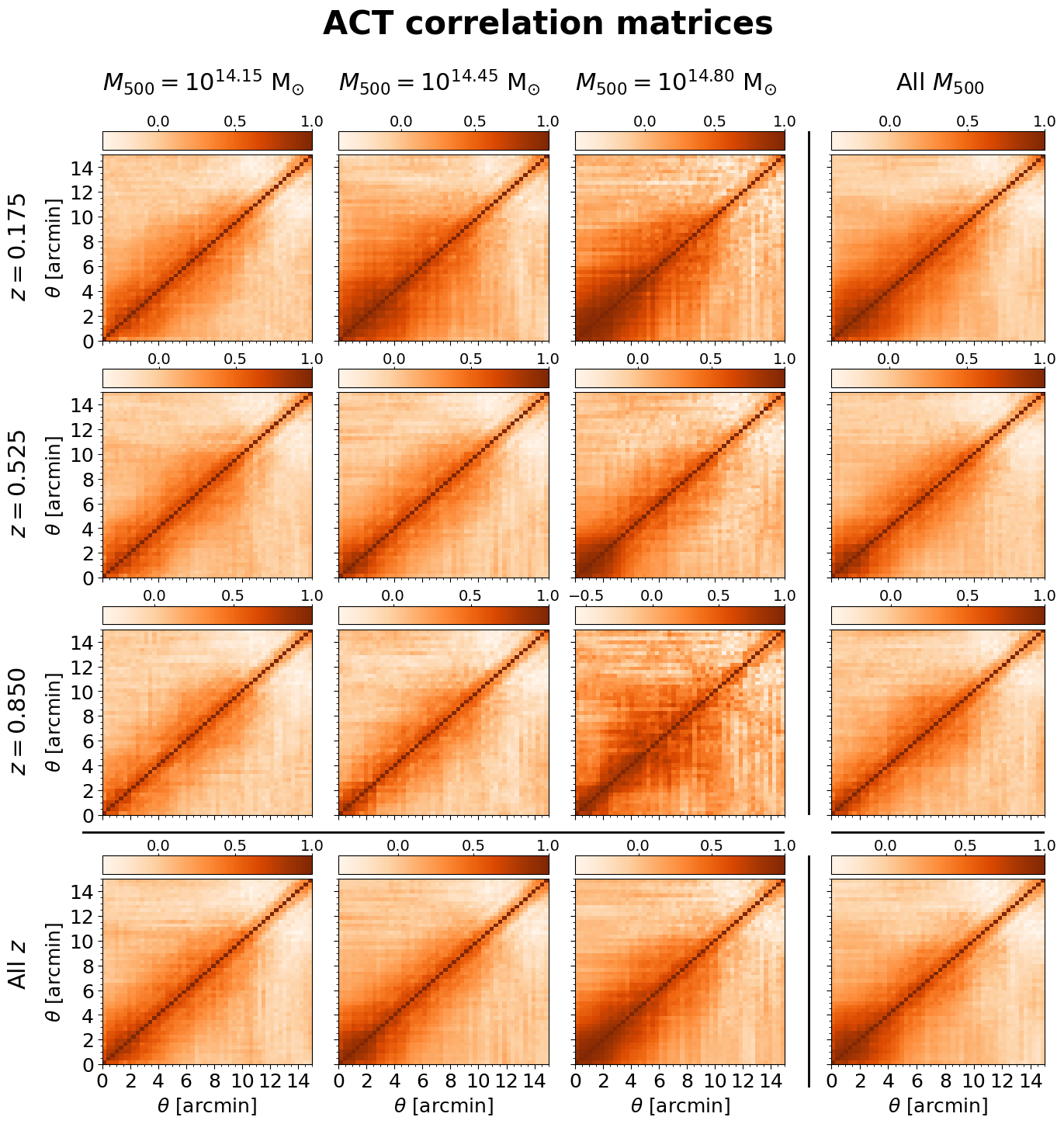}
\caption{Correlation matrices computed from the randomized stacks as described in 
	Section~\ref{ssec:stacks}. Results are shown for \textit{Planck} on the left and 
	ACT on the right, using the same $M_{500}, z$ binning scheme adopted in Figs.~\ref{fig:planck_stacks} and~\ref{fig:act_stacks}.\label{fig:corrmatr}}
\end{figure*}  % +++++++++++++++++++++++++++++++++++++++++++++++++++++++++++++++

%=================================================================================================
\subsection{Cluster stacks and angular profiles}
\label{ssec:stacks}

For each of our 16 cluster samples we obtain an independent stacked map. Let $N_{\rm cl}$ be 
the number of clusters in one $M$-$z$ bin; for the generic $i$-th cluster we trim a local sub-map 
centered on its nominal coordinates, with size $30'\times30'$ in the case of \textit{Planck} 
and $15'\times15'$ in the case of ACT\footnote{These different values are a result of the different 
beam size for the two maps, and they were chosen to enable a reconstruction of the whole cluster 
signal and part of the neighboring background, see Figs.~\ref{fig:planck_stacks} and~\ref{fig:act_stacks}.}. 
We label $S_i$ such a map, which carries the information 
on the $y$ signal around the cluster; we also obtain an equivalent weight map $W_i$ extracted 
in the same way from the corresponding survey mask. The $y$ stack map $S$ for the considered 
sample is then obtained as:
\begin{equation}
	\label{eq:stack}
	S = \frac{1}{W} \sum_{i=1}^{N_{\rm cl}} \,S_i\, W_i,
\end{equation}
where the total weight is given by:
\begin{equation}
	W = \sum_{i=1}^{N_{\rm cl}} \, W_i.
\end{equation}
The resulting 16 stacks for each $M$-$z$ bin are shown in Fig.~\ref{fig:planck_stacks} for 
the \textit{Planck} map and in Fig.~\ref{fig:act_stacks} for the ACT map. For each stack, the 
axes show the angular separation from the center, which is also marked with a pair of white 
dashed lines. 

From each of these stacks we can derive the associated Compton parameter profile, $y(\theta)$, 
as a function of the angular separation $\theta$ from the cluster center. The profile 
is built by splitting the pixels over a set of $N_{\rm b}$ bins in the angular separation 
from the map center and taking their mean value in each bin. As we are interested 
in the contrast of the local cluster signal, we first quantify a mean background value to be 
subtracted from each map\footnote{The stacks plotted in Fig.~\ref{fig:planck_stacks} 
and~\ref{fig:act_stacks} are not background-subtracted.}. The latter is estimated by repeating, for each $M$-$z$ bin, the 
Compton $y$ stack using 500 different replicas of the associated cluster sample, where in each 
case the cluster coordinates are randomly shuffled within the allowed ACT footprint. This results
in 500 radial profiles per $M$-$z$ bin; the resulting mean is used to evaluate the background 
contribution as a function of $\theta$, and is subtracted from the corresponding $M$-$z$ stack obtained 
using the real cluster sample. The results are shown in Fig.~\ref{fig:shuffled} for both ACT 
and \textit{Planck}. We notice that in all cases the mean profile does not show any strong dependence on 
the angular separation from the center; we then quote in each panel the mean value $\bar{y}$ 
of the average profile over the considered $\theta$ range (i.e., up to 30 arcmin for \textit{Planck} and 
10 arcmin for ACT). We stress that clusters from different 
$M$-$z$ bins actually span the same area, which is set by the ACT patch boundaries. Hence, we 
would expect to obtain the same estimate for the mean background level using different cluster 
samples. Fig.~\ref{fig:shuffled} shows that this is in general verified, with the exception of the 
top mass bin which displays larger deviations. This is due to the much lower number of clusters 
in this mass range, which produces a considerably larger spread of the individual shuffled profiles around 
the mean (the effect is more relevant for ACT). However, for all these $M$-$z$ bins the stack 
amplitude from Fig.~\ref{fig:planck_stacks} and~\ref{fig:act_stacks} is high enough that this zero-level 
correction is negligible ($\lesssim 1\%$ at the profile peak). 
For the lower mass bins, instead, the estimate of the mean background level converges to the values 
of $\bar{y}=6.3\times10^{-8}$ for \textit{Planck} and $\bar{y}=9.6\times10^{-8}$ for ACT.

Even after the removal of the background contribution as described above, the $y$ profiles do not 
always reach a null amplitude for large $\theta$ values. We then subtract from each profile an additional 
zero-level contribution measured as the average of the $y$ values from pixels bounded by a circular annulus 
outside the cluster outskirts. The inner and outer annulus radii are $[22,28]\,\text{arcmin}$ 
for \textit{Planck}, and $[10,15]\,\text{arcmin}$ for ACT, chosen in such a way as to leave out 
any cluster signal contribution (see again Figs.~\ref{fig:planck_stacks} and~\ref{fig:act_stacks} for 
a reference). Notice that this operation will be also 
applied to our theoretical predictions of the cluster profile, see Eq.~\eqref{eq:y_bkg}. 
The final, background-subtracted profiles are shown in Fig.~\ref{fig:profiles} for both \textit{Planck} 
and ACT (left and right panels, respectively). These figures focus on the comparison 
between the profiles from different redshift samples, for each bin in cluster mass.  

To complete the statistical characterization of our measurements, we compute the covariance 
matrices associated with the profiles. Again, for each $M$-$z$ bin, we perform the stacks of 
$N_{\rm rand}=500$ replicas of the associated cluster sample obtained via bootstrap resampling 
(each randomized catalog is obtained by randomly selecting clusters from the original sample,
up to the same number $N_{\rm cl}$, with possibility of repetition). The covariance between two 
angular separations $\theta_i$ and $\theta_j$ can then be computed as:
\begin{eqnarray}
\label{eq:covmat}
C(\theta_i,\theta_j) &=& \dfrac{1}{N_{\rm rand}} \nonumber \\ 
& \times & \sum_{k=1}^{N_{\rm rand}} \left[y_k(\theta_i)-\bar{y}(\theta_i)\right]\left[y_k(\theta_j)-\bar{y}(\theta_j) \right],
\end{eqnarray}
where $y_k$ denotes the $k$-th random profile and $\bar{y}$ is the average profile out of 
the random realizations:
\begin{equation}
\bar{y}(\theta_i) = \dfrac{1}{N_{\rm rand}} \sum_{k=1}^{N_{\rm rand}} y_k(\theta_i).
\end{equation}
By construction, the diagonal of each covariance matrix quantifies the profile variance at the 
corresponding angular separation, $\sigma^2(\theta_i)=C(\theta_i,\theta_i)$; these $\sigma(\theta_i)$ 
values are the effective uncertainties in our profile measurements, and are shown as shaded regions 
in Fig.~\ref{fig:profiles}. In Fig.~\ref{fig:corrmatr} we show, for both \textit{Planck} and ACT, 
the associated 16 correlation matrices, which are obtained from the covariance matrices as:
\begin{equation}
	\text{Corr}(\theta_i,\theta_j) = \frac{C(\theta_i,\theta_j)}{\sigma(\theta_i)\,\sigma(\theta_j)}.
\end{equation}

\begin{table*}
\centering
\caption{Significances per bin $\chi^2_{\rm b}$ on the mean $y$ profile measurements for all the considered $M$-$z$ bins, for both \textit{Planck} and ACT, computed as in equation~\eqref{eq:chi2meas}. The table also reports the number of clusters stacked in each case.}
\label{tab:stats}
\setlength{\tabcolsep}{1.2em}
\renewcommand{\arraystretch}{1.2}
\begin{tabular}{c|c|cccc}
\hline
 & \diagbox[width=10em]{$z$}{$M_{500}$} & $[10^{14.0},10^{14.3}] \,\text{M}_{\odot}$ & $[10^{14.3},10^{14.6}] \,\text{M}_{\odot}$ & $[10^{14.6},10^{15.1}] \,\text{M}_{\odot}$ & All $M_{500}$\\
\hline
 $\mathbf{N_{\textbf{cl}}}$ & \multirow{3}{8em}{$\quad\quad[0.00,0.35]$} & $\mathbf{3652}$ & $\mathbf{673}$ & $\mathbf{96}$ & $\mathbf{4421}$ \\  $\chi^2_{\rm b}$ (\textit{Planck}) &  & 62.4 & 107.0 & 91.4 & 157.7 \\  $\chi^2_{\rm b}$ (ACT) &  &  5.8 &  6.2 &  5.1 &  7.8 \\
\hline
 $\mathbf{N_{\textbf{cl}}}$ & \multirow{3}{8em}{$\quad\quad[0.35,0.70]$} & $\mathbf{13518}$ & $\mathbf{2125}$ & $\mathbf{243}$ & $\mathbf{15886}$ \\  $\chi^2_{\rm b}$ (\textit{Planck}) &  & 33.0 & 76.3 & 125.5 & 82.0 \\  $\chi^2_{\rm b}$ (ACT) &  &  7.6 &  9.4 &  6.0 & 10.0 \\
\hline
 $\mathbf{N_{\textbf{cl}}}$ & \multirow{3}{8em}{$\quad\quad[0.70,1.00]$} & $\mathbf{2840}$ & $\mathbf{620}$ & $\mathbf{53}$ & $\mathbf{3513}$ \\  $\chi^2_{\rm b}$ (\textit{Planck}) &  & 13.3 & 19.0 & 23.5 & 30.1 \\  $\chi^2_{\rm b}$ (ACT) &  &  4.1 &  4.3 &  3.0 &  6.4 \\
\hline
 $\mathbf{N_{\textbf{cl}}}$ & \multirow{3}{8em}{$\quad\quad\quad$All $z$} & $\mathbf{20010}$ & $\mathbf{3418}$ & $\mathbf{392}$ & $\mathbf{23820}$ \\  $\chi^2_{\rm b}$ (\textit{Planck}) &  & 94.2 & 160.5 & 179.6 & 229.0 \\  $\chi^2_{\rm b}$ (ACT) &  &  9.7 & 11.4 &  7.7 & 12.4 \\
\hline
\end{tabular}
\end{table*}
 %--------------------------------------------------------

The covariance matrices also allow us to compute the significance of the measured $y$ profile 
for each $M$-$z$ bin, defined as the chi-square:
\begin{equation}
	\label{eq:chi2meas}
	\chi^2 = \sum_{i=1}^{N_{\rm b}}\sum_{j=1}^{N_{\rm b}} y(\theta_i)\,I(\theta_i,\theta_j)\,y(\theta_j),
\end{equation}
where $I$ is the inverse of the covariance matrix 
corrected by an overall scaling factor to yield an unbiased estimator~\citep{hartlap07}:
\begin{equation}
	I(\theta_i,\theta_j) = \dfrac{N_{\rm rand}-N_{\rm b}-2}{N_{\rm rand}-1}\,C^{-1}(\theta_i,\theta_j).
\end{equation}
The significance for all the stacks is reported in Table~\ref{tab:stats}, together with 
the number of clusters in each sample. As the quantity from Eq.~\eqref{eq:chi2meas} depends on the 
number of considered bins, we quote the significance per bin, defined as $\chi^2_{\rm b}=\chi^2/N_{\rm b}$, 
where $N_{\rm b}$ is the number of angular bins until the inner radius of the annulus employed to estimate
the background value (33 for ACT and 24 for \textit{Planck}). Since we 
are not using the $\chi^2$ estimate to fit for a model, the number of degrees of freedom is equal to 
the number of bins. Hence, our $\chi^2_{\rm b}$ measurement is in fact a reduced-$\chi^2$ measurement 
for the null model $y(\theta)=0$, which can be used to evaluate the significance of the detection. 

%=================================================================================================
\subsection{Discussion of profile measurements}
\label{ssec:discussion1}

Figs.~\ref{fig:planck_stacks} and~\ref{fig:act_stacks} show that the cluster signal is 
clearly detected in all cases; in order to better show its contrast with respect to the 
background, the color scale is chosen to saturate the signal in each stack. The main difference 
between \textit{Planck} and ACT results is the resolution of the 
reconstructed signal, as expected. The much larger \textit{Planck} beam implies not only that 
the cluster signal is artificially broadened to larger angular separations from the stack center, 
but also that most of the features on scales below a few arcminutes are smoothed out. The ACT 
stacks, instead, allow for a more accurate reconstruction of the same signal at smaller scales.
A larger beam also implies a more severe dilution of the cluster signal, as it is evident by 
comparing the amplitude of any stack between the two $y$ maps. 

As expected, for each survey we see that the significance of the stacks tends to be lower for 
decreasing mass and increasing redshift. This is much clearer for the ACT stacks, which are 
less affected by the smoothing effect of the beam convolution. A reduction in mass determines 
the highest variations in the signal amplitude and yields in general more irregular 
stacks. The much higher statistics available for the lowest mass bin (first panel in 
Fig.~\ref{fig:distrib}) ensures the cluster signal is still detected despite the significant 
drop in the $S/N$ expected for individual objects in this $M_{500}$ range. An increase in redshift also 
produces a decrease in the signal amplitude (although to a lesser extent compared to the mass change), 
with the highest redshift bin showing stronger background fluctuations. The small ACT beam also allows us to appreciate 
the shrinking of the cluster signal for increasing redshift, as the angular diameter distance 
is monotonically increasing with $z$ in the redshift range we explore. The decrease of the profile amplitude with 
$z$ could be a result of higher redshift clusters being in earlier stages of their evolution,
and as such possibly not yet fully virialized, which would lower the associated SZ signal (we remind that our cluster sample is selected on the 
basis of optical observations, which cannot probe directly the ICM). Besides, higher $z$ clusters 
subtend smaller angular scales and are therefore more severely affected by the beam smoothing 
effect; this could account for the higher $z$ dependence of the profile amplitude observed in \textit{Planck}
stacks compared to ACT stacks.
The marginalized cases 
over $M_{500}$ typically show features in between the ones of the two lowest mass bins, as 
those encompass the vast majority of clusters in our samples. The marginalized cases over 
$z$ are instead quite similar to the mid-redshift bin, as the cluster number in different 
$z$ bins is comparable. Finally, the marginalized case over both $M_{500}$ and $z$ shows the 
highest $S/N$ and the most regular, rounded cluster emission, as it is expected from the 
exploitation of the full cluster statistics. 

Most of these considerations can also be inferred from the angular profiles in 
Fig.~\ref{fig:profiles}. Notice that, because each panel compares the profiles from different 
redshift bins, it does not include the associated instrumental beam profile to avoid 
excessive clutter. A direct comparison between each measured $y$ profile and the beam 
profile (scaled to the measured peak amplitude) will be shown later in Figs.~\ref{fig:planck_bfits} 
and~\ref{fig:act_bfits}. Those figures prove that the measured profiles extend to higher 
angular separations than the instrumental beam, i.e. the profiles are resolved in all cases; 
it is then meaningful to employ them to fit for the underlying pressure profile parameters 
(Section~\ref{sec:estimation}), even for the case of~\textit{Planck}. Going back to 
Fig.~\ref{fig:profiles}, we see the profiles for \textit{Planck} always show a regular trend, 
also thanks to the smoothing effect from the large instrumental beam: for a chosen $M_{500}$ bin, 
the profiles have a similar shape with the amplitude steadily decreasing for increasing $z$, 
and with the marginalized case over all redshifts being generally close to the mid-$z$ sample. 
The higher resolution of ACT allows instead to reveal more irregularities in the profiles. 
The lowest mass and redshift bin case shows a decrease of the profile amplitude towards the 
stack center; this results from the peak of the stacked signal being located off-center, as 
it is also clear from the corresponding stack panel in Fig.~\ref{fig:act_stacks}. This could 
be due to a possible contamination from neighboring high mass clusters, or to a mismatch 
between the sample cluster nominal and real central positions; the latter possibility will be taken into 
account in our modeling in Sec.~\ref{ssec:miscentering}. We also notice that for the top mass bin the 
top and mid redshift bins have a comparable peak amplitude; once again, this is probably due to 
a miscentering effect, which artificially dilutes the signal amplitude and becomes more relevant for lower redshifts.  
We conclude this section by commenting on the correlation matrices plotted in Fig.~\ref{fig:corrmatr}. 
The angular $y$ profile is expected to introduce correlations between neighboring $\theta$ bins. 
This is indeed reflected in the plots, especially for the lowest angular separations. Once again 
the situation is remarkably different depending 
on the considered survey, with the correlation extending up to a scale of $\sim 1$-$2\,{\rm arcmin}$ 
in the case of ACT and up to $\sim10\,\text{arcmin}$ in the case of \textit{Planck}, as expected from 
the different FWHMs. For each survey, the correlation matrices do not show an appreciable 
dependence on the chosen mass and redshift bins. The diagonal values of the covariance matrices are 
the square of the uncertainties overplotted to the radial profiles as shaded areas 
in Fig.~\ref{fig:profiles}. We see that the number of clusters stacked in each case partially affects 
the magnitude of the uncertainties, with the marginalized cases generally showing the smaller 
error bars. This is also reflected in the significance values quoted in Table~\ref{tab:stats}.
As expected, the detection significance increases for high masses and low redshifts, for both 
\textit{Planck} and ACT, but is also partially affected by the number of clusters in the 
associated sample. We can quote the square root of the $\chi^2_{\rm b}$ values as the detection 
significances per bin in $\sigma$ units. The significance per bin is typically above 
$\sim3.6\,\sigma$ for \textit{Planck} and $\sim1.7\,\sigma$ for ACT, and reaches the top 
values $15.1\,\sigma$ and $3.5\,\sigma$ respectively for the stack of the 
full cluster sample. The higher significances associated with \textit{Planck}'s measurements 
are a result of the lower uncertainties in the reconstructed profiles, which in turn are due 
to the smoothing effect of the larger beam. 

%=================================================================================================
%=================================================================================================

%=================================================================================================
%======================================	  FORMALISM   ============================================
%=================================================================================================

\section{Theoretical modeling}
\label{sec:theory}

The following details the formalism we adopt to predict theoretically the Compton profiles presented in the previous section.

%=================================================================================================
\subsection{The pressure profile for individual clusters}
\label{ssec:clusterpress}
The electron pressure profile $P_{\rm e}$ for a galaxy cluster of mass $M_{500}$ at 
redshift $z$ can be written as:
\begin{eqnarray}
\label{eq:pressprof}
P_{\rm e}(r;M_{500},z) &=& P_{500}(M_{500},z)\,f(M_{500},z) \nonumber \\
&\times & \mathbb{P}(x=r/\tilde{R}_{500}).
\end{eqnarray}
Here, $\mathbb{P}(x)$ is the UPP already defined in Eq.~\eqref{eq:upp} 
and $P_{500}$ is the characteristic cluster pressure expected in the self-similar model:
\begin{eqnarray}
\label{eq:p500}
\frac{P_{500}(M_{500},z)}{\text{keV}\text{cm}^{-3}} &=& 1.65\times10^{-3} E(z)^{8/3}\,\nonumber\\ 
& \times & \left[\dfrac{(1-b_{\rm h})\,M_{500}}{3\times10^{14}h_{70}^{-1}\text{M}_{\odot}}\right]^{2/3}\,h_{70}^2,
\end{eqnarray}
where $E(z) \equiv H(z)/H_0=\sqrt{\Omega_{\rm m}(1+z)^{3}+\Omega_{\Lambda}}$, 
with $\Omega_{\Lambda}=1-\Omega_{\rm m}$ for flatness, $h_{70}=h/0.7$ and $b_{\rm h}$ is the hydrostatic mass bias. 
A similar expression, but without the inclusion of the $(1-b_{\rm h})$ factor, 
was also employed in~\citet{arnaud10}; in that work  
cluster masses were derived from scaling relations based on the quantity $Y_{\rm X}$, 
defined as the product between the gas mass within $R_{500}$ and the spectral temperature 
$T_{\rm X}$~\citep[see also][]{nagai07}. Such mass estimates rely on the assumption of local 
hydrostatic equilibrium in the ICM, and can therefore be biased with respect to the true 
cluster masses. As in this paper we employ mass definitions based on lensing observations, which 
probe the true cluster mass content, we account for this bias by explicitly introducing 
the $(1-b_{\rm h})$ factor to scale our $M_{500}$ values in Eq.~\eqref{eq:p500}. 
The bias $b_{\rm h}$ will be let free in our analysis and estimated together with the other 
UPP parameters. We remind that the introduction of the hydrostatic bias
also affects the definition of the scale radius in the computation of the pressure profile, i.e. 
$x \equiv r/\tilde{R}_{500}$, with $\tilde{R}_{500}=R_{500}(1-b_{\rm h})^{1/3}$. 

We notice that Eq.~\eqref{eq:pressprof} differs from the expression in 
Eq.~\eqref{eq:upp}, which was the initial \textit{ansatz} proposed by~\citet{nagai07}, 
by an additional factor $f(M_{500},z)$. The latter determines a break in the self-similarity 
and was introduced by~\citet{arnaud10} to accommodate a residual mass dependence found in 
the scaled X-ray pressure profiles. When parametrized around the same pivot mass 
$M_{500}=3\times10^{14}\,h_{70}^{-1}\text{M}_{\odot}$, it reads:
\begin{equation}
\label{eq:ssbreak}
	f(x;M_{500},z) = \left[\dfrac{(1-b_{\rm h})\,M_{500}}{3\times10^{14}h_{70}^{-1}\text{M}_{\odot}}\right]^{\alpha_{\rm p}},
\end{equation}
with $\alpha_{\rm p }=0.12$ and where again we explicitly introduced the bias correction factor $(1-b_{\rm h})$. 

%=================================================================================================
\subsection{The mean Compton profile for a population of clusters}

The profiles obtained from our stacks contain the contribution from a large number of clusters
with different mass and redshift values. Formally, such a merged profile can be evaluated as the 
two-point correlation between the Compton $y$ map and the cluster sample, which in turn can 
be obtained by inverse Fourier transform of the associated cross-correlation power 
spectrum $C^{y\rm{c}}_{\ell}$:
\begin{equation}
\label{eq:ytheo}
	y(\theta) = \int \text{d}\ell \frac{\ell}{2\pi} J_{0}(\ell\theta) C^{y\rm{c}}_{\ell} B_{\ell},
\end{equation}
with $\ell$ the multipole and $J_{0}$ the zeroth-order Bessel function. In order to account for 
the instrumental smoothing of the reconstructed Compton map, the expression also includes the
beam window function $B_{\ell}$:
\begin{equation}
	B_{\ell} = \exp{\left[-\dfrac{1}{2}\ell(\ell+1)\sigma_{\rm b}^2\right]},
\end{equation}
where $\sigma_{\rm b}=\theta_{\rm FWHM}/\sqrt{8\ln{2}}$. 
The cross-correlation power spectrum $C^{y\rm{c}}_{\ell}$ can be computed using 
a halo model framework, which considers the contribution of both a one-halo 
and a two-halo term as~\citep{komatsu99}:
\begin{equation}
\label{eq:halomodel}
	C^{y\rm{c}}_{\ell} = C^{y\rm{c},1}_{\ell} + C^{y\rm{c},2}_{\ell}.
\end{equation}

The one-halo term $C^{y\rm{c},1}_{\ell}$ quantifies the integrated contribution of 
individual clusters, and is computed as:
\begin{eqnarray}
\label{eq:onehalo}
	C^{y\rm{c},1}_{\ell} &=& \frac{1}{\bar{n}_{\text{c}}} \int\text{d}z \frac{\text{d}^2V}{\text{d}\Omega\text{d}z}(z) \int\text{d}M \frac{\text{d}n}{\text{d}M}(M,z)  \nonumber \\
	& \times & S(M,z)\,\tilde{y}_{\ell}(M,z),
\end{eqnarray}
where the integrals are weighted by the comoving volume element $\text{d}^2V/\text{d}\Omega\text{d}z=c\chi^{2}/H(z)$ ($\chi$ is the comoving distance to redshift $z$) 
and the halo mass function $\text{d}n/\text{d}M$~\citep[e.g.][]{tinker08}. 
The selection function $S(M,z)$ depends on the particular cluster sample we consider; it 
encodes any deviations from the theoretical mass function, due to observational selection 
effects and other constraints applied to our catalogs. The entire expression is normalized 
by the mean angular number density of halos $\bar{n}_{\rm c}$, given by (see also, e.g.,~\citealt{fang12}):
\begin{equation}
\label{eq:n2d}
	\bar{n}_{\text{c}} = \int\text{d}z \frac{\text{d}^2V}{\text{d}\Omega\text{d}z} \int\text{d}M \frac{\text{d}n}{\text{d}M}(M,z) \,S(M,z).
\end{equation}
The quantity $\tilde{y}_{\ell}$ in Eq.~\eqref{eq:onehalo} is the Fourier transform of 
the Compton parameter profile, which can be computed as:
\begin{eqnarray}
	\label{eq:ytilde}
	\tilde{y}_{\ell}(M,z) &=&\dfrac{\sigma_{\rm T}}{m_{\rm e}c^2}\dfrac{4\pi \tilde{R}_{500}}{\ell_{\rm s}^2} \nonumber \\ 
	& \times & \int \text{d}x \,x^2 \dfrac{\sin{(\ell x/\ell_{\rm s})}}{\ell x/\ell_{\rm s}} P_{\rm e}(x;M,z),
\end{eqnarray}
where $\ell_{\rm s} = d_{\rm A}/R_{500}$ and $d_{\rm A}$ is the angular diameter distance; the 
expression contains the cluster electron pressure profile $P_{\rm e}$ defined in Eq.~\eqref{eq:pressprof} and the radius $\tilde{R}_{500}$ corrected for the hydrostatic bias.

The two-halo term $C^{y\rm{c},2}_{\ell}$ quantifies the correlation between different clusters, 
and is computed as:
\begin{eqnarray}
\label{eq:twohalo}
	C^{y\rm{c},2}_{\ell} &=& \frac{1}{\bar{n}_{\text{c}}} \int \text{d}z \frac{\text{d}^2V}{\text{d}\Omega\text{d}z} \,W_{\ell}^{y}(z)\, P_{\rm m}\left(k=\dfrac{\ell+0.5}{\chi(z)},z\right) \nonumber \\
	 &\times & \int\text{d}M \frac{\text{d}n}{\text{d}M}(M,z) \,S(M,z)\,b(M,z),
\end{eqnarray}
where $P_{\rm m}(k,z)$ is the linear matter power spectrum, $b(M,z)$ is the linear halo bias, and:
\begin{equation}
\label{eq:wy}
	W_{\ell}^{y}(z) = \int\text{d}M\, \frac{\text{d}n}{\text{d}M}(M,z)\, b(M,z) \,\tilde{y}_{\ell}(M,z).
\end{equation}
We compute the linear halo bias adopting the parametrization from~\citet{tinker10}.

%=================================================================================================
\subsection{Application to our sample}
\label{ssec:application}

In our case the selection function $S(M,z)$ cannot be simply expressed as a combined cut in 
mass and redshift, depending on the chosen $M$-$z$ bin. Due to the extended processing 
of the cluster catalogs prior to the stacking analysis, as described in Sections~\ref{ssec:mcal}
and~\ref{ssec:merging}, 
the selection function cannot be modeled analytically. A solution proposed in~\citetalias{gong19} 
consists in splitting the mass and redshift ranges into a set of smaller $N_{M}$ and $N_{z}$
intervals\footnote{To avoid confusion we keep using the word ``bin'' for denoting each of the 
redshift and mass choices that generate the 16 $M$-$z$ cluster samples we stacked on 
the $y$ map, and the word ``interval'' to designate 
this further splitting of each bin in smaller separations in mass and redshift.}, respectively. 
We call $\bar{M}_i$ and $\bar{z}_j$ the mean values within the generic 
$i$-th mass and $j$-th redshift intervals. Within each interval, the integral evaluation can be 
reasonably approximated as the product of the interval width and the integrand function evaluated at 
the interval mean value. 
For the generic $i$-th mass and $j$-th redshift interval, the one-halo term then reads:
\begin{equation}
	\label{eq:1happrox}
	\mathcal{C}_{\ell}^{y\rm{c}, 1}(\bar{M}_i,\bar{z}_j) = \tilde{y}_{\ell}(\bar{M}_i,\bar{z}_j),
\end{equation}
i.e., it is simply equal to the Compton Fourier transform evaluated at the mean redshift and mass. 
For the two-halo term we have:
\begin{eqnarray}
	\label{eq:2happrox}	
	\mathcal{C}_{\ell}^{y\rm{c}, 2}(\bar{M}_i,\bar{z}_j) &=& b(\bar{M}_i,\bar{z}_j) \nonumber \\
	& \times & P_{\rm m}\left(k=\dfrac{\ell+0.5}{\chi(\bar{z}_j)},\bar{z}_j\right)\,W_{\ell}^y(\bar{z}_j).
\end{eqnarray}
In this case, the quantity $W_{\ell}^y(\bar{z}_j)$ still requires to be evaluated via a full mass 
integral as in Eq.~\eqref{eq:wy}. In summary, when working with intervals, Eqs.~\eqref{eq:1happrox} 
and~\eqref{eq:2happrox} replace Eqs.~\eqref{eq:onehalo} and~\eqref{eq:twohalo}, respectively.

For both the one- and two-halo terms, the full quantity over the chosen cluster sample can be 
recovered as:
\begin{equation}
	\label{eq:approx}
	\mathcal{C}_{\ell}^{y\rm{c},X} = \frac{1}{N_{\rm cl}} \sum_{i}^{N_{M}} \sum_{j}^{N_{z}} n_{ij} \,\mathcal{C}_{\ell}^{y\rm{c}, X}(\bar{M}_i,\bar{z}_j),
\end{equation}
where $X$ is either 1 or 2, or even the sum of both halo terms. 
The factor $n_{ij}$ is the number of clusters with mass and redshift within the   
$i$-th mass and $j$-th redshift interval, such that the total number of clusters $N_{\rm cl}$ in the chosen 
$M$-$z$ bin is recovered as:
\begin{equation}
	N_{\rm cl} = \sum_{i=1}^{N_M} \sum_{j=1}^{N_z}  n_{ij}.
\end{equation}
The expression in Eq.~(\ref{eq:approx}) ensures that, whatever selection is applied to build the 
considered cluster sample, it will also be accounted for in the theoretical modeling of the 
associated mean Compton profile. 

We stress that a direct approach based on computing the individual 
$y(\theta;M_{500},z)$ profile for each stacked cluster and considering the resulting mean value, 
would not be adequate for our analysis. Firstly, such an approach would not take into account the 
inter-cluster correlations and only consider the contribution from the one-halo term; secondly, 
the computation of $\sim10^4$ profiles would be unpractical for the parameter estimation 
methodology described in Section~\ref{sec:estimation}. The use of Eqs.~\eqref{eq:1happrox} 
to~\eqref{eq:approx} allows us to solve both these issues. In our implementation we adopt 
$N_{M}=5$ and $N_{z}=5$ when dealing with individual $M$-$z$ cross-bins, and 
$N_{M}=15$ and $N_{z}=15$ when dealing respectively with mass- or redshift-marginalized cases.
We checked that this choice yields deviations $<1\%$ from the theoretical profile obtained adopting 
the full formalism from Eqs.~\eqref{eq:onehalo} and~\eqref{eq:twohalo} (with the $n_{ij}$ factors
computed integrating the halo mass function over the chosen mass and redshift intervals), which is within 
the error bars of our measurements (Fig.~\ref{fig:profiles}).

%=================================================================================================
\subsection{Miscentering and zero-level}
\label{ssec:miscentering}

In our modeling we consider possible offsets of the real cluster centers 
with respect to their quoted coordinates~\citep{yan20}. For the stacking analysis this would imply 
that the reconstructed profile is artificially diluted, so that we actually measure a mean 
``offset'' profile $\bar{y}_{\rm off}(\theta)$ instead of the true intrinsic profile $y(\theta)$. 
We follow the approach presented in~\citet{bellagamba18} and~\citet{giocoli21}, and divide the cluster population 
into a fraction $f_{\rm off}$ which is affected by miscentering and a fraction $1-f_{\rm off}$ 
whose cluster true positions coincide with the nominal ones. The resulting observed, miscentered profile $y_{\rm msc}$ is then:
\begin{equation}
	y_{\rm msc}(\theta) = f_{\rm off}\,\bar{y}_{\rm off}(\theta) + (1-f_{\rm off})\,y(\theta).
\end{equation}
\\

The problem reduces therefore to the computation of the mean offset profile $\bar{y}_{\rm off}(\theta)$. 
For a known angular offset $\theta_{\rm off}$, the miscentered profile $y_{\rm off}(\theta|\theta_{\rm off})$ 
can be computed starting from the centered profile $y(\theta)$ as~\citep{yang06}:
\begin{align}
	&y_{\rm off}(\theta|\theta_{\rm off}) = \dfrac{1}{2\pi} \times \nonumber \\
	& \int_0^{\infty} \text{d}\varphi \, y\left( \sqrt{(\theta^2+\theta^2_{\rm off}+2\theta\,\theta_{\rm off}\cos{\varphi})} \right),
\end{align}
i.e., integrating a set of profiles whose center is located $\theta_{\rm off}$ away from the 
nominal position, over all possible directions.

The value of the miscentering offset $\theta_{\rm off}$ is generally not known \textit{a priori}. 
Previous work found it reasonably follows a Rayleigh distribution with parameter 
$\sigma_{\rm off}$~\citep{johnston07}:
\begin{equation}
P(\theta_{\rm off}, \sigma_{\rm off}) = \dfrac{\theta_{\rm off}}{\sigma^2_{\rm off}}\exp{\left[-\dfrac{1}{2}\left( \dfrac{\theta_{\rm off}}{\sigma_{\rm off}}\right)^2\right]},
\end{equation}
so that the expected value for the offset is\footnote{Hereafter we shall refer to the parameter 
$\sigma_{\rm off}$ as the miscentering offset, although the real angular displacement is 
quantified by $\theta_{\rm off}$.} $\theta_{\rm off}\simeq 1.25\,\sigma_{\rm off}$. 
The mean miscentered profile $\bar{y}_{\rm off}(\theta)$ can then be evaluated by averaging over all 
possible values of the miscentering, weighted by its probability distribution:
\begin{equation}
	\bar{y}_{\rm off}(\theta|\sigma_{\rm off}) = \int_{0}^{\infty} \text{d}\theta_{\rm off}\, P(\theta_{\rm off},\sigma_{\rm off})\, y_{\rm off}(\theta|\theta_{\rm off}).
\end{equation}
In our analysis both the miscentering offset $\sigma_{\rm off}$ and the fraction of 
miscentered profiles $f_{\rm off}$ are taken as free parameters.

As a very last step in our theoretical prediction, we subtract from the profile its zero level:
\begin{equation}
	\label{eq:y_bkg}
	y_{\rm theo}(\theta) =  y_{\rm msc}(\theta) - \bar{y}_{\rm bkg}.
\end{equation}
The background value $\bar{y}_{\rm bkg}$ is computed as the mean of the profile amplitude over the same $\theta$ range 
that was considered in Section~\ref{ssec:stacks} when computing the zero-level for the profiles measured from the 
$y$ stacks.

%=================================================================================================
%=================================================================================================

%=================================================================================================
%=======================================  ESTIMATION  ============================================
%=================================================================================================

\section{Parameter estimation}
\label{sec:estimation}
In this section we present the results on the estimates of the pressure profile parameters.

\begin{figure*} % +++++++++++++++++++++++++++++++++++++++++++++++++++++++++++++++
	\includegraphics[trim= 0mm 0mm 0mm 0mm, scale=0.42]{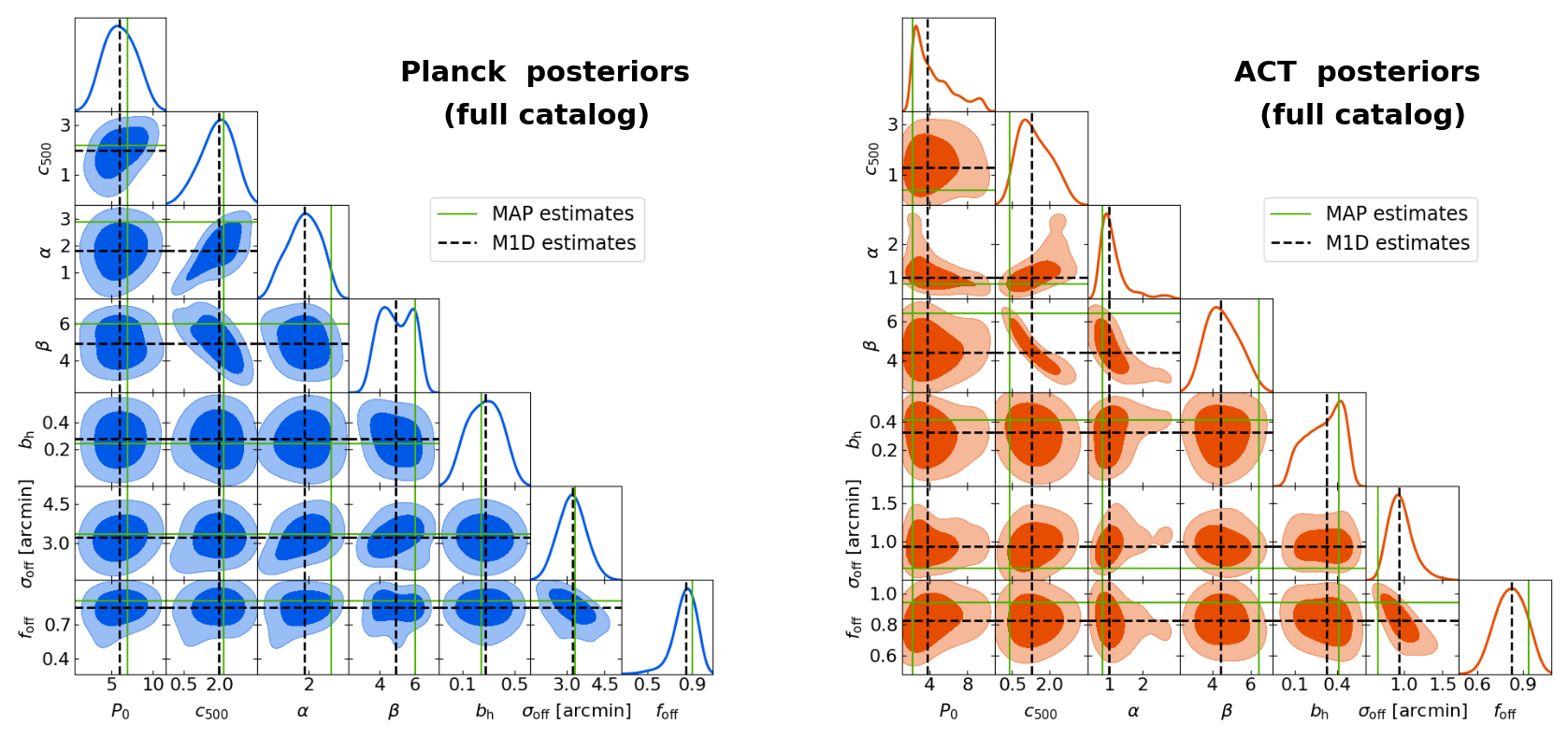}
	\caption{Posterior probability distributions on the model parameters, obtained fitting our theoretical model 
	on the profile corresponding to the stack of the full cluster catalog on \textit{Planck} (\textit{left}) and 
	ACT (\textit{right}). Posteriors are plotted as 68\% and 95\% C.L. contours. The solid green lines mark the best-fit, 
	highest-probability (MAP) values; the dashed black lines mark instead the M1D estimates computed from the 
	marginalized 1-dimensional distributions.}
	\label{fig:posteriors}
\end{figure*} % +++++++++++++++++++++++++++++++++++++++++++++++++++++++++++++++

\begin{table*}
\centering
\caption{Summary of the M1D parameter estimates for all the considered mass and redshift sub-samples, fitted on \textit{Planck} Compton profiles. Values are computed as the marginalized distribution medians; error bars quantify the 68\% C.L.}
\label{tab:planck_bestfits}
\renewcommand{\arraystretch}{1.4}
\begin{tabular}{cc|ccccccc}
\hline
\multicolumn{9}{c}{\textbf{\textit{Planck} parameter estimates}} \\
\hline
\multicolumn{2}{c}{Priors:} &\quad $[2.0, 10.0]$ \quad &\quad $[0.0, 3.0]$ \quad &\quad $[0.0, 3.0]$ \quad &\quad $[2.0, 6.5]$ \quad &\quad $[0.1, 0.5]$ \quad &\quad $[0.0, 12.0]$ \quad &\quad $[0.0, 1.0]$ \quad \\
\hline
 $z$ & $M_{500} [M_{\odot}]$  & $P_0$ & $c_{500}$ & $\alpha$ & $\beta$ & $b_{\rm h}$ & $\sigma_{\rm off} [\text{arcmin}]$ & $f_{\rm off}$ \\
\hline
0.175 & $10^{14.15}$ & $6.4^{+2.2}_{-2.7}$ & $2.2^{+0.5}_{-0.6}$ & $2.1^{+0.6}_{-0.7}$ & $5.1^{+1.0}_{-1.0}$ & $0.3^{+0.2}_{-0.1}$ & $2.9^{+0.9}_{-0.7}$ & $0.7^{+0.2}_{-0.2}$ \\
0.175 & $10^{14.45}$ & $5.2^{+3.3}_{-2.4}$ & $1.7^{+0.7}_{-0.5}$ & $1.7^{+0.8}_{-0.6}$ & $4.8^{+1.2}_{-0.9}$ & $0.2^{+0.2}_{-0.1}$ & $3.2^{+0.9}_{-0.7}$ & $0.8^{+0.1}_{-0.2}$ \\
0.175 & $10^{14.80}$ & $5.1^{+3.0}_{-2.1}$ & $1.9^{+0.7}_{-0.6}$ & $1.9^{+0.7}_{-0.7}$ & $4.7^{+1.2}_{-1.1}$ & $0.3^{+0.2}_{-0.2}$ & $3.4^{+0.8}_{-0.7}$ & $0.8^{+0.1}_{-0.1}$ \\
0.175 & All $M_{500}$ & $6.2^{+2.6}_{-2.6}$ & $2.1^{+0.5}_{-0.5}$ & $2.2^{+0.5}_{-0.7}$ & $5.3^{+0.8}_{-0.9}$ & $0.3^{+0.1}_{-0.1}$ & $3.1^{+0.5}_{-0.5}$ & $0.9^{+0.1}_{-0.1}$ \\
\hline
0.525 & $10^{14.15}$ & $6.1^{+3.0}_{-2.7}$ & $2.2^{+0.5}_{-0.6}$ & $1.9^{+0.8}_{-0.9}$ & $4.2^{+1.2}_{-0.9}$ & $0.3^{+0.1}_{-0.1}$ & $4.4^{+1.0}_{-1.0}$ & $0.9^{+0.1}_{-0.1}$ \\
0.525 & $10^{14.45}$ & $4.7^{+3.2}_{-2.0}$ & $1.9^{+0.7}_{-0.7}$ & $1.1^{+1.5}_{-0.5}$ & $3.6^{+0.7}_{-0.4}$ & $0.2^{+0.2}_{-0.1}$ & $7.2^{+2.9}_{-2.8}$ & $0.5^{+0.2}_{-0.4}$ \\
0.525 & $10^{14.80}$ & $6.4^{+2.0}_{-2.8}$ & $1.8^{+0.7}_{-0.7}$ & $1.4^{+0.9}_{-0.6}$ & $4.9^{+0.9}_{-0.8}$ & $0.2^{+0.2}_{-0.1}$ & $2.1^{+2.1}_{-1.4}$ & $0.3^{+0.3}_{-0.2}$ \\
0.525 & All $M_{500}$ & $5.5^{+2.8}_{-2.3}$ & $1.9^{+0.7}_{-0.8}$ & $1.6^{+0.8}_{-0.7}$ & $4.1^{+1.6}_{-0.7}$ & $0.3^{+0.1}_{-0.2}$ & $4.3^{+0.8}_{-1.2}$ & $0.8^{+0.1}_{-0.1}$ \\
\hline
0.850 & $10^{14.15}$ & $5.9^{+2.8}_{-2.4}$ & $2.0^{+0.6}_{-0.7}$ & $1.8^{+0.9}_{-0.7}$ & $4.8^{+1.2}_{-1.0}$ & $0.3^{+0.2}_{-0.2}$ & $2.9^{+3.5}_{-1.4}$ & $0.5^{+0.3}_{-0.3}$ \\
0.850 & $10^{14.45}$ & $6.0^{+2.9}_{-2.8}$ & $2.1^{+0.6}_{-0.7}$ & $1.5^{+1.1}_{-0.7}$ & $4.2^{+1.3}_{-0.9}$ & $0.3^{+0.1}_{-0.1}$ & $4.6^{+2.4}_{-2.0}$ & $0.6^{+0.2}_{-0.3}$ \\
0.850 & $10^{14.80}$ & $5.6^{+3.1}_{-2.5}$ & $2.2^{+0.5}_{-0.7}$ & $1.6^{+0.9}_{-0.7}$ & $5.0^{+1.1}_{-0.9}$ & $0.3^{+0.1}_{-0.2}$ & $4.1^{+5.9}_{-2.7}$ & $0.3^{+0.3}_{-0.2}$ \\
0.850 & All $M_{500}$ & $6.0^{+2.6}_{-2.3}$ & $2.0^{+0.6}_{-0.8}$ & $1.6^{+0.9}_{-0.7}$ & $5.0^{+0.9}_{-1.1}$ & $0.3^{+0.2}_{-0.2}$ & $2.6^{+1.5}_{-0.8}$ & $0.7^{+0.2}_{-0.3}$ \\
\hline
All $z$ & $10^{14.15}$ & $5.7^{+2.6}_{-2.3}$ & $2.0^{+0.6}_{-0.6}$ & $1.9^{+0.8}_{-0.6}$ & $5.0^{+0.9}_{-0.9}$ & $0.3^{+0.2}_{-0.1}$ & $3.3^{+0.5}_{-0.4}$ & $0.9^{+0.1}_{-0.1}$ \\
All $z$ & $10^{14.45}$ & $6.0^{+2.5}_{-2.2}$ & $1.8^{+1.0}_{-1.0}$ & $1.4^{+0.9}_{-0.7}$ & $4.1^{+1.4}_{-0.9}$ & $0.3^{+0.1}_{-0.1}$ & $5.0^{+1.9}_{-1.7}$ & $0.6^{+0.1}_{-0.2}$ \\
All $z$ & $10^{14.80}$ & $6.8^{+2.8}_{-3.6}$ & $1.9^{+0.6}_{-0.8}$ & $1.6^{+1.0}_{-0.8}$ & $5.2^{+0.7}_{-1.0}$ & $0.2^{+0.2}_{-0.1}$ & $2.3^{+0.6}_{-0.8}$ & $0.7^{+0.2}_{-0.3}$ \\
All $z$ & All $M_{500}$ & $5.9^{+2.3}_{-2.0}$ & $2.0^{+0.7}_{-0.7}$ & $1.8^{+0.7}_{-0.7}$ & $4.9^{+1.2}_{-1.0}$ & $0.3^{+0.1}_{-0.2}$ & $3.2^{+0.5}_{-0.6}$ & $0.8^{+0.1}_{-0.1}$ \\
\hline
\end{tabular}
\end{table*}
 %--------------------------------------------------------
\begin{table*}
\centering
\caption{Summary of the M1D parameter estimates for all the considered mass and redshift sub-samples, fitted on ACT Compton profiles. Values are computed as the marginalized distribution medians; error bars quantify the 68\% C.L.}
\label{tab:act_bestfits}
\renewcommand{\arraystretch}{1.4}
\begin{tabular}{cc|ccccccc}
\hline
\multicolumn{9}{c}{\textbf{ACT parameter estimates}} \\
\hline
\multicolumn{2}{c}{Priors:} &\quad $[2.0, 10.0]$ \quad &\quad $[0.0, 3.0]$ \quad &\quad $[0.0, 3.0]$ \quad &\quad $[2.0, 6.5]$ \quad &\quad $[0.1, 0.5]$ \quad &\quad $[0.0, 3.0]$ \quad &\quad $[0.0, 1.0]$ \quad \\
\hline
 $z$ & $M_{500} [M_{\odot}]$  & $P_0$ & $c_{500}$ & $\alpha$ & $\beta$ & $b_{\rm h}$ & $\sigma_{\rm off} [\text{arcmin}]$ & $f_{\rm off}$ \\
\hline
0.175 & $10^{14.15}$ & $5.1^{+3.0}_{-2.2}$ & $2.0^{+0.7}_{-0.7}$ & $1.6^{+0.8}_{-0.5}$ & $4.2^{+1.0}_{-0.6}$ & $0.3^{+0.1}_{-0.2}$ & $1.8^{+0.3}_{-0.2}$ & $1.0^{+0.0}_{-0.0}$ \\
0.175 & $10^{14.45}$ & $4.2^{+3.1}_{-1.7}$ & $1.9^{+0.6}_{-0.6}$ & $1.9^{+0.6}_{-0.6}$ & $4.5^{+1.1}_{-0.8}$ & $0.3^{+0.1}_{-0.1}$ & $1.8^{+0.5}_{-0.4}$ & $0.7^{+0.1}_{-0.1}$ \\
0.175 & $10^{14.80}$ & $5.0^{+3.0}_{-2.2}$ & $2.0^{+0.6}_{-0.7}$ & $1.5^{+0.7}_{-0.4}$ & $4.2^{+1.0}_{-0.7}$ & $0.3^{+0.1}_{-0.2}$ & $1.7^{+0.5}_{-0.5}$ & $0.6^{+0.1}_{-0.2}$ \\
0.175 & All $M_{500}$ & $4.3^{+3.1}_{-1.7}$ & $1.8^{+0.7}_{-0.6}$ & $1.5^{+0.8}_{-0.5}$ & $4.2^{+1.0}_{-0.6}$ & $0.3^{+0.1}_{-0.2}$ & $1.5^{+0.3}_{-0.2}$ & $0.9^{+0.1}_{-0.1}$ \\
\hline
0.525 & $10^{14.15}$ & $4.1^{+3.0}_{-1.6}$ & $1.6^{+0.8}_{-0.7}$ & $1.1^{+0.8}_{-0.4}$ & $4.4^{+1.2}_{-0.9}$ & $0.3^{+0.1}_{-0.2}$ & $1.2^{+0.4}_{-0.3}$ & $0.8^{+0.1}_{-0.1}$ \\
0.525 & $10^{14.45}$ & $4.2^{+3.5}_{-1.7}$ & $1.7^{+0.8}_{-0.7}$ & $1.5^{+1.0}_{-0.6}$ & $4.5^{+1.1}_{-0.8}$ & $0.3^{+0.1}_{-0.2}$ & $1.2^{+0.4}_{-0.5}$ & $0.6^{+0.1}_{-0.1}$ \\
0.525 & $10^{14.80}$ & $4.0^{+3.0}_{-1.5}$ & $1.7^{+0.8}_{-0.6}$ & $1.6^{+0.6}_{-0.4}$ & $4.5^{+1.1}_{-0.8}$ & $0.3^{+0.1}_{-0.2}$ & $1.1^{+0.5}_{-0.4}$ & $0.5^{+0.2}_{-0.2}$ \\
0.525 & All $M_{500}$ & $4.2^{+3.0}_{-1.7}$ & $1.3^{+0.9}_{-0.6}$ & $1.0^{+0.6}_{-0.2}$ & $4.6^{+1.1}_{-0.8}$ & $0.3^{+0.1}_{-0.1}$ & $1.0^{+0.3}_{-0.3}$ & $0.7^{+0.1}_{-0.1}$ \\
\hline
0.850 & $10^{14.15}$ & $5.0^{+3.1}_{-2.1}$ & $1.8^{+0.8}_{-0.9}$ & $1.3^{+0.8}_{-0.4}$ & $4.3^{+1.4}_{-0.9}$ & $0.3^{+0.1}_{-0.2}$ & $0.9^{+0.5}_{-0.3}$ & $0.8^{+0.2}_{-0.2}$ \\
0.850 & $10^{14.45}$ & $5.4^{+2.9}_{-2.5}$ & $2.0^{+0.7}_{-0.9}$ & $1.2^{+0.7}_{-0.4}$ & $4.2^{+1.0}_{-0.7}$ & $0.3^{+0.1}_{-0.2}$ & $0.7^{+0.5}_{-0.4}$ & $0.5^{+0.3}_{-0.3}$ \\
0.850 & $10^{14.80}$ & $4.6^{+3.0}_{-2.0}$ & $2.3^{+0.5}_{-0.8}$ & $1.4^{+0.8}_{-0.5}$ & $4.6^{+1.1}_{-1.0}$ & $0.3^{+0.1}_{-0.2}$ & $0.8^{+1.4}_{-0.3}$ & $0.6^{+0.3}_{-0.4}$ \\
0.850 & All $M_{500}$ & $5.1^{+3.2}_{-2.5}$ & $1.8^{+0.8}_{-0.7}$ & $1.2^{+0.7}_{-0.3}$ & $4.2^{+1.2}_{-0.7}$ & $0.3^{+0.1}_{-0.2}$ & $0.8^{+0.3}_{-0.2}$ & $0.8^{+0.2}_{-0.2}$ \\
\hline
All $z$ & $10^{14.15}$ & $4.5^{+2.9}_{-2.0}$ & $1.8^{+0.8}_{-1.0}$ & $1.1^{+0.6}_{-0.3}$ & $4.0^{+1.4}_{-0.5}$ & $0.3^{+0.1}_{-0.2}$ & $1.1^{+0.2}_{-0.2}$ & $0.9^{+0.1}_{-0.1}$ \\
All $z$ & $10^{14.45}$ & $3.7^{+3.0}_{-1.3}$ & $1.7^{+0.8}_{-0.6}$ & $1.3^{+0.7}_{-0.4}$ & $4.1^{+0.8}_{-0.5}$ & $0.3^{+0.1}_{-0.2}$ & $1.2^{+0.5}_{-0.4}$ & $0.5^{+0.1}_{-0.1}$ \\
All $z$ & $10^{14.80}$ & $3.4^{+4.3}_{-1.1}$ & $1.6^{+0.7}_{-0.6}$ & $1.4^{+0.6}_{-0.3}$ & $4.4^{+1.1}_{-0.7}$ & $0.3^{+0.1}_{-0.2}$ & $0.9^{+0.4}_{-0.4}$ & $0.5^{+0.2}_{-0.2}$ \\
All $z$ & All $M_{500}$ & $3.8^{+2.9}_{-1.4}$ & $1.3^{+0.9}_{-0.6}$ & $1.0^{+0.4}_{-0.2}$ & $4.4^{+1.1}_{-0.8}$ & $0.3^{+0.1}_{-0.2}$ & $0.9^{+0.2}_{-0.2}$ & $0.8^{+0.1}_{-0.1}$ \\
\hline
\end{tabular}
\end{table*}
 %--------------------------------------------------------

%=================================================================================================
\subsection{Methodology}
\label{ssec:emcee}

The ultimate goal of this study is to provide novel estimates of the parameters entering the 
expression of the universal pressure profile in Eq.~(\ref{eq:upp}). As our analysis is based on 
tSZ measurements alone, without the inclusion of numerical simulations or X-ray data, we do not 
fit for the central slope of the profile, which we fix to the fiducial value $\gamma=0.31$, as it 
was done in similar tSZ-based works (Table~\ref{tab:upp_summary}). In addition, we fit 
for the hydrostatic mass bias $b_{\rm h}$ and for the values of the miscentering offset 
$\sigma_{\rm off}$ and fraction $f_{\rm off}$. Our parameter space is then 7-dimensional, 
with a generic parameter state $\Theta$ defined as the list of values:
\begin{equation}
	\Theta = \{P_0,c_{500},\alpha,\beta,b_{\rm h},\sigma_{\rm off},f_{\rm off}\}.
\end{equation}
For a given set of the aforementioned parameters, the formalism described in 
Section~\ref{sec:theory} allows us to compute the associated profile $y(\Theta)$. 

For each of the 16 cluster samples, we fit the theoretical prediction $y(\Theta)$ 
against the profile $y^{\rm obs}$ extracted from the stack. The best-fit parameters 
are defined as the set $\Theta_{\rm bf}$ that maximizes the likelihood:
\begin{align}
\label{eq:likelihood}
	&\mathcal{L}(\Theta) = \nonumber \\
	&\exp\left[ -\dfrac{1}{2} (y(\Theta)-y^{\rm obs})^{\text{T}}  
	\,C^{-1} \,(y(\Theta)-y^{\rm obs}) \right],
\end{align}
where $C$ is the covariance matrix 
measured for the chosen sample as described in Section~\ref{ssec:stacks}. In fact, due to 
the relative high number of parameters, a direct maximization of the likelihood is not feasible; 
we adopt instead a Markov chain Monte Carlo (MCMC) approach to explore the parameter space. 
Specifically, we employ the \texttt{Python emcee} package, which is an implementation of the 
affine invariant ensamble sampler from~\citet{goodman10}. We adopt flat, uninformative priors 
on all parameters; the associated ranges are quoted in Tables~\ref{tab:planck_bestfits} 
and~\ref{tab:act_bestfits}. After burn-in removal and chain thinning, for each mass and redshift 
bin we are left with $\sim 5000$ samples of the posterior distribution on each parameter. 
The resulting joint probability contours for all parameter pairs, and the one-dimensional distributions 
for individual parameters, are shown in Fig.~\ref{fig:posteriors} for both \textit{Planck} and ACT. 
The figure shows the results for the parameter estimation on the full catalog; the posteriors for 
all the others $M$-$z$ bins are qualitatively similar, and are shown in Appendix~\ref{app:systematics}. 

We consider two different approaches to determine the associated parameter estimates. Firstly, we retrieve 
the 7-tuple of parameters yielding the maximum value for the posterior probability distribution; 
we shall refer to these values as the MAP (maximum \textit{a posteriori}) estimates. Since we are using flat 
priors, the maximum of the posterior distribution also corresponds to the maximum of our likelihood from 
Eq.~\eqref{eq:likelihood}, or in other words, the MAPs are by definition our best-fit values $\Theta_{\rm bf}$. 
Secondly, we consider estimates computed for each individual parameter as the 50\% percentile over its 
marginalized 1-dimensional posterior distribution (i.e. the distributions plotted along the diagonal of 
the triangular plots in Fig.~\ref{fig:posteriors} and similar); the associated lower and upper error 
bars are evaluated as the separations from the 16\% and 84\% percentiles of the distribution, respectively. 
We shall label the resulting values as the M1D (marginalized 1-dimensional) estimates. 

In general, the MAP and M1D estimates do not necessarily agree. In fact, in situations involving a 
large number of parameters with non-linear degeneracies, as it is the case with our model, they are 
expected to show important differences~\citep[for an extensive discussion see for example Section 6 
in][]{joachimi21}. This is clearly visible in Fig.~\ref{fig:posteriors}, where MAPs are shown as solid 
green lines and M1Ds as dashed black lines. The two sets of parameters carry different information: 
the MAPs are our best-fit values, 
and are employed to compute the associated best-fit predictions for the Compton parameter profiles. The 
comparison with our measurements is shown for all $M$-$z$ bins in Fig.~\ref{fig:planck_bfits} for 
\textit{Planck} and in Fig.~\ref{fig:act_bfits} for ACT. The M1Ds, instead, quantify our fiducial 68\% 
confidence level on individual parameters according to their final probability distributions, and can be 
compared with estimates from other works. The M1Ds with associated error bars are quoted in 
Table~\ref{tab:planck_bestfits} for \textit{Planck} and in Table~\ref{tab:act_bestfits} for ACT; they are
also plotted in Fig.~\ref{fig:bestfits}, where it is possible to 
visualize their dependence on the selected mass and redshift bin.

As already anticipated in Section~\ref{ssec:mcal}, we also conducted an analogous parameter estimation
analysis on the $y$ profiles obtained from the merged catalog without any explicit mass rescaling. 
Hereafter, we shall refer to such profiles as ``unscaled'', while we shall refer to the profiles shown in 
Fig.~\ref{fig:profiles} as ``fiducial''.
In Figs.~\ref{fig:planck_profcmp} and~\ref{fig:act_profcmp} we show, for \textit{Planck} and ACT respectively, 
a comparison between the fiducial and unscaled profiles, together with the theoretical predictions
obtained using the parameters fitted on the latter. Figs.~\ref{fig:posteriors_cmp} to~\ref{fig:act_contours_marg}
show the posterior distributions for both the fiducial and the unscaled profiles. As in these figures 
the focus is on the comparison between two different sets of contours, we do not mark the locations of 
the associated MAP and M1D estimates to avoid excessive cluttering.

%=================================================================================================
\subsection{Discussion on MCMC results}
\label{ssec:discussion2}

We begin by commenting the posterior probability contours plotted in Fig.~\ref{fig:posteriors} 
and in Figs.~\ref{fig:posteriors_cmp} to~\ref{fig:act_contours_marg}. The main feature of these plots is 
the asymmetry in the shape of most probability contours. This is a direct result of the complexity 
of the model we are fitting, and of the strong degeneracy existing between the UPP parameters. Even though
the chains are converged (in MCMC terms, the thinned chains are longer than 50 times their autocorrelation 
length), existing correlations between different parameters produce strong elongations of the contours and yield 
rather asymmetric posteriors for most of them. Notice that, when fitting this kind of model, this is 
not an unusual situation, as it can be seen i.e. from Figs. 5 and 7 in~\citetalias{gong19}; especially in the 
case of ACT, the shape of our posterior distributions on $P_0$, $c_{500}$, $\alpha$ and $\beta$ mimics the ones
presented in that work. In the case of \textit{Planck}, however, the contours are slightly larger and their shape
is more regular, which is most likely a result of the smoothing effect induced by the beam 
and of the resulting extended correlation between neighboring $\theta$ bins. The higher resolution of 
ACT results instead in tighter contours and overall in a better quality fit. 

\begin{figure*} % +++++++++++++++++++++++++++++++++++++++++++++++++++++++++++++++
	\includegraphics[trim= 0mm 0mm 0mm 0mm, scale=0.37]{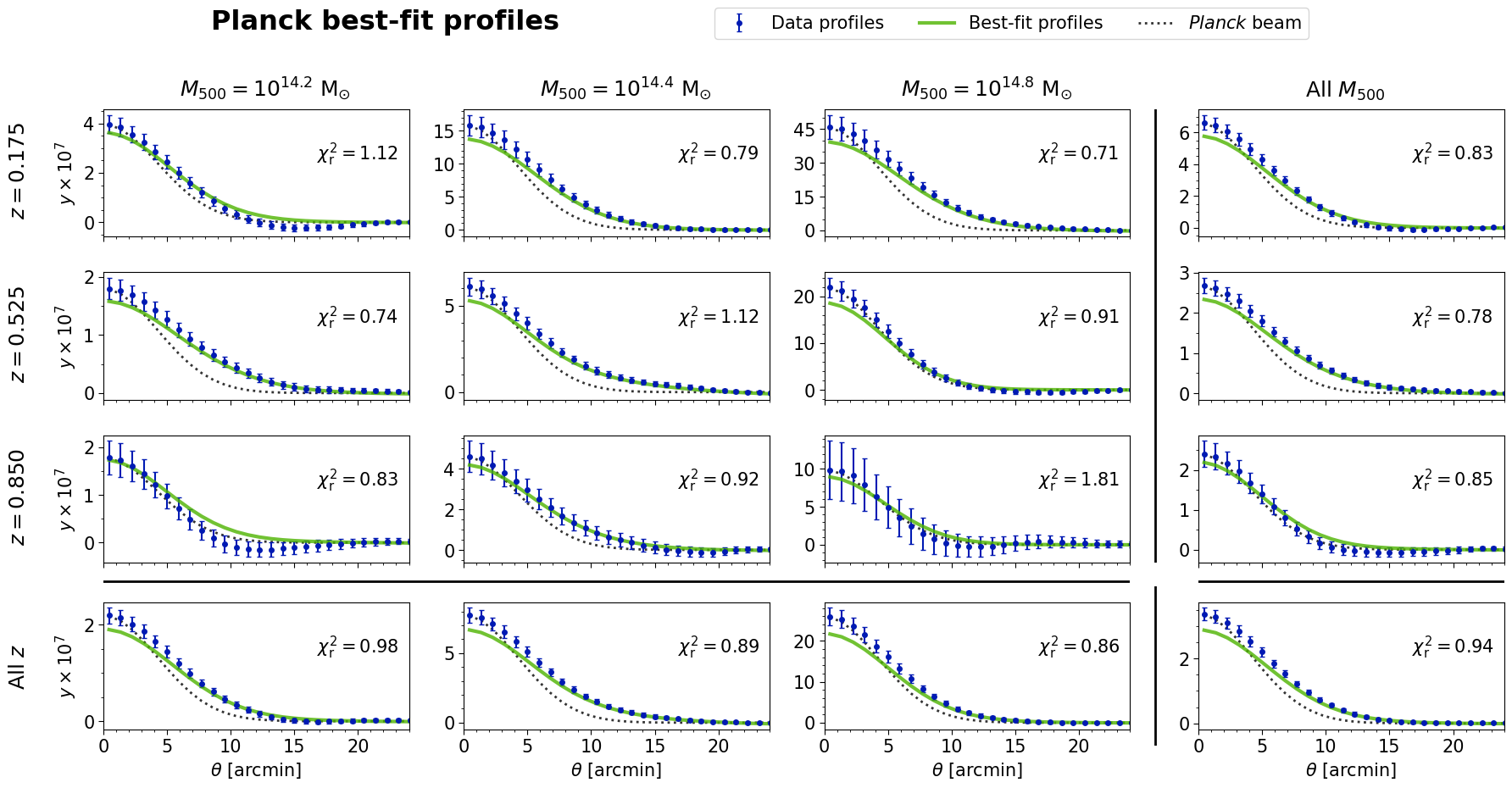}
	\caption{Comparison between the Compton profiles measured on the \textit{Planck} $y$ map 
		(data points) and the theoretical profiles (solid line) computed adopting the best-fit 
		(MAP) parameter values, for all the $(M_{500},z)$ cluster 
		samples considered in this analysis. The plots also show, for reference, the 
		\textit{Planck} beam profile (dashed line) and quote the reduced chi-square values.\label{fig:planck_bfits}}
\end{figure*} % +++++++++++++++++++++++++++++++++++++++++++++++++++++++++++++++

\begin{figure*} % +++++++++++++++++++++++++++++++++++++++++++++++++++++++++++++++
	\includegraphics[trim= 0mm 0mm 0mm 0mm, scale=0.37]{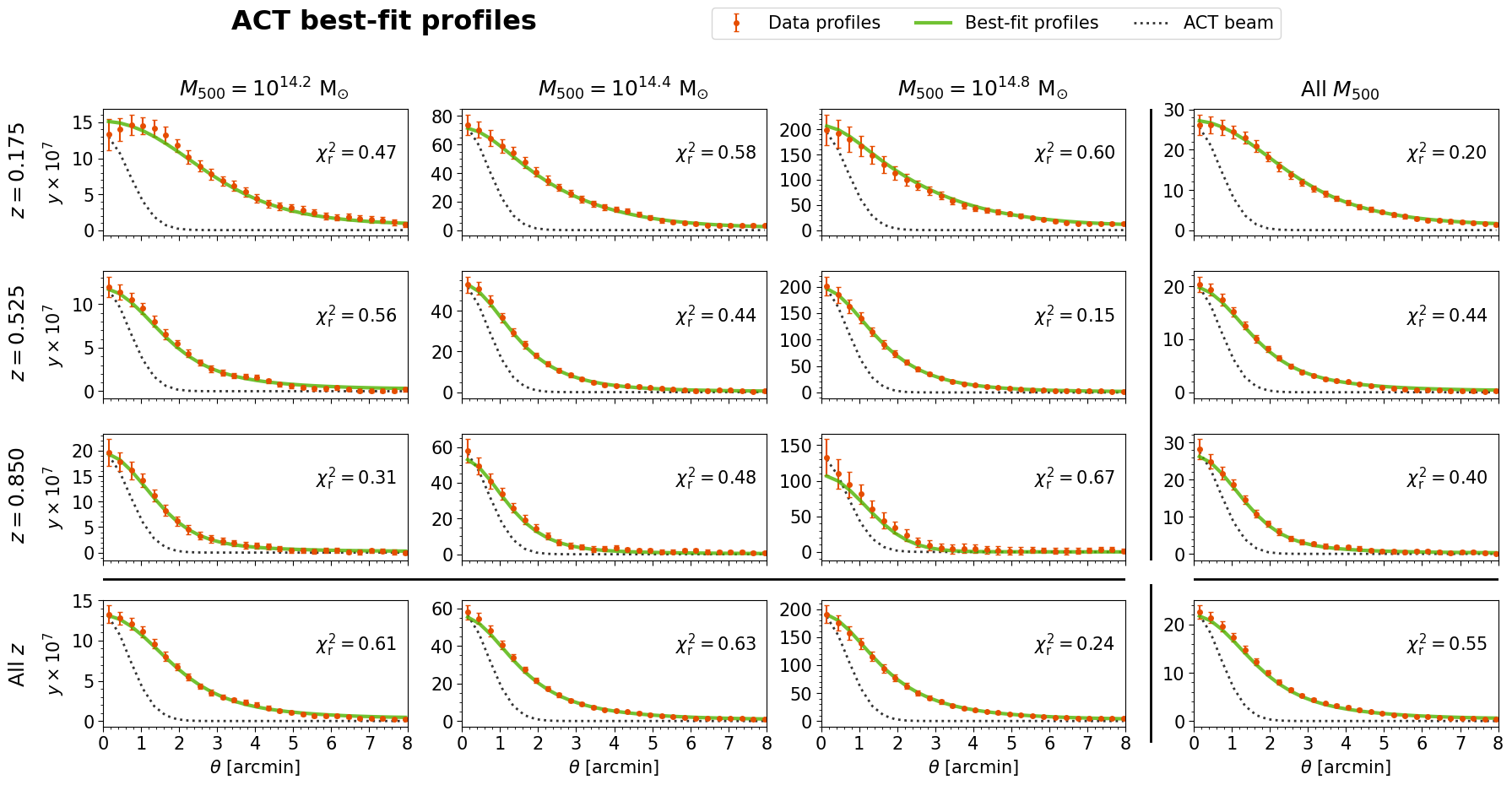}
	\caption{Same as in Fig.~\ref{fig:planck_bfits} but for the ACT profiles.\label{fig:act_bfits}}
\end{figure*} % +++++++++++++++++++++++++++++++++++++++++++++++++++++++++++++++

\begin{figure*} % +++++++++++++++++++++++++++++++++++++++++++++++++++++++++++++++
	\includegraphics[trim= 0mm 0mm 0mm 0mm, scale=0.4]{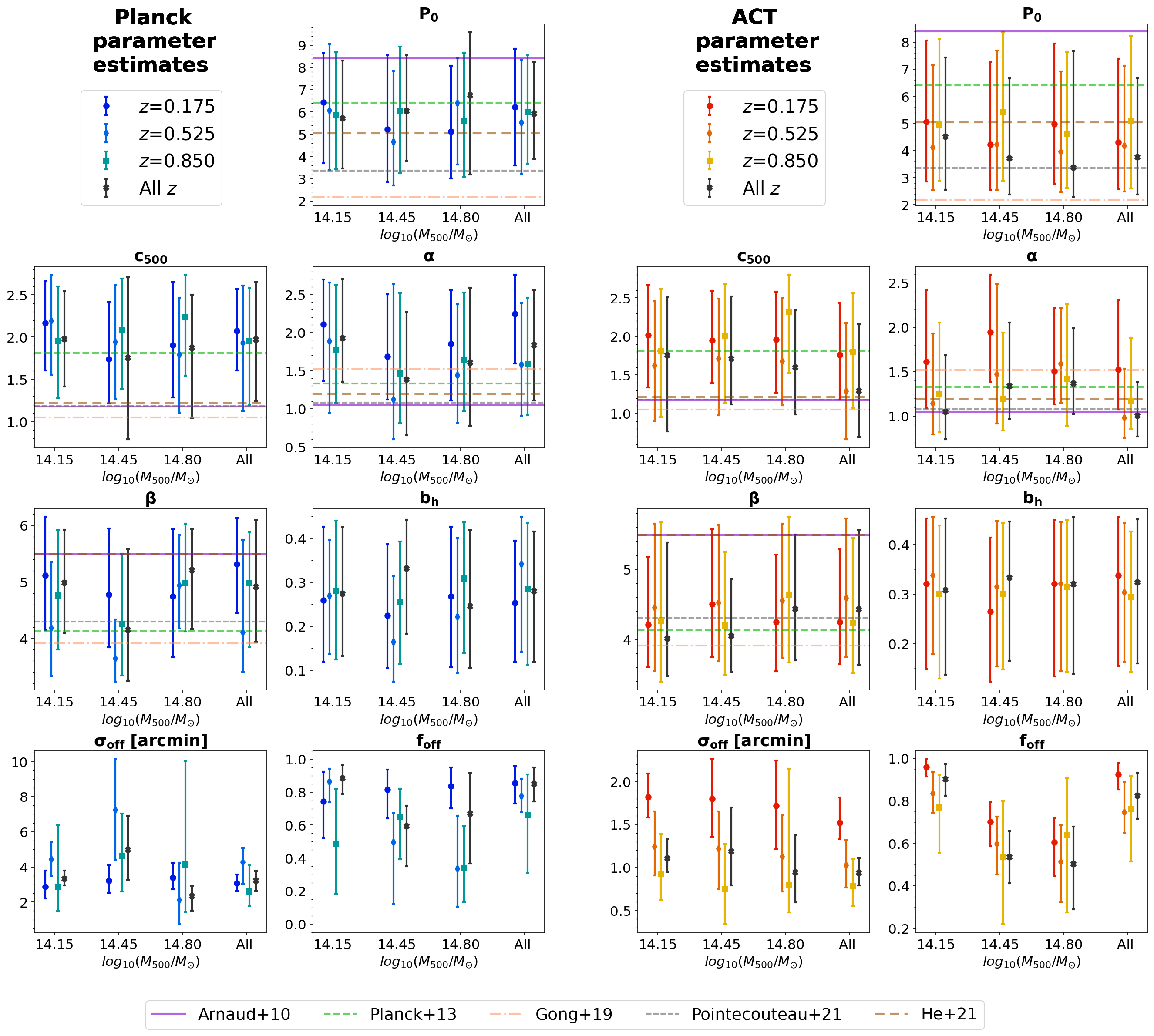}\qquad\qquad
	\caption{Final estimates on our model parameters, computed as the M1D values for the marginalized distributions,
		obtained from the \textit{Planck} (\textit{left}) and ACT (\textit{right}) stacks. Each panel shows the results for 
		an individual parameter, grouping the estimates 
		by mass bin, and showing different redshift bins with different colors and markers 
		as detailed in the plot legend. We also include, in each panel, estimates on the considered parameter obtained by previous works, marked 
		by horizontal solid lines, color-coded as detailed in the bottom of the figure.
		\label{fig:bestfits}}
\end{figure*} % +++++++++++++++++++++++++++++++++++++++++++++++++++++++++++++++

In this context the difference between the MAP and M1D estimates becomes evident. We verified that the choice 
of larger priors would yield poorer fitting results for both \textit{Planck} and ACT: for very extended priors, the 
complex nature of the model leads the chains to encounter local likelihood maxima in positions of the parameter space 
where the actual values of individual parameters are rather unphysical. As a result, not only would the chains 
not reach a proper convergence, but also the final posteriors, obtained by the merged contribution of these 
local maxima, would be artificially broadened and lead to unrealistic M1D estimates. The MAPs, by construction, would 
still yield best-fit profiles that match the data, but they would not necessarily have a real physical meaning;  
their combined numerical values would simply be effective in generating a prediction that matches our measurement, 
but due to the degeneracies in our model, different numerical sets could also provide a good agreement with the data. 
A physical interpretation of the results requires then our M1D estimates and their posterior distributions. 
In principle, in order to improve the chain convergence, we could impose Gaussian priors on some of the parameters 
(e.g., on $c_{500}$, which typically shows the strongest correlations) based on previous results. We verified that  
Gaussian priors with the typical uncertainties found in the literature would override the constraining power of the chains 
and yield posteriors that resemble the chosen prior distributions. In order to keep our analysis independent from 
previous works we then maintain flat priors, chosen with a reasonable width in order to encompass other estimates 
from the literature and at the same time ensure a good convergence of our Markov chains. Clearly, this choice 
has an effect on the resulting M1D estimates, but it is still less obvious than imposing explicit Gaussian 
priors. 

The plots in Figs.~\ref{fig:planck_bfits} and~\ref{fig:act_bfits} show that the models computed using our MAP parameters 
are effective in reproducing the measured profiles across the considered $\theta$ range. The figures also report, 
for each case, the associated reduced chi-square, defined as $\chi_{\rm r}^2=\chi^2/N_{\rm dof}$, where 
$\chi^2=-2\,\ln{\mathcal{L}}$ with $\mathcal{L}$ from Eq.~\eqref{eq:likelihood}, and the number of degrees of 
freedom $N_{\rm dof}$ is computed as the number of angular bins employed in the fit minus the number of free 
parameters. In the case of ACT we always have $\chi_{\rm r}^2<1$, which proves the good agreement between predictions 
and data. In the case of \textit{Planck} we generally have higher values and for some bins $\chi_{\rm r}^2>1$; 
a slight offset of the prediction with respect to the data is also visible in some of the plots. Once more, this 
effect has already been observed in the literature~\citepalias[e.g., Fig. 6 in][]{gong19}. In our case, the offset 
could be a result of the larger \textit{Planck} beam; the latter, indeed, tends to regularize the  
bootstrap profiles and reduce their scatter around the mean, thus probably leading to an underestimation of the 
errors (for the same reason, the significances quoted for our measurements in Table~\ref{tab:stats} are always 
higher for \textit{Planck} than for ACT). Besides, a larger beam implies more important correlations between 
neighbouring angular bins. These two effects result in decreasing the relative importance of the diagonal elements 
in the covariance matrix compared to the off-diagonal entries (as it can also be appreciated from the 
correlation plots in Fig.~\ref{fig:corrmatr}). At the parameter estimation level, this can produce the 
observed offsets in \textit{Planck}'s best-fit profiles. 

We turn now our attention to the M1D estimates, and to the effect of mass and redshift on their values. Fig.~\ref{fig:bestfits}
provides a comprehensive summary, which shows the trend of each parameter when changing $M_{500}$ (different 
positions inside each panel) and $z$ (different colors and marker shapes for the points). The most striking feature 
in these plots is the fact that, 
regardless of the choice of a mass or redshift bin, all estimates on the UPP parameters are compatible within 
$1\,\sigma$. This observation, combined with the good agreement between the best-fit predictions and  
our measurements, suggests that the universal pressure profile is indeed successful in modeling the ICM 
electron pressure over the mass and redshift ranges spanned by our cluster sample. Still, Fig.~\ref{fig:bestfits} 
shows that there are indeed mild variations for the parameters when changing $M_{500}$ or $z$. Such variations, 
however, lack in general a consistent trend, so they are no evidence of an effective mass or redshift residual 
dependence on the UPP parameters. Most likely, this is again a result of the model degeneracy, which also produces the rather large 
error bars on our estimates. Another feature corroborating this hypothesis is the fact that the marginalized cluster samples 
have often values that are not in between the ones obtained from individual bins. We then conclude that the 
observed variations in the estimates for different $M$-$z$ bins are just a result of our parameter estimation, 
and not the evidence of a real residual dependence of the UPP parameters on mass or redshift. We also find a 
substantial agreement (within the error bars) between \textit{Planck}-based and ACT-based estimates of the UPP parameters.

\subsection{Discussion on best-fit estimates}
\label{ssec:discussion3}

We can now compare our findings with the values obtained from previous studies. Fig.~\ref{fig:bestfits} overplots to 
each panel a few horizontal stripes to mark the location of the results obtained from some of the works listed 
in Table~\ref{tab:upp_summary}. We observe that our estimates are in general compatible with these previous results. 
The only exception is the slope at intermediate radii $\alpha$, for which in the case of \textit{Planck} we obtain 
somewhat larger values. We stress however that in~\citetalias{gong19} some of the considered redshift bins yielded even larger values for $\alpha$, up to ~6 in the most extreme case. This large scatter of the estimates 
obtained from different studies is again the result of the existing degeneracies between the UPP parameters. We conclude 
that our estimates confirm the results from previous works; the size of our error bars is also comparable with 
the uncertainties quoted in the literature (when available; see again Table~\ref{tab:upp_summary}). 

We comment that studies based on hydrodynamical simulations have also provided insights into 
possible mass and redshift evolutions of the UPP parameters. The work presented 
in~\citet{battaglia12} suggested that $P_0$, $c_{500}$ and $\beta$ require an explicit mass 
and redshift dependence in order to fit their simulation results (see Table 1 in the reference for details); 
the analysis in~\citet{lebrun15} also confirmed the mass 
dependence of $P_0$ and $c_{500}$, finding both parameters increase with $M_{500}$ (Table 2 in the reference). 
The mass range explored in these works encompasses the one 
we probe with our cluster sample. We stress, however, that the mass and redshift dependences
detected in those studies are in all cases very mild, and can be measured only with the high 
resolution provided by numerical simulations. In fact, the works in~\citet{battaglia12}
and~\citet{lebrun15} probe the cluster pressure profile down to radial separations of $r\sim 0.04\,R_{200}$ (Figs. 1 and 2 in the reference) and 
$r\sim 0.1\,R_{500}$ (Fig. 3 in the reference), respectively; for our lowest-redshift and highest-mass clusters these 
values translate into angular scales of the order of $\sim1\,\text{arcmin}$, which even for 
ACT are comparable with the beam, and where we typically have the largest uncertainties in our 
profile measurements. Hence, those mass and redshift variations in the UPP parameters are 
definitely subdominant in the context of reconstructing 
the pressure profile from real data, due to a series of factors such as the instrumental beam, 
residuals in the $y$ map, systematics in the cluster mass estimates and miscentering effects. 
As far as our measurements are concerned, if an effective dependence of the UPP parameters on 
$M_{500}$ and $z$ does exist, it is well below the uncertainties in our final estimates. 

We move on now to comment our results on the hydrostatic mass bias $b_{\rm h}$. Our estimates are typically in the range 
from 0.2 to 0.3, with somewhat larger values obtained with ACT; the confidence intervals are in any case quite broad, with 
typical error bars up to $\sim 0.2$, and no evident strong degeneracies with the other parameters. A number of 
different estimates for $b_{\rm h}$ have been provided in the literature; for a summary, see e.g. table~3 
in~\citet{ibitoye22}. Our results are again in agreement with 
previous findings. Numerical simulation results tend to agree in that the hydrostatic bias has a mass dependence, and can 
accommodate values for $b_{\rm h}$ as large as 0.3 for massive clusters~\citep{pearce20, barnes21}. Our findings are also 
in agreement with cross-correlation analyses~\citep{makiya20, rotti21, ibitoye22} and studies of variously
selected cluster samples~\citep{von_der_linden14, hoekstra15, sereno17, ferragamo21, aguado_barahona22}. We stress, however, 
that our MCMC analysis does not provide strong constraints on the hydrostatic bias.

Finally, the miscentering parameters are the ones that show the largest scatter across different $M$-$z$ bins. 
In this case a comparison between \textit{Planck} and ACT is not meaningful, as we set different priors on 
$\sigma_{\rm off}$ to account for the considerably different beam size. We find values for $\sigma_{\rm off}$
around 4 arcmin in the case of \textit{Planck} and around 1.5 arcmin in the case of ACT; for the latter, the 
miscentering offset is then comparable with the beam FWHM, and is therefore a relatively more important effect, 
as expected. As for the miscentering fraction $f_{\rm off}$, we find that in general more than 50\%
of the clusters in each sample are offset from their nominal position; a mild anti-correlation can be observed 
between $\sigma_{\rm off}$ and $f_{\rm off}$ in Fig.~\ref{fig:posteriors}, which is easily understandable as 
these parameters produce opposite effects in quantifying the mean miscentering. We also notice from 
Fig.~\ref{fig:bestfits} that this time it is possible to recognize a trend in the M1D values, especially for ACT,
with lower redshifts and lower masses requiring a higher miscentering offset. This could have been anticipated by  
looking at the stacks plotted in Fig.~\ref{fig:act_stacks}, where the lowest masses and redshifts tend to have 
broader and more irregular profiles. These results confirm that the miscentering is a necessary inclusion in our 
theoretical modeling, without which the final estimates on the other UPP parameters would most likely be biased. 
For example we tested that, without including the miscentering in our theoretical prediction, the resulting 
estimates on the parameter $P_0$ are typically very low and unphysical ($P_0\lesssim 2$); this can be 
understood as $P_0$ is the parameter that most directly controls the amplitude of the measured signal, 
and as such most readily absorbs any dilution effect produced by the miscentering on the profiles. 

Before closing this section, we comment on the effect that the mass rescaling we applied to WHL and DESI clusters
has on our parameter estimates. Figs.~\ref{fig:posteriors_cmp} to~\ref{fig:act_contours_marg} show the comparison
between the final contours obtained from our fiducial profiles and from the unscaled profiles, for all $M$-$z$ bins
and for both \textit{Planck} and ACT. The resulting best-fit predictions for the unscaled profiles (computed based on the 
associated MAP estimates) are overplotted to the profiles themselves in Figs.~\ref{fig:planck_profcmp} 
and~\ref{fig:act_profcmp}. Regarding the agreement between 
predictions and measurements for the unscaled profiles, similar considerations as for the case of the fiducial profiles 
apply. As for the final M1D parameter constraints, it is clear from the contour 
plots that the posteriors for the unscaled profiles are almost indistinguishable from the posteriors obtained 
for the fiducial profiles. Even for the case of the lowest redshift bin, where the two sets of profiles show the 
largest tensions, the final parameter estimates are rather compatible. This shows once more that the final contour 
sizes are largely determined by our prior choice and by the correlation between the parameters entering our model;
such correlations also result in large error bars and make the estimates overall in agreement. Clearly, the actual, 
final M1D values are not exactly the same, and 
one could quote the difference between them as a systematic error component to be included in our final parameter 
uncertainties. In our case, however, such systematic contribution will be much smaller than the statistical uncertainties 
quoted in Tables~\ref{tab:planck_bestfits} and~\ref{tab:act_bestfits}. We then conclude that the mass rescaling presented 
in Section~\ref{ssec:mcal} has a negligible impact on the final conclusions of this study.

%=================================================================================================
%=================================================================================================

%=================================================================================================
%========================================  CONCLUSIONS ===========================================
%=================================================================================================

\section{Conclusions}
\label{sec:conclusions}

The cluster pressure profile is one of the primary tools to explore the physical state of the 
ICM; as it is clear from Eq.~(\ref{eq:y}), the Compton parameter $y$ is a very direct 
probe of the ICM electron pressure, and it has been exploited in this sense by several 
works in the past decade. 
In this paper, we explored possible mass and redshift dependences of the parameters governing 
the shape of the universal pressure profile in galaxy clusters, by analyzing the $y$ 
profiles obtained from cluster stacks on Compton parameter maps. 

We employed the   
 $y$-maps delivered by both the \textit{Planck} satellite and the Atacama Cosmology Telescope, 
the latter limiting the analysis to an effective sky area of $\sim2,000\,\text{deg}^2$ but 
at the same time providing a considerably higher angular resolution in the reconstructed 
tSZ signal. We built a large cluster sample by merging existing galaxy cluster catalogs based on 
observations from KiDS, SDSS (WHL) and DESI. As cluster masses from these catalogs were estimated 
following different methodologies, we first homogenized the mass definition by 
scaling the WHL and DESI cluster masses to the KiDS definition, which is based on weak lensing
measurements and is as such less affected by \textit{ad hoc} assumptions on the ICM physical state. 
The scaling parameters were obtained by comparing the masses from common clusters across 
pairs of catalogs, for different redshift intervals. After applying a lower mass cut of   
$10^{14}\,\text{M}_{\odot}$ (below which we found the stacks would become too noisy) and removing
repeated clusters, we merged the three catalogs obtaining a final sample of 23,820 clusters 
overlapping with the ACT map footprint.

We split these clusters into three mass and three redshift bins, also considering the 
respective marginalized cases for a total of 16 different cluster samples. We stacked these 
samples on both the \textit{Planck} and the ACT maps, in all cases obtaining a clear 
detection of the cluster signal against the background. We extracted a circularly symmetric 
radial angular profile from each stack map, and computed the associated covariance matrix 
by repeating the stacks with a set of 500 replicas of the catalog obtained via bootstrap 
resampling. The covariance matrices allowed us
to determine the uncertainties to be assigned to the angular profiles and to compute 
the significance per bin in their measurements, which is always larger than 13 for 
\textit{Planck} and 3 for ACT. 

We modeled theoretically the mean $y$ profile with a halo-model approach, taking into account 
the effective cluster mass and redshift distributions in each sample. The theoretical 
predictions depend not only on the UPP parameters, but also on the hydrostatic bias on the 
cluster mass, and on two parameters quantifying the magnitude and occurrence of possible 
miscentering of the clusters from their nominal positions. We then employed the MCMC method to 
reconstruct the posterior distributions on these parameters with initial flat priors, where the likelihood compares 
the theoretical prediction with the observed profile using the covariance matrix measured
for each cluster sample. In all cases we fixed the pressure profile central slope to $\gamma=0.31$, 
as it is customary in other works based on tSZ data alone. From the MCMC runs we extracted two 
sets of parameters, the MAPs from the full 7-dimensional likelihood, and the M1Ds from the marginalized 1-dimensional
posteriors.

The profile predictions computed with the MAPs provide indeed a good fit to our measurements, yielding 
$\chi^2_{\rm r}<1$ for almost all cases. The M1D estimates show good agreement between \textit{Planck} and ACT, 
and with constraints obtained by previous works. The results also do not show any compelling evidence 
for a residual dependence of the UPP parameters with either $M_{500}$ or $z$. Although marginal differences are 
visible, there is no clear trend and the values are largely compatible within the recovered error bars. The main 
conclusion from this work is that the adopted UPP functional form is effective in 
describing the ICM electron pressure profile for clusters in the mass range 
($10^{14.0}\,\text{M}_{\odot}<M_{500}<10^{15.1}\,\text{M}_{\odot}$) and in the redshift range 
($0.02<z<0.97$) explored with our clusters. This is the first time the UPP is tested over such a large 
cluster sample, which is mostly complete within the chosen $M_{500}$ and $z$ limits.
We also obtain loose constraints on the hydrostatic mass 
bias in agreement with previous works based on both numerical simulations and analyses of 
cluster samples. We prove that miscentering is an important piece of the cluster profile modeling, 
with more than 50\% of the clusters being offset from their nominal position by an amount 
commensurate with the FWHM value of the corresponding $y$ map. Finally, we showed that possible 
systematic errors induced by our explicit mass rescaling are well below the statistical uncertainties 
obtained for each parameter from the MCMC analysis. 

%=================================================================================================
%=================================================================================================

\section*{Acknowledgements}

We would like to thank Yan Gong, Zhong-Lue Wen, Joachim Harnois Deraps and Xiaohu Yang for their useful discussions. Based on data products from observations made with ESO Telescopes at the La Silla Paranal Observatory under programme IDs 177.A-3016, 177.A-3017 and 177.A-3018, and on data products produced by Target/OmegaCEN, INAF-OACN, INAF-OAPD and the KiDS production team, on behalf of the KiDS consortium. OmegaCEN and the KiDS production team acknowledge support by NOVA and NWO-M grants. Members of INAF-OAPD and INAF-OACN also acknowledge the support from the Department of Physics \& Astronomy of the University of Padova, and of the Department of Physics of Univ. Federico II (Naples).
DT acknowledges the support from the Chinese Academy of Sciences (CAS) President's International Fellowship Initiative (PIFI) with Grant N. 2020PM0042, and from the National Natural Science Foundation of China (NSFC) Research Fund for International Scientists (RFIS) with Grant N. 12150410315. YZM acknowledges the support of National Research Foundation with grant no. 120385 and 120378, and SARAO group grant. Project 12047503 is supported by National Natural Science Foundation of China. This work was part of the research programme ``New Insights into Astrophysics and Cosmology with Theoretical Models confronting Observational Data'' of the National Institute for Theoretical and Computational Sciences of South Africa. ZY acknowledges support from the Max Planck Society and the Alexander
von Humboldt Foundation in the framework of the Max Planck-Humboldt Research Award endowed by the Federal Ministry of Education and Research (Germany). CG and LM acknowledge the support from the grant PRIN-MIUR 2017 WSCC32 ZOOMING, and the support from the grant ASI n.2018-23-HH.0. CG acknowledges funding from the Italian National Institute of Astrophysics under the grant "Bando PrIN 2019", PI: Viola Allevato. AHW is supported by an European Research Council Consolidator Grant (No. 770935). MS acknowledges financial contributions from contract ASI-INAF n.2017-14-H.0 and contract INAF mainstream project 1.05.01.86.10.

\appendix

\section{Summary of previous works}
\label{app:upp_summary}
We provide here further information about previous results for the UPP parameters, 
as an integration of the discussion presented in Section~\ref{sec:introduction}. We focus in 
particular on the cluster-based studies which are listed in Table~\ref{tab:upp_summary}.

The work in~\citet{arnaud10} considered 33 clusters with 
$M_{500}\in[1,10]\times10^{14}\,\text{M}_{\odot}$ at $z<0.2$ observed with 
\textit{XMM-Newton}, and compared their individual pressure profiles, each scaled by the 
characteristic pressure $P_{500}$; these clusters were selected from the REFLEX Cluster Survey 
imposing a lower X-ray luminosity threshold of $0.4\times10^{44}\,h^{-2}\,\text{erg}\,\text{s}^{-1}$ 
in the $0.1-2.4\,\text{keV}$ band~\citep[the REXCESS sample,][]{bohringer07}. REXCESS is by 
construction a representative sample of an X-ray flux-limited cluster population, which does not privilege
specific morphologies or dynamical states in its member clusters.
Although finding deviations in the central region between cool-core and morphologically disturbed 
systems, the scaled profiles in~\citet{arnaud10} showed a good 
agreement at larger radii, up to $R_{500}$. The UPP parameters were fitted on the combination 
of the mean X-ray data profile with the mean profile obtained from numerical simulations, 
the latter allowing to extend the profile reconstruction beyond $R_{500}$. This work has been traditionally 
taken as a reference in all subsequent studies of cluster pressure profiles. 

A similar analysis was conducted in~\citet{planck_ir_v} on a set of 62 tSZ-detected clusters with 
$M_{500}\in[2,20]\times10^{14}\,\text{M}_{\odot}$ at $z<0.45$; this cluster sample had already 
been used to calibrate the tSZ-mass scaling relation in~\citet{planck_er_xi}, where it was selected 
from the \textit{Planck} early SZ source catalog~\citep[ESZ,][]{planck_er_viii} 
on the basis of existing high quality \textit{XMM-Newton} observations. This time, 
information on the pressure profile was obtained from the reconstructed cluster Compton profile, 
which allowed the authors to explore the ICM gas out to $\sim 3\,R_{500}$; the average, derived pressure profile 
was combined with the average profile obtained from X-ray data. The inclusion of X-ray data allowed 
to reconstruct the pressure profile down to $0.02\,R_{500}$, yielding substantial 
agreement with SZ data in the overlap range. Again, marginal differences were observed between cool-core 
and non cool-core clusters, but within the statistical error bars. Compared with results from numerical 
simulations, the profile in the cluster outskirts was found to be rather flatter, 
while providing a good agreement at low radii with simulations that implement AGN feedback. 

The work in~\citet{sayers16} considered a set of 
47 clusters with $M_{500} \in [3,25]\times10^{14}\,\text{M}_{\odot}$ at $z<0.9$, chosen from observations with 
\textit{Chandra} and Bolocam~\citep{sayers11} on the basis of their redshift and high X-ray 
temperature; this sample slightly extended the one already studied in~\citet{czakon15} with the inclusion of 
two additional clusters. This work focused on the reconstruction of the pressure profile in the cluster 
outskirts, based on measurements of the integrated $y$ profile from Bolocam and \textit{Planck} data. 
More precisely, it kept all UPP parameters fixed to the~\citet{arnaud10} estimates with 
the exception of the normalization $P_0$ and the profile slope $\beta$ at large radii. The best-fit values
were found in agreement with results from numerical simulations over the same mass and redshift span of the 
considered cluster sample. The authors also found evidence for a residual mild dependence of the profile slope on 
the cluster mass, with more massive clusters favoring higher values of $\beta$. Finally, the work acknowledged 
how these results can be affected by systematics such as sample selection and calibration of cluster masses.

\begin{figure*} % +++++++++++++++++++++++++++++++++++++++++++++++++++++++++++++++
	\includegraphics[trim= 0mm 0mm 0mm 0mm, scale=0.4]{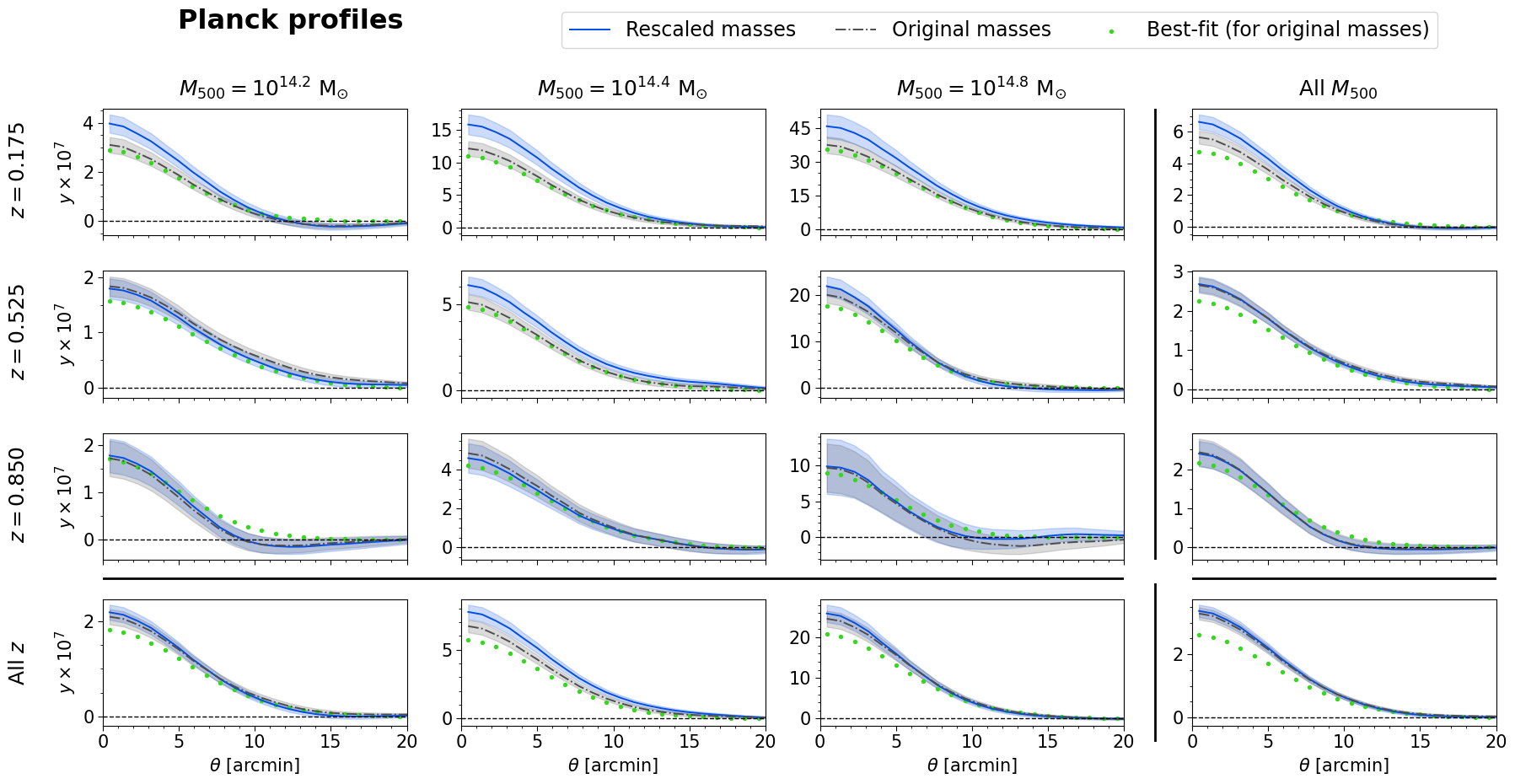}
	\caption{Comparison between the angular $y$ profiles measured on the \textit{Planck} map with the fiducial sample (blue)
	and the unscaled sample (gray). For both cases the shaded areas quantify the associated 
	uncertainties. We also show with green dots the model predictions for the unscaled profiles, computed 
	using the MAP estimates from the associated MCMC runs.
	\label{fig:planck_profcmp}}
\end{figure*} % +++++++++++++++++++++++++++++++++++++++++++++++++++++++++++++++

\begin{figure*} % +++++++++++++++++++++++++++++++++++++++++++++++++++++++++++++++
	\includegraphics[trim= 0mm 0mm 0mm 0mm, scale=0.4]{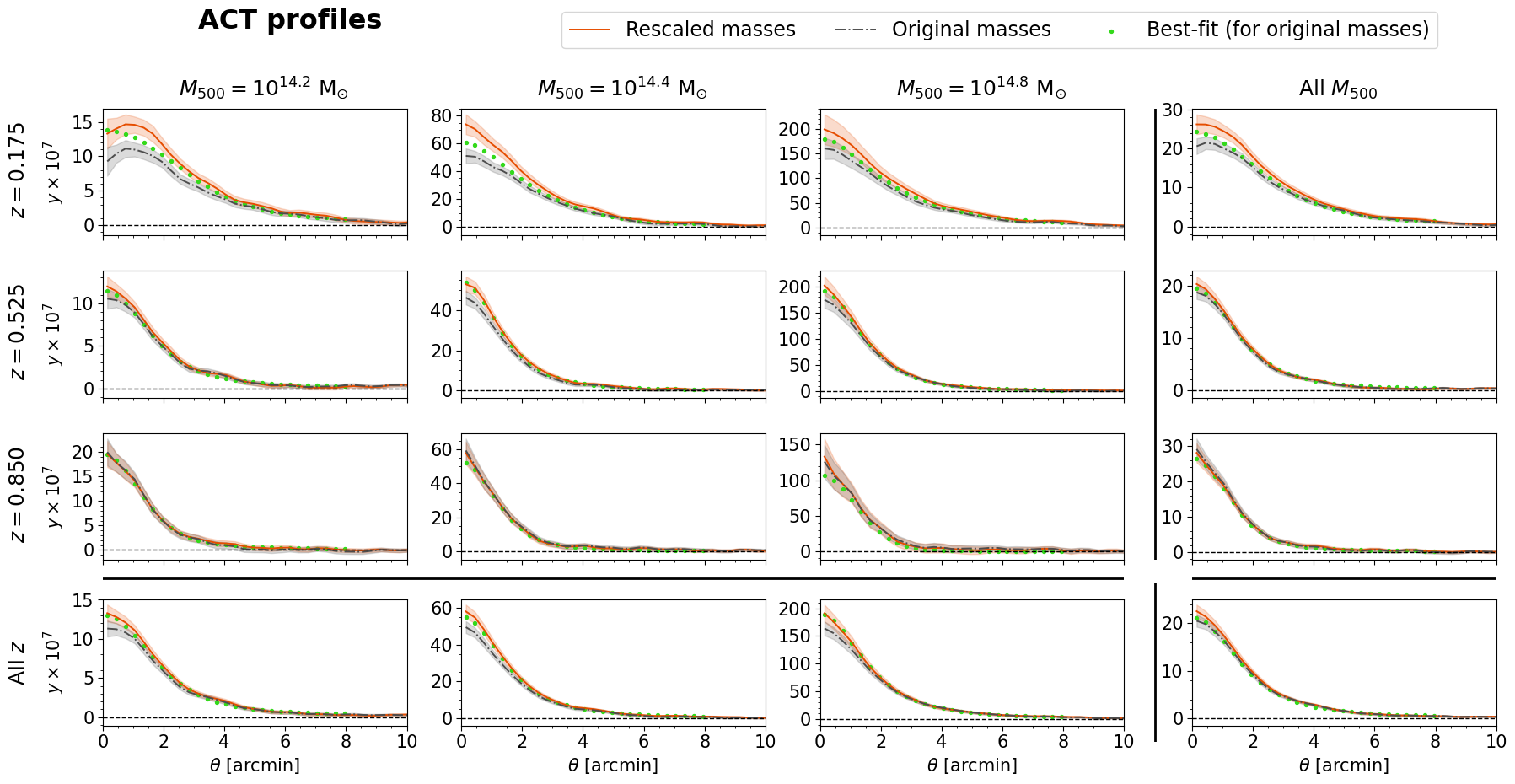}
	\caption{Same as in Fig.~\ref{fig:planck_profcmp}, but for the profiles measured 
	on the ACT map. The profiles obtained with the fiducial sample are this time plotted 
	in red color.\label{fig:act_profcmp}}
\end{figure*} % +++++++++++++++++++++++++++++++++++++++++++++++++++++++++++++++

In~\citet{pointecouteau21} the cluster sample consisted of 31 clusters with 
$M_{500}\in[3.4,13.1]\times10^{14}\,\text{M}_{\odot}$ at $z<0.71$, which had been previously listed as SZ sources 
in both \textit{Planck}- and ACT-based catalogs. A major novelty of this work is that the SZ signal was 
measured on a joint \textit{Planck}-ACT Compton parameter map, built from the linear combination of individual 
frequency maps from the two surveys as described in~\citet{aghanim19}. The authors extracted $y$ profiles (and derived the 
associated pressure profiles) for individual clusters in the sample and fitted the resulting mean pressure profile
with a UPP model. As the fitting results were strongly affected by parameter degeneracies, the authors fixed both 
$\gamma$ and $c_{500}$ to the values from~\citet{arnaud10}. Estimates for the remaining UPP parameters showed 
broad agreement with previous findings, particularly when it comes to the profile amplitude at the outer radii, 
while a somewhat larger tension was found at the intermediate radii; the authors mentioned the higher relevance of 
ACT data (a novelty in this analysis) in this radial range as a possible explanation, while stressing again 
the limitations inherent to the use of a relatively small, non-representative sample. 

Finally, \citet{he21} employed the same cluster sample (REXCESS) as in~\citet{arnaud10} and focused on  
assessing the effect on scaling relations and UPP parameter estimates deriving from the bias between the
true cluster mass and the hydrostatic cluster mass sampled by X-ray and SZ observations (see also 
Section~\ref{ssec:clusterpress}), an issue that was already acknowledged in~\citet{arnaud10}. The authors fitted
the scaling between the two mass definitions on hydrodynamic simulations and employed it to correct cluster 
masses in the REXCESS sample, finding the initial hydrostatic mass values were underestimated by 7\% on 
average, the effect being larger for higher masses. The scaling was then quantified via a hydrostatic mass bias
and incorporated in the formalism by scaling the UPP normalization pressure $P_0$ and concentration $c_{500}$ 
(this is equivalent to our treatment of the bias described in Section~\ref{ssec:clusterpress}, with the difference 
that we kept the UPP functional form unchanged and scaled the values of $M_{500}$ and $R_{500}$ instead). The 
authors fitted this modified UPP model on the new mean pressure profile, finding the resulting prediction to yield 
a reduction in the deviation (quantified by the term in Eq.~\eqref{eq:ssbreak}) from the self-similar model, 
compared to the original UPP profile.

The list of studies described in this section, and also the summary reported in Table~\ref{tab:upp_summary}, 
are by no means exhaustive; as the present paper is not intended to be a review on the subject, we redirect to the 
additional references cited in those works for further reading. We chose to present and discuss 
this particular selection of papers to highlight the novel aspects introduced by each of them, namely the 
systematic application of the UPP to cluster pressure profiles derived from X-ray data~\citep{arnaud10}, 
the extension of a similar study to profiles reconstructed from SZ data~\citep{planck_ir_v}, 
considerations on possible mass and redshift dependences of UPP parameters~\citep{sayers16},
the introduction of ACT data~\citep{pointecouteau21} and the importance of the hydrostatic mass 
bias~\citep{he21}. All of these aspects are also considered in our data analysis and theoretical 
modeling; the change in 
the approach from considering a reduced number of clusters to reconstructing the statistical properties of an 
extended sample, as in~\citet{gong19}, is instead particularly relevant and is then described in the main text 
in Section~\ref{sec:introduction}.

\begin{figure*} % +++++++++++++++++++++++++++++++++++++++++++++++++++++++++++++++
	\includegraphics[trim= 0mm 0mm 0mm 0mm, scale=0.42]{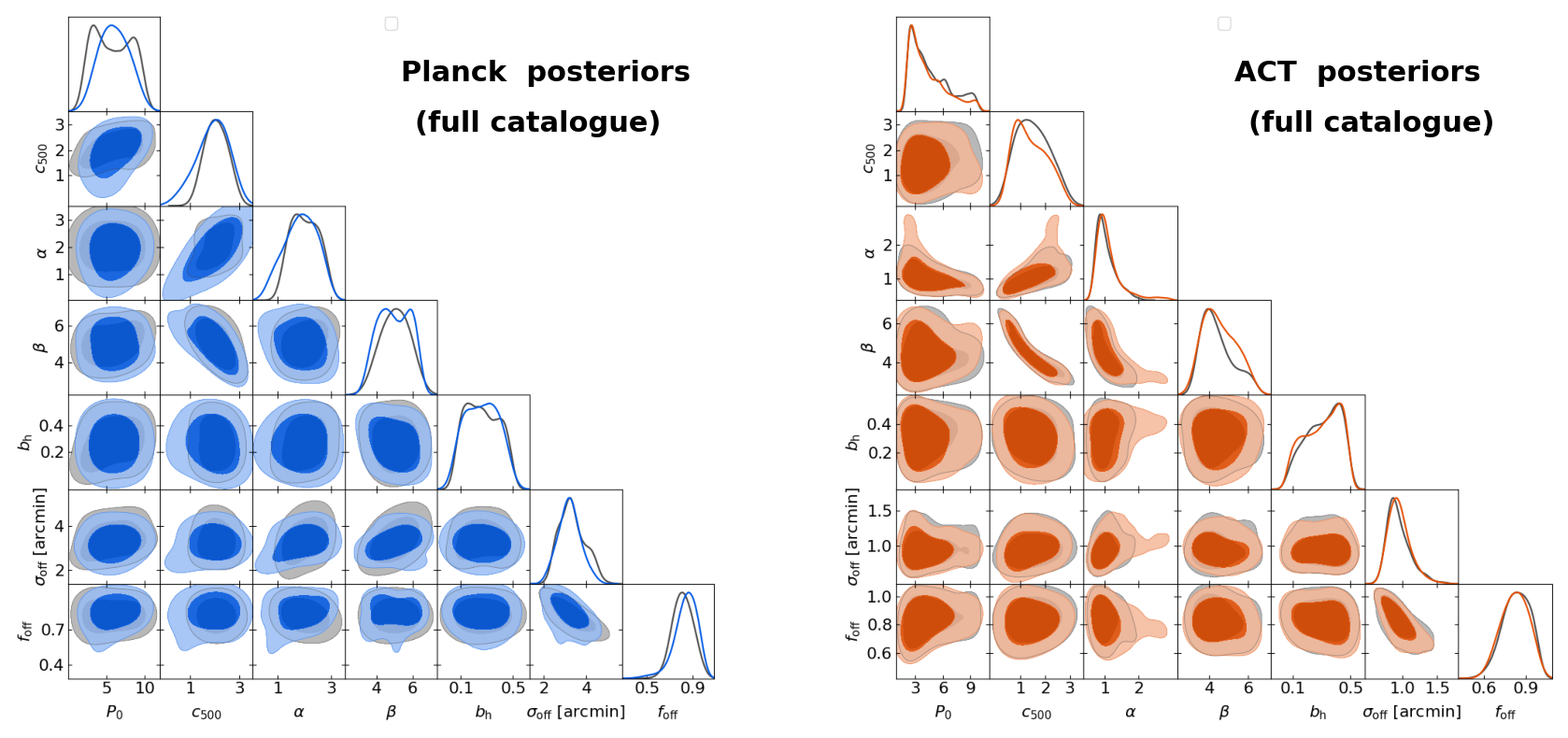}
	\caption{Comparison between the final posteriors (plotted as 68\% and 95\% C.L. contours) 
	obtained using the cluster sample with and 
	without mass rescaling, for the case of \textit{Planck} (\textit{left}) and ACT (\textit{right}). The colored 
	contours were obtained with the mass-rescaled catalog, and are the same as the ones shown 
	in Fig.~\ref{fig:posteriors}. Gray contours were derived from the catalog obtained with 
	no mass-rescaling.\label{fig:posteriors_cmp}}
\end{figure*} % +++++++++++++++++++++++++++++++++++++++++++++++++++++++++++++++

Looking back at Table~\ref{tab:upp_summary}, the best-fit numerical values resulting from these studies 
show a large scatter. Two main factors contribute to the observed differences. First of all, the UPP parameters are 
intrinsically degenerate, as it is clear from the functional form in Eq.~\ref{eq:upp}; this implies that 
different combinations of parameter values can provide equally effective predictions for the observed pressure 
profiles. A possible way to circumvent this issue, adopted for example by~\citet{sayers16} and~\citet{pointecouteau21}, 
is to keep some of the parameters fixed in the analysis, which comes at the price of a slight loss in the 
generality of the fitted model. In our study we chose, in a more bias-free approach as in~\citet{gong19}, 
to keep all parameters free, with the exception of the inner slope $\gamma$ as it is customary in purely 
SZ-based analyses. Still, we acknowledge that parameter degeneracy is an important issue, and as such it 
is extensively addressed in our discussion of the fit results in Sections~\ref{ssec:discussion2} and~\ref{ssec:discussion3}.

The second main reason for the scatter observed in Table~\ref{tab:upp_summary} is that, 
whenever the chosen sample is restricted to a handful of clusters which are well resolved and 
characterized by observations at different wavelengths, the results on the UPP estimates are 
inevitably subject to potential selection biases. This is in generally acknowledged in the studies described
in this section. In particular, it is worth stressing that cluster samples constructed on the basis of existing 
high-quality data in ancillary studies are non-representative, which prevents the conclusions 
from the corresponding studies to be extended to a generic cluster with mass and redshift in the sample span. 
The only exceptions from the above list are the works based on REXCESS, which is purposely built as a representative sample. 
The cluster sample used in this work is not only representative (as it includes all clusters with estimated mass above 
a common threshold) but also complete (typically $>90\%$, see Section~\ref{sec:catalogs}), which is a 
fundamental difference compared to the previous studies. Possible dependences of the UPP parameters on cluster mass and redshift is 
actually one of the core topic of our analysis, and is explored by binning our cluster sample in different $M_{500}$ and $z$ bins as detailed in Section~\ref{ssec:binning}.  

\section{Systematic errors from mass re-scaling}
\label{app:systematics}

We consider an alternative version of our reference cluster sample, obtained by merging KiDS, WHL and 
DESI only after imposing the lower mass cut $M_{500}>10^{14}\,\text{M}_{\odot}$, but without applying
any mass rescaling to WHL and DESI. We remove cluster repetitions following the same criteria as in 
Section~\ref{ssec:merging}, i.e., we always keep KiDS clusters while for the remaining repetitions between 
WHL and DESI we randomly choose which one to discard. In the end, KiDS, WHL and DESI contribute with 
806, 15,114 and 9,684 clusters respectively, for a total of 25,604 clusters spanning again the redshift 
range $0.02<z<0.97$ and the mass range $14.0<\log{\left(M_{500}/\text{M}_{\odot}\right)}<15.2$. The cluster 
redshift and mass distributions are qualitatively similar to the ones shown in Fig.~\ref{fig:distrib}. 
We then split the clusters over the same set of $M$-$z$ bins considered for our fiducial sample, and proceed
with the measurement of the stacked signal, the angular profiles and the covariance matrices following 
the same methodology described in Section~\ref{ssec:stacks}. We do not include plots of the resulting 
stacks and correlation matrices, as they are similar to the ones already shown for our fiducial sample. 
Instead, we show a comparison between the measured profiles against the ones from the fiducial sample
in Fig.~\ref{fig:planck_profcmp} for \textit{Planck} and~\ref{fig:act_profcmp} for ACT. We notice that 
for the lowest redshift bin there is a clear offset in the profile amplitude; this difference is instead 
practically negligible for the highest redshift bin, which is mostly dominated by DESI clusters that 
(for $z>0.80$) did not undergo any explicit mass rescaling. In general, the difference is also quite mild
for the marginalized cases, with the profiles obtained from the full samples showing compatibility over 
the whole $\theta$ range.

\begin{figure*} % +++++++++++++++++++++++++++++++++++++++++++++++++++++++++++++++
	\includegraphics[trim= 0mm 0mm 0mm 0mm, scale=0.3]{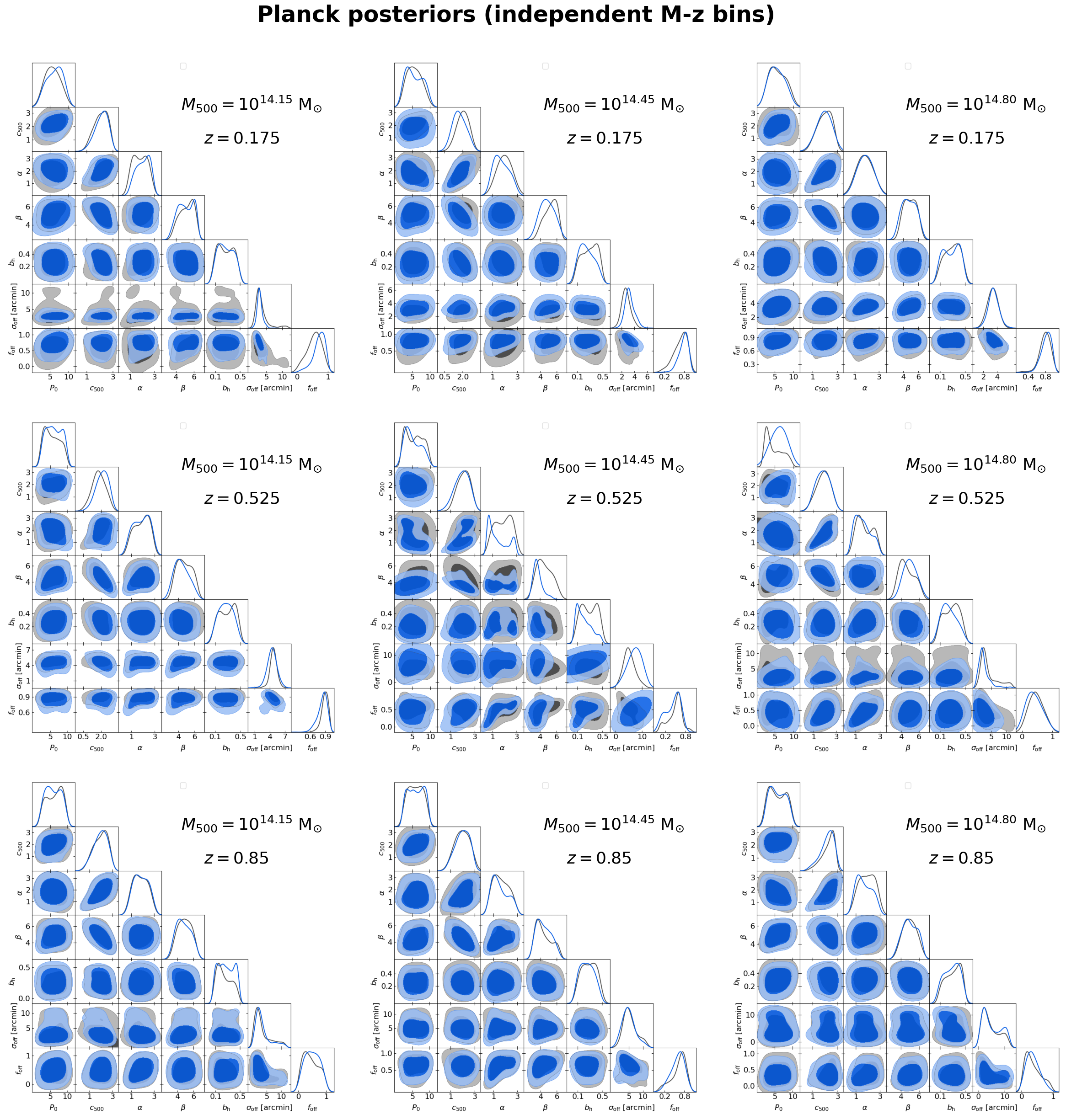}
	\caption{\textit{Planck} posterior distributions (68\% and 95\% C.L. contours) on the fitted parameters for each independent 
	$M$-$z$ bin, obtained from the mass-rescaled catalog (blue) and the one with no mass 
	rescaling (gray).\label{fig:planck_contours}}
\end{figure*}

\begin{figure*} % +++++++++++++++++++++++++++++++++++++++++++++++++++++++++++++++
	\includegraphics[trim= 0mm 0mm 0mm 0mm, scale=0.3]{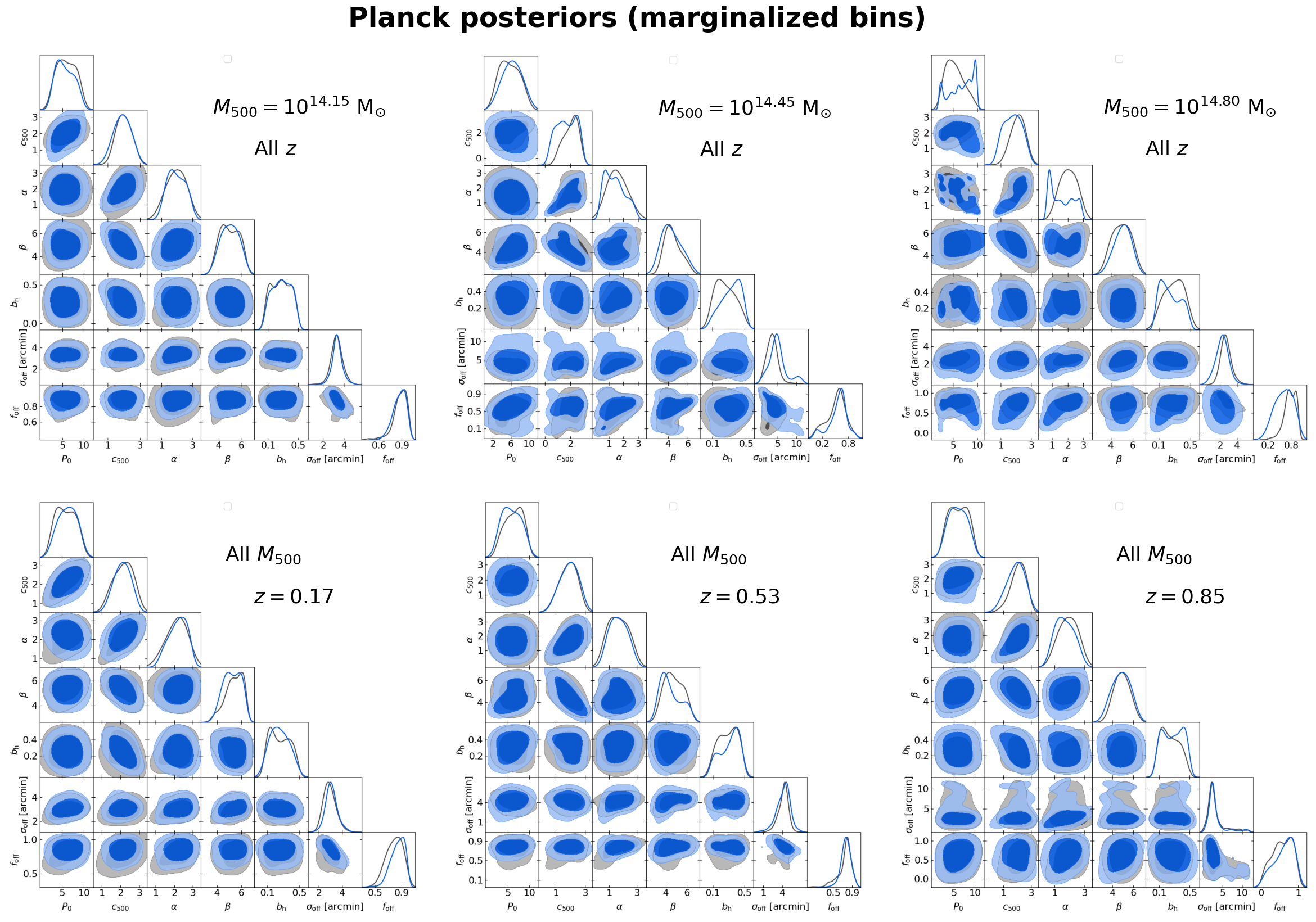}
	\caption{Same as in Fig.~\ref{fig:planck_contours} but showing this time the bins marginalized 
	over $M_{500}$ or $z$.\label{fig:planck_contours_marg}}
\end{figure*}

\begin{figure*} % +++++++++++++++++++++++++++++++++++++++++++++++++++++++++++++++
	\includegraphics[trim= 0mm 0mm 0mm 0mm, scale=0.3]{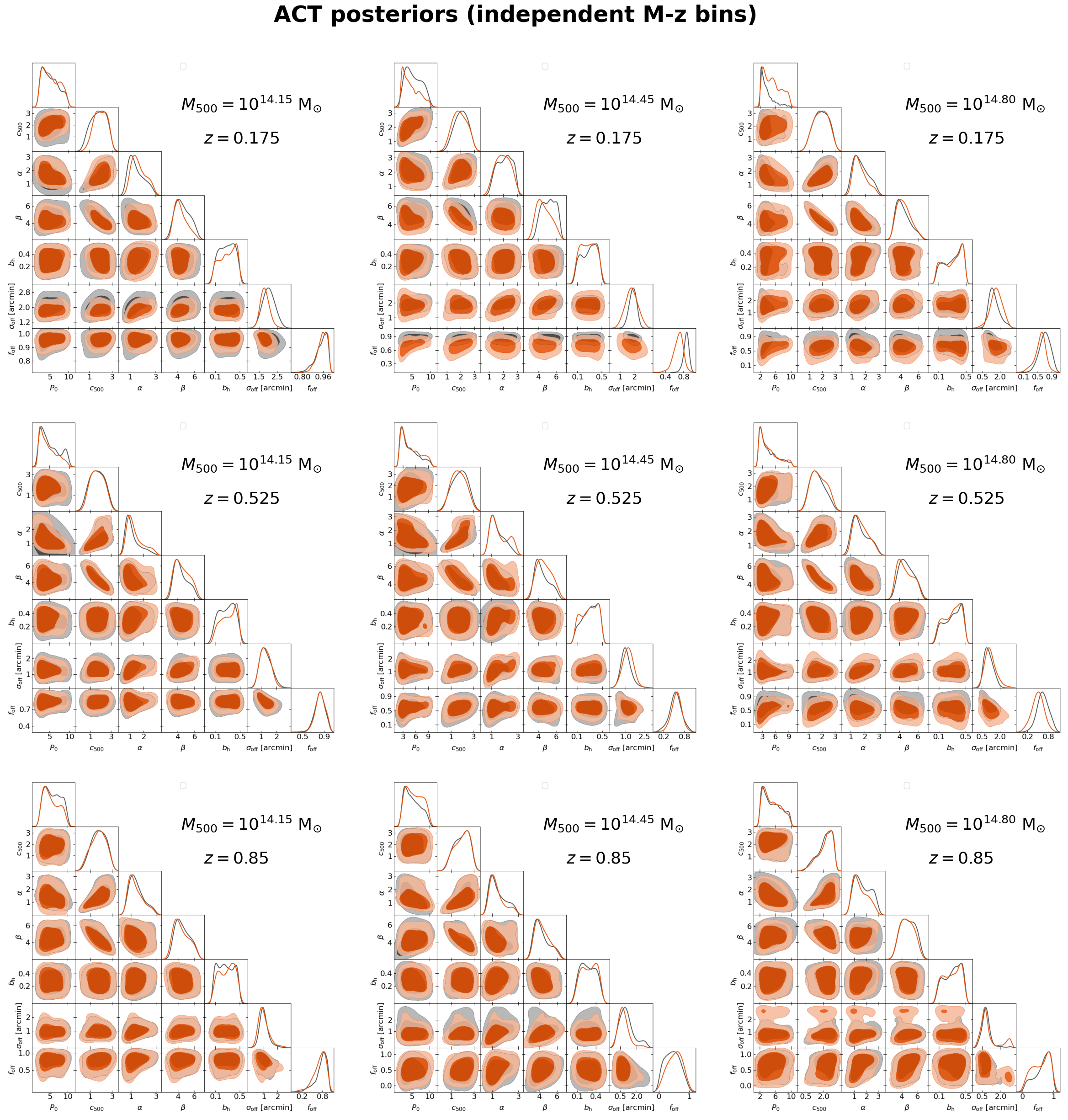}
	\caption{ACT posterior distributions (68\% and 95\% C.L. contours) on the fitted parameters for each independent 
	$M$-$z$ bin, obtained from the mass-rescaled catalog (red) and the one with no mass 
	rescaling (gray).\label{fig:act_contours}}
\end{figure*}

\begin{figure*} % +++++++++++++++++++++++++++++++++++++++++++++++++++++++++++++++
	\includegraphics[trim= 0mm 0mm 0mm 0mm, scale=0.3]{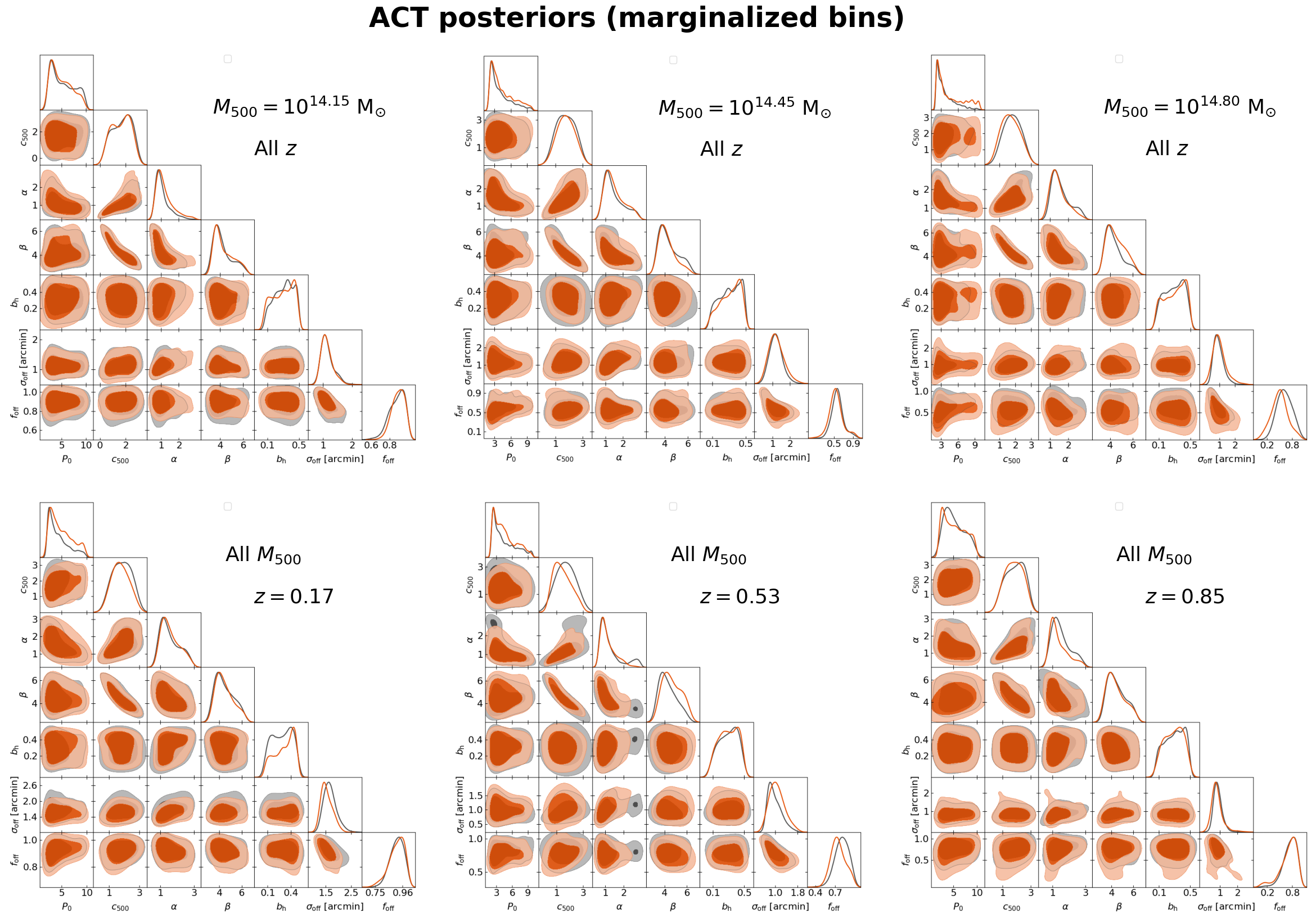}
	\caption{Same as in Fig.~\ref{fig:act_contours} but showing this time the bins marginalized 
	over $M_{500}$ or $z$.\label{fig:act_contours_marg}}
\end{figure*}

As commented in Section~\ref{ssec:emcee}, we perform our MCMC parameter estimation also on this 
new set of unscaled profiles. The resulting contours are plotted in Figs.~\ref{fig:posteriors_cmp} 
to~\ref{fig:act_contours_marg}, for all $M$-$z$ bins. These plots are also used to show the 
posterior distributions obtained with our fiducial sample; given the large number of resulting 
contours, we include them here in order to keep the main paper text lighter. As already discussed 
in Section~\ref{ssec:discussion3}, in general the contours obtained from the unscaled sample resemble 
the ones obtained from the fiducial sample; any systematic errors deriving from our mass rescaling are 
then of second order compared to the final statistical errors obtained on the parameters, which 
in turn are mostly driven by the degeneracy of the model.

 % +++++++++++++++++++++++++++++++++++++++++++++++++++++++++++++++

 % +++++++++++++++++++++++++++++++++++++++++++++++++++++++++++++++

 % +++++++++++++++++++++++++++++++++++++++++++++++++++++++++++++++

 % +++++++++++++++++++++++++++++++++++++++++++++++++++++++++++++++

%=================================================================================================
%=================================================================================================

\bibliography{bibliography}{}

\begin{thebibliography}{}
\expandafter\ifx\csname natexlab\endcsname\relax\def\natexlab#1{#1}\fi
\providecommand{\url}[1]{\href{#1}{#1}}
\providecommand{\dodoi}[1]{doi:~\href{http://doi.org/#1}{\nolinkurl{#1}}}
\providecommand{\doeprint}[1]{\href{http://ascl.net/#1}{\nolinkurl{http://ascl.net/#1}}}
\providecommand{\doarXiv}[1]{\href{https://arxiv.org/abs/#1}{\nolinkurl{https://arxiv.org/abs/#1}}}

\bibitem[{{Aghanim} {et~al.}(2019){Aghanim}, {Douspis}, {Hurier}, {Crichton},
  {Diego}, {Hasselfield}, {Macias-Perez}, {Marriage}, {Pointecouteau},
  {Remazeilles}, \& {Soubri{\'e}}}]{aghanim19}
{Aghanim}, N., {Douspis}, M., {Hurier}, G., {et~al.} 2019, \aap, 632, A47,
  \dodoi{10.1051/0004-6361/201935271}

\bibitem[{{Aguado-Barahona} {et~al.}(2022){Aguado-Barahona},
  {Rubi{\~n}o-Mart{\'\i}n}, {Ferragamo}, {Barrena}, {Streblyanska}, \&
  {Tramonte}}]{aguado_barahona22}
{Aguado-Barahona}, A., {Rubi{\~n}o-Mart{\'\i}n}, J.~A., {Ferragamo}, A.,
  {et~al.} 2022, \aap, 659, A126, \dodoi{10.1051/0004-6361/202039980}

\bibitem[{{Aihara} {et~al.}(2011){Aihara}, {Allende Prieto}, {An}, {Anderson},
  {Aubourg}, {Balbinot}, {Beers}, {Berlind}, {Bickerton}, {Bizyaev}, {Blanton},
  {Bochanski}, {Bolton}, {Bovy}, \& {Brandt}}]{aihara11}
{Aihara}, H., {Allende Prieto}, C., {An}, D., {et~al.} 2011, \apjs, 193, 29,
  \dodoi{10.1088/0067-0049/193/2/29}

\bibitem[{{Alam} {et~al.}(2015){Alam}, {Albareti}, {Allende Prieto}, {Anders},
  {Anderson}, {Anderton}, {Andrews}, {Armengaud}, {Aubourg}, {Bailey}, {Basu},
  {Bautista}, {Beaton}, {Beers}, \& {Bender}}]{alam15}
{Alam}, S., {Albareti}, F.~D., {Allende Prieto}, C., {et~al.} 2015, \apjs, 219,
  12, \dodoi{10.1088/0067-0049/219/1/12}

\bibitem[{{Allen} {et~al.}(2011){Allen}, {Evrard}, \& {Mantz}}]{allen11}
{Allen}, S.~W., {Evrard}, A.~E., \& {Mantz}, A.~B. 2011, \araa, 49, 409,
  \dodoi{10.1146/annurev-astro-081710-102514}

\bibitem[{{Arnaud} {et~al.}(2010){Arnaud}, {Pratt}, {Piffaretti},
  {B{\"o}hringer}, {Croston}, \& {Pointecouteau}}]{arnaud10}
{Arnaud}, M., {Pratt}, G.~W., {Piffaretti}, R., {et~al.} 2010, \aap, 517, A92,
  \dodoi{10.1051/0004-6361/200913416}

\bibitem[{{Barnes} {et~al.}(2021){Barnes}, {Vogelsberger}, {Pearce}, {Pop},
  {Kannan}, {Cao}, {Kay}, \& {Hernquist}}]{barnes21}
{Barnes}, D.~J., {Vogelsberger}, M., {Pearce}, F.~A., {et~al.} 2021, \mnras,
  506, 2533, \dodoi{10.1093/mnras/stab1276}

\bibitem[{{Battaglia} {et~al.}(2012){Battaglia}, {Bond}, {Pfrommer}, \&
  {Sievers}}]{battaglia12}
{Battaglia}, N., {Bond}, J.~R., {Pfrommer}, C., \& {Sievers}, J.~L. 2012, \apj,
  758, 75, \dodoi{10.1088/0004-637X/758/2/75}

\bibitem[{{Bellagamba} {et~al.}(2018){Bellagamba}, {Roncarelli}, {Maturi}, \&
  {Moscardini}}]{bellagamba18}
{Bellagamba}, F., {Roncarelli}, M., {Maturi}, M., \& {Moscardini}, L. 2018,
  \mnras, 473, 5221, \dodoi{10.1093/mnras/stx2701}

\bibitem[{{Bellagamba} {et~al.}(2019){Bellagamba}, {Sereno}, {Roncarelli},
  {Maturi}, {Radovich}, {Bardelli}, {Puddu}, {Moscardini}, {Getman},
  {Hildebrandt}, \& {Napolitano}}]{bellagamba19}
{Bellagamba}, F., {Sereno}, M., {Roncarelli}, M., {et~al.} 2019, \mnras, 484,
  1598, \dodoi{10.1093/mnras/stz090}

\bibitem[{{Birkinshaw}(1999)}]{birkinshaw99}
{Birkinshaw}, M. 1999, \physrep, 310, 97, \dodoi{10.1016/S0370-1573(98)00080-5}

\bibitem[{{Bleem} {et~al.}(2020){Bleem}, {Bocquet}, {Stalder}, {Gladders},
  {Ade}, {Allen}, {Anderson}, {Annis}, {Ashby}, {Austermann}, {Avila}, {Avva},
  {Bayliss}, {Beall}, {Bechtol}, {Bender}, {Benson}, {Bertin}, {Bianchini}, \&
  {Blake}}]{bleem20}
{Bleem}, L.~E., {Bocquet}, S., {Stalder}, B., {et~al.} 2020, \apjs, 247, 25,
  \dodoi{10.3847/1538-4365/ab6993}

\bibitem[{{B{\"o}hringer} {et~al.}(2007){B{\"o}hringer}, {Schuecker}, {Pratt},
  {Arnaud}, {Ponman}, {Croston}, {Borgani}, {Bower}, {Briel}, {Collins},
  {Donahue}, {Forman}, {Finoguenov}, {Geller}, {Guzzo}, {Henry}, {Kneissl},
  {Mohr}, {Matsushita}, {Mullis}, {Ohashi}, {Pedersen}, {Pierini}, {Quintana},
  {Raychaudhury}, {Reiprich}, {Romer}, {Rosati}, {Sabirli}, {Temple}, {Viana},
  {Vikhlinin}, {Voit}, \& {Zhang}}]{bohringer07}
{B{\"o}hringer}, H., {Schuecker}, P., {Pratt}, G.~W., {et~al.} 2007, \aap, 469,
  363, \dodoi{10.1051/0004-6361:20066740}

\bibitem[{{Carlstrom} {et~al.}(2002){Carlstrom}, {Holder}, \&
  {Reese}}]{carlstrom02}
{Carlstrom}, J.~E., {Holder}, G.~P., \& {Reese}, E.~D. 2002, \araa, 40, 643,
  \dodoi{10.1146/annurev.astro.40.060401.093803}

\bibitem[{{Czakon} {et~al.}(2015){Czakon}, {Sayers}, {Mantz}, {Golwala},
  {Downes}, {Koch}, {Lin}, {Molnar}, {Moustakas}, {Mroczkowski}, {Pierpaoli},
  {Shitanishi}, {Siegel}, \& {Umetsu}}]{czakon15}
{Czakon}, N.~G., {Sayers}, J., {Mantz}, A., {et~al.} 2015, \apj, 806, 18,
  \dodoi{10.1088/0004-637X/806/1/18}

\bibitem[{{Davis} {et~al.}(1985){Davis}, {Efstathiou}, {Frenk}, \&
  {White}}]{davis85}
{Davis}, M., {Efstathiou}, G., {Frenk}, C.~S., \& {White}, S.~D.~M. 1985, \apj,
  292, 371, \dodoi{10.1086/163168}

\bibitem[{{de Jong} {et~al.}(2017){de Jong}, {Verdoes Kleijn}, {Erben},
  {Hildebrandt}, {Kuijken}, {Sikkema}, {Brescia}, {Bilicki}, {Napolitano},
  {Amaro}, {Begeman}, {Boxhoorn}, {Buddelmeijer}, {Cavuoti}, {Getman}, {Grado},
  {Helmich}, {Huang}, {Irisarri}, {La Barbera}, {Longo}, {McFarland},
  {Nakajima}, {Paolillo}, {Puddu}, {Radovich}, {Rifatto}, {Tortora},
  {Valentijn}, {Vellucci}, {Vriend}, {Amon}, {Blake}, {Choi}, {Conti}, {Gwyn},
  {Herbonnet}, {Heymans}, {Hoekstra}, {Klaes}, {Merten}, {Miller}, {Schneider},
  \& {Viola}}]{dejong17}
{de Jong}, J. T.~A., {Verdoes Kleijn}, G.~A., {Erben}, T., {et~al.} 2017, \aap,
  604, A134, \dodoi{10.1051/0004-6361/201730747}

\bibitem[{{Dey} {et~al.}(2019){Dey}, {Schlegel}, {Lang}, {Blum}, {Burleigh},
  {Fan}, {Findlay}, {Finkbeiner}, {Herrera}, \& {Juneau}}]{dey19}
{Dey}, A., {Schlegel}, D.~J., {Lang}, D., {et~al.} 2019, \aj, 157, 168,
  \dodoi{10.3847/1538-3881/ab089d}

\bibitem[{{Fang} {et~al.}(2012){Fang}, {Kadota}, \& {Takada}}]{fang12}
{Fang}, W., {Kadota}, K., \& {Takada}, M. 2012, \prd, 85, 023007,
  \dodoi{10.1103/PhysRevD.85.023007}

\bibitem[{{Ferragamo} {et~al.}(2021){Ferragamo}, {Barrena},
  {Rubi{\~n}o-Mart{\'\i}n}, {Aguado-Barahona}, {Streblyanska}, {Tramonte},
  {G{\'e}nova-Santos}, {Hempel}, \& {Lietzen}}]{ferragamo21}
{Ferragamo}, A., {Barrena}, R., {Rubi{\~n}o-Mart{\'\i}n}, J.~A., {et~al.} 2021,
  \aap, 655, A115, \dodoi{10.1051/0004-6361/202140382}

\bibitem[{{Giocoli} {et~al.}(2021){Giocoli}, {Marulli}, {Moscardini}, {Sereno},
  {Veropalumbo}, {Gigante}, {Maturi}, {Radovich}, {Bellagamba}, {Roncarelli},
  {Bardelli}, {Contarini}, {Covone}, {Harnois-D{\'e}raps}, {Ingoglia}, {Lesci},
  {Nanni}, \& {Puddu}}]{giocoli21}
{Giocoli}, C., {Marulli}, F., {Moscardini}, L., {et~al.} 2021, \aap, 653, A19,
  \dodoi{10.1051/0004-6361/202140795}

\bibitem[{{Gong} {et~al.}(2019){Gong}, {Ma}, \& {Tanimura}}]{gong19}
{Gong}, Y., {Ma}, Y.-Z., \& {Tanimura}, H. 2019, \mnras, 486, 4904,
  \dodoi{10.1093/mnras/stz1177}

\bibitem[{{Goodman} \& {Weare}(2010)}]{goodman10}
{Goodman}, J., \& {Weare}, J. 2010, Communications in Applied Mathematics and
  Computational Science, 5, 65, \dodoi{10.2140/camcos.2010.5.65}

\bibitem[{{G{\'o}rski} {et~al.}(2005){G{\'o}rski}, {Hivon}, {Banday}, {Wand
  elt}, {Hansen}, {Reinecke}, \& {Bartelmann}}]{gorski05}
{G{\'o}rski}, K.~M., {Hivon}, E., {Banday}, A.~J., {et~al.} 2005, \apj, 622,
  759, \dodoi{10.1086/427976}

\bibitem[{{Hartlap} {et~al.}(2007){Hartlap}, {Simon}, \&
  {Schneider}}]{hartlap07}
{Hartlap}, J., {Simon}, P., \& {Schneider}, P. 2007, \aap, 464, 399,
  \dodoi{10.1051/0004-6361:20066170}

\bibitem[{{Hasselfield} {et~al.}(2013){Hasselfield}, {Hilton}, {Marriage},
  {Addison}, {Barrientos}, {Battaglia}, {Battistelli}, {Bond}, {Crichton},
  {Das}, {Devlin}, {Dicker}, {Dunkley}, {D{\"u}nner}, {Fowler}, {Gralla},
  {Hajian}, {Halpern}, {Hincks}, {Hlozek}, {Hughes}, {Infante}, {Irwin},
  {Kosowsky}, {Marsden}, {Menanteau}, {Moodley}, {Niemack}, {Nolta}, {Page},
  {Partridge}, {Reese}, {Schmitt}, {Sehgal}, {Sherwin}, {Sievers}, {Sif{\'o}n},
  {Spergel}, {Staggs}, {Swetz}, {Switzer}, {Thornton}, {Trac}, \&
  {Wollack}}]{hasselfield13}
{Hasselfield}, M., {Hilton}, M., {Marriage}, T.~A., {et~al.} 2013, \jcap, 2013,
  008, \dodoi{10.1088/1475-7516/2013/07/008}

\bibitem[{{He} {et~al.}(2021){He}, {Mansfield}, {Rau}, {Trac}, \&
  {Battaglia}}]{he21}
{He}, Y., {Mansfield}, P., {Rau}, M.~M., {Trac}, H., \& {Battaglia}, N. 2021,
  \apj, 908, 91, \dodoi{10.3847/1538-4357/abd0ff}

\bibitem[{{Hicks} {et~al.}(2008){Hicks}, {Ellingson}, {Bautz}, {Cain},
  {Gilbank}, {Gladders}, {Hoekstra}, {Yee}, \& {Garmire}}]{hicks08}
{Hicks}, A.~K., {Ellingson}, E., {Bautz}, M., {et~al.} 2008, \apj, 680, 1022,
  \dodoi{10.1086/587682}

\bibitem[{{Hilton} {et~al.}(2021){Hilton}, {Sif{\'o}n}, {Naess},
  {Madhavacheril}, {Oguri}, {Rozo}, {Rykoff}, {Abbott}, {Adhikari}, {Aguena},
  \& et~al.}]{hilton21}
{Hilton}, M., {Sif{\'o}n}, C., {Naess}, S., {et~al.} 2021, \apjs, 253, 3,
  \dodoi{10.3847/1538-4365/abd023}

\bibitem[{{Hoekstra} {et~al.}(2015){Hoekstra}, {Herbonnet}, {Muzzin}, {Babul},
  {Mahdavi}, {Viola}, \& {Cacciato}}]{hoekstra15}
{Hoekstra}, H., {Herbonnet}, R., {Muzzin}, A., {et~al.} 2015, \mnras, 449, 685,
  \dodoi{10.1093/mnras/stv275}

\bibitem[{{Hojjati} {et~al.}(2015){Hojjati}, {McCarthy}, {Harnois-Deraps},
  {Ma}, {Van Waerbeke}, {Hinshaw}, \& {Le Brun}}]{hojjati15}
{Hojjati}, A., {McCarthy}, I.~G., {Harnois-Deraps}, J., {et~al.} 2015, \jcap,
  2015, 047, \dodoi{10.1088/1475-7516/2015/10/047}

\bibitem[{{Hojjati} {et~al.}(2017){Hojjati}, {Tr{\"o}ster},
  {Harnois-D{\'e}raps}, {McCarthy}, {van Waerbeke}, {Choi}, {Erben}, {Heymans},
  {Hildebrandt}, {Hinshaw}, {Ma}, {Miller}, {Viola}, \& {Tanimura}}]{hojjati17}
{Hojjati}, A., {Tr{\"o}ster}, T., {Harnois-D{\'e}raps}, J., {et~al.} 2017,
  \mnras, 471, 1565, \dodoi{10.1093/mnras/stx1659}

\bibitem[{{Hurier} {et~al.}(2013){Hurier}, {Mac{\'{\i}}as-P{\'e}rez}, \&
  {Hildebrandt}}]{hurier13}
{Hurier}, G., {Mac{\'{\i}}as-P{\'e}rez}, J.~F., \& {Hildebrandt}, S. 2013,
  \aap, 558, A118, \dodoi{10.1051/0004-6361/201321891}

\bibitem[{{Ibitoye} {et~al.}(2022){Ibitoye}, {Tramonte}, {Ma}, \&
  {Dai}}]{ibitoye22}
{Ibitoye}, A., {Tramonte}, D., {Ma}, Y.-Z., \& {Dai}, W.-M. 2022, arXiv
  e-prints, arXiv:2206.05689.
\newblock \doarXiv{2206.05689}

\bibitem[{{Ishiyama} {et~al.}(2021){Ishiyama}, {Prada}, {Klypin}, {Sinha},
  {Metcalf}, {Jullo}, {Altieri}, {Cora}, {Croton}, {de la Torre},
  {Mill{\'a}n-Calero}, {Oogi}, {Ruedas}, \& {Vega-Mart{\'\i}nez}}]{ishiyama21}
{Ishiyama}, T., {Prada}, F., {Klypin}, A.~A., {et~al.} 2021, \mnras, 506, 4210,
  \dodoi{10.1093/mnras/stab1755}

\bibitem[{{Joachimi} {et~al.}(2021){Joachimi}, {Lin}, {Asgari}, {Tr{\"o}ster},
  {Heymans}, {Hildebrandt}, {K{\"o}hlinger}, {S{\'a}nchez}, {Wright},
  {Bilicki}, {Blake}, {van den Busch}, {Crocce}, {Dvornik}, {Erben}, {Getman},
  {Giblin}, {Hoekstra}, {Kannawadi}, {Kuijken}, {Napolitano}, {Schneider},
  {Scoccimarro}, {Sellentin}, {Shan}, {von Wietersheim-Kramsta}, \&
  {Zuntz}}]{joachimi21}
{Joachimi}, B., {Lin}, C.~A., {Asgari}, M., {et~al.} 2021, \aap, 646, A129,
  \dodoi{10.1051/0004-6361/202038831}

\bibitem[{{Johnston} {et~al.}(2007){Johnston}, {Sheldon}, {Wechsler}, {Rozo},
  {Koester}, {Frieman}, {McKay}, {Evrard}, {Becker}, \& {Annis}}]{johnston07}
{Johnston}, D.~E., {Sheldon}, E.~S., {Wechsler}, R.~H., {et~al.} 2007, arXiv
  e-prints, arXiv:0709.1159.
\newblock \doarXiv{0709.1159}

\bibitem[{{Klein} {et~al.}(2021){Klein}, {Oguri}, {Mohr}, {Grandis},
  {Ghirardini}, {Liu}, {Liu}, {Bulbul}, {Wolf}, {Comparat}, {Ramos-Ceja},
  {Buchner}, {Chiu}, {Clerc}, {Merloni}, {Miyatake}, {Miyazaki}, {Okabe},
  {Ota}, {Pacaud}, {Salvato}, \& {Driver}}]{klein21}
{Klein}, M., {Oguri}, M., {Mohr}, J.~J., {et~al.} 2021, arXiv e-prints,
  arXiv:2106.14519.
\newblock \doarXiv{2106.14519}

\bibitem[{{Komatsu} \& {Kitayama}(1999)}]{komatsu99}
{Komatsu}, E., \& {Kitayama}, T. 1999, \apjl, 526, L1, \dodoi{10.1086/312364}

\bibitem[{{Le Brun} {et~al.}(2015){Le Brun}, {McCarthy}, \& {Melin}}]{lebrun15}
{Le Brun}, A. M.~C., {McCarthy}, I.~G., \& {Melin}, J.-B. 2015, \mnras, 451,
  3868, \dodoi{10.1093/mnras/stv1172}

\bibitem[{{Ma} {et~al.}(2021){Ma}, {Gong}, {Tr{\"o}ster}, \& {Van
  Waerbeke}}]{ma21}
{Ma}, Y.-Z., {Gong}, Y., {Tr{\"o}ster}, T., \& {Van Waerbeke}, L. 2021, \mnras,
  500, 1806, \dodoi{10.1093/mnras/staa3369}

\bibitem[{{Madhavacheril} {et~al.}(2020){Madhavacheril}, {Hill}, {N{\ae}ss},
  {Addison}, {Aiola}, {Baildon}, {Battaglia}, {Bean}, {Bond}, {Calabrese},
  {Calafut}, {Choi}, {Darwish}, {Datta}, {Devlin}, {Dunkley}, {D{\"u}nner},
  {Ferraro}, {Gallardo}, {Gluscevic}, {Halpern}, {Han}, {Hasselfield},
  {Hilton}, {Hincks}, {Hlo{\v{z}}ek}, {Ho}, {Huffenberger}, {Hughes},
  {Koopman}, {Kosowsky}, {Lokken}, {Louis}, {Lungu}, {MacInnis}, {Maurin},
  {McMahon}, {Moodley}, {Nati}, {Niemack}, {Page}, {Partridge}, {Robertson},
  {Sehgal}, {Schaan}, {Schillaci}, {Sherwin}, {Sif{\'o}n}, {Simon}, {Spergel},
  {Staggs}, {Storer}, {van Engelen}, {Vavagiakis}, {Wollack}, \&
  {Xu}}]{madhavacheril20}
{Madhavacheril}, M.~S., {Hill}, J.~C., {N{\ae}ss}, S., {et~al.} 2020, \prd,
  102, 023534, \dodoi{10.1103/PhysRevD.102.023534}

\bibitem[{{Makiya} {et~al.}(2020){Makiya}, {Hikage}, \& {Komatsu}}]{makiya20}
{Makiya}, R., {Hikage}, C., \& {Komatsu}, E. 2020, \pasj, 72, 26,
  \dodoi{10.1093/pasj/psz147}

\bibitem[{{Mantz} {et~al.}(2010){Mantz}, {Allen}, {Ebeling}, {Rapetti}, \&
  {Drlica-Wagner}}]{mantz10}
{Mantz}, A., {Allen}, S.~W., {Ebeling}, H., {Rapetti}, D., \& {Drlica-Wagner},
  A. 2010, \mnras, 406, 1773, \dodoi{10.1111/j.1365-2966.2010.16993.x}

\bibitem[{{Maturi} {et~al.}(2019){Maturi}, {Bellagamba}, {Radovich},
  {Roncarelli}, {Sereno}, {Moscardini}, {Bardelli}, \& {Puddu}}]{maturi19}
{Maturi}, M., {Bellagamba}, F., {Radovich}, M., {et~al.} 2019, \mnras, 485,
  498, \dodoi{10.1093/mnras/stz294}

\bibitem[{{Mehrtens} {et~al.}(2012){Mehrtens}, {Romer}, {Hilton},
  {Lloyd-Davies}, {Miller}, {Stanford}, {Hosmer}, {Hoyle}, {Collins}, {Liddle},
  {Viana}, {Nichol}, {Stott}, {Dubois}, {Kay}, {Sahl{\'e}n}, {Young}, {Short},
  {Christodoulou}, {Watson}, {Davidson}, {Harrison}, {Baruah}, {Smith},
  {Burke}, {Mayers}, {Deadman}, {Rooney}, {Edmondson}, {West}, {Campbell},
  {Edge}, {Mann}, {Sabirli}, {Wake}, {Benoist}, {da Costa}, {Maia}, \&
  {Ogando}}]{mehrtens12}
{Mehrtens}, N., {Romer}, A.~K., {Hilton}, M., {et~al.} 2012, \mnras, 423, 1024,
  \dodoi{10.1111/j.1365-2966.2012.20931.x}

\bibitem[{{Nagai} {et~al.}(2007){Nagai}, {Kravtsov}, \& {Vikhlinin}}]{nagai07}
{Nagai}, D., {Kravtsov}, A.~V., \& {Vikhlinin}, A. 2007, \apj, 668, 1,
  \dodoi{10.1086/521328}

\bibitem[{{Navarro} {et~al.}(1997){Navarro}, {Frenk}, \& {White}}]{navarro97}
{Navarro}, J.~F., {Frenk}, C.~S., \& {White}, S.~D.~M. 1997, \apj, 490, 493,
  \dodoi{10.1086/304888}

\bibitem[{{Pearce} {et~al.}(2020){Pearce}, {Kay}, {Barnes}, {Bower}, \&
  {Schaller}}]{pearce20}
{Pearce}, F.~A., {Kay}, S.~T., {Barnes}, D.~J., {Bower}, R.~G., \& {Schaller},
  M. 2020, \mnras, 491, 1622, \dodoi{10.1093/mnras/stz3003}

\bibitem[{{Piffaretti} {et~al.}(2011){Piffaretti}, {Arnaud}, {Pratt},
  {Pointecouteau}, \& {Melin}}]{piffaretti11}
{Piffaretti}, R., {Arnaud}, M., {Pratt}, G.~W., {Pointecouteau}, E., \&
  {Melin}, J.~B. 2011, \aap, 534, A109, \dodoi{10.1051/0004-6361/201015377}

\bibitem[{{Planck Collaboration} {et~al.}(2011{\natexlab{a}}){Planck
  Collaboration}, {Ade}, {Aghanim}, {Arnaud}, {Ashdown}, {Aumont}, \&
  {Baccigalupi}}]{planck_er_viii}
{Planck Collaboration}, {Ade}, P.~A.~R., {Aghanim}, N., {et~al.}
  2011{\natexlab{a}}, \aap, 536, A8, \dodoi{10.1051/0004-6361/201116459}

\bibitem[{{Planck Collaboration} {et~al.}(2011{\natexlab{b}}){Planck
  Collaboration}, {Ade}, {Aghanim}, {Arnaud}, {Ashdown}, {Aumont},
  {Baccigalupi}, {Balbi}, {Banday}, {Barreiro}, {Bartelmann}, {Bartlett},
  {Battaner}, {Benabed}, {Beno{\^\i}t}, {Bernard}, \&
  {Bersanelli}}]{planck_er_xi}
---. 2011{\natexlab{b}}, \aap, 536, A11, \dodoi{10.1051/0004-6361/201116458}

\bibitem[{{Planck Collaboration} {et~al.}(2013){Planck Collaboration}, {Ade},
  {Aghanim}, {Arnaud}, {Ashdown}, {Atrio-Barandela}, {Aumont}, {Baccigalupi},
  {Balbi}, {Banday}, {Barreiro}, {Bartlett}, {Battaner}, \&
  {Benabed}}]{planck_ir_v}
---. 2013, \aap, 550, A131, \dodoi{10.1051/0004-6361/201220040}

\bibitem[{{Planck Collaboration} {et~al.}(2016{\natexlab{a}}){Planck
  Collaboration}, {Ade}, {Aghanim}, {Arnaud}, {Ashdown}, {Aumont},
  {Baccigalupi}, {Banday}, {Barreiro}, {Barrena}, \&
  {Bartlett}}]{planck_15_xxvii}
---. 2016{\natexlab{a}}, \aap, 594, A27, \dodoi{10.1051/0004-6361/201525823}

\bibitem[{{Planck Collaboration} {et~al.}(2016{\natexlab{b}}){Planck
  Collaboration}, {Aghanim}, {Arnaud}, {Ashdown}, {Aumont}, {Baccigalupi},
  {Banday}, {Barreiro}, {Bartlett}, {Bartolo}, \& {Battaner}}]{planck_15_xxii}
{Planck Collaboration}, {Aghanim}, N., {Arnaud}, M., {et~al.}
  2016{\natexlab{b}}, \aap, 594, A22, \dodoi{10.1051/0004-6361/201525826}

\bibitem[{{Planck Collaboration} {et~al.}(2020){Planck Collaboration},
  {Aghanim}, {Akrami}, {Ashdown}, {Aumont}, {Baccigalupi}, {Ballardini},
  {Banday}, {Barreiro}, {Bartolo}, \& {Basak}}]{planck_18_vi}
{Planck Collaboration}, {Aghanim}, N., {Akrami}, Y., {et~al.} 2020, \aap, 641,
  A6, \dodoi{10.1051/0004-6361/201833910}

\bibitem[{{Pointecouteau} {et~al.}(2021){Pointecouteau}, {Santiago-Bautista},
  {Douspis}, {Aghanim}, {Crichton}, {Diego}, {Hurier}, {Macias-Perez},
  {Marriage}, {Remazeilles}, {Caretta}, \& {Bravo-Alfaro}}]{pointecouteau21}
{Pointecouteau}, E., {Santiago-Bautista}, I., {Douspis}, M., {et~al.} 2021,
  \aap, 651, A73, \dodoi{10.1051/0004-6361/202040213}

\bibitem[{{Remazeilles} {et~al.}(2011){Remazeilles}, {Delabrouille}, \&
  {Cardoso}}]{remazeilles11}
{Remazeilles}, M., {Delabrouille}, J., \& {Cardoso}, J.-F. 2011, \mnras, 410,
  2481, \dodoi{10.1111/j.1365-2966.2010.17624.x}

\bibitem[{{Rotti} {et~al.}(2021){Rotti}, {Bolliet}, {Chluba}, \&
  {Remazeilles}}]{rotti21}
{Rotti}, A., {Bolliet}, B., {Chluba}, J., \& {Remazeilles}, M. 2021, \mnras,
  \dodoi{10.1093/mnras/stab469}

\bibitem[{{Sarazin}(1988)}]{sarazin88}
{Sarazin}, C.~L. 1988, \skytel, 76, 639

\bibitem[{{Sayers} {et~al.}(2011){Sayers}, {Golwala}, {Ameglio}, \&
  {Pierpaoli}}]{sayers11}
{Sayers}, J., {Golwala}, S.~R., {Ameglio}, S., \& {Pierpaoli}, E. 2011, \apj,
  728, 39, \dodoi{10.1088/0004-637X/728/1/39}

\bibitem[{{Sayers} {et~al.}(2016){Sayers}, {Golwala}, {Mantz}, {Merten},
  {Molnar}, {Naka}, {Pailet}, {Pierpaoli}, {Siegel}, \& {Wolman}}]{sayers16}
{Sayers}, J., {Golwala}, S.~R., {Mantz}, A.~B., {et~al.} 2016, \apj, 832, 26,
  \dodoi{10.3847/0004-637X/832/1/26}

\bibitem[{{Sereno} {et~al.}(2017){Sereno}, {Covone}, {Izzo}, {Ettori},
  {Coupon}, \& {Lieu}}]{sereno17}
{Sereno}, M., {Covone}, G., {Izzo}, L., {et~al.} 2017, \mnras, 472, 1946,
  \dodoi{10.1093/mnras/stx2085}

\bibitem[{{Sunyaev} \& {Zeldovich}(1972)}]{sunyaev72}
{Sunyaev}, R.~A., \& {Zeldovich}, Y.~B. 1972, Comments on Astrophysics and
  Space Physics, 4, 173

\bibitem[{{Takey} {et~al.}(2014){Takey}, {Schwope}, \& {Lamer}}]{takey14}
{Takey}, A., {Schwope}, A., \& {Lamer}, G. 2014, \aap, 564, A54,
  \dodoi{10.1051/0004-6361/201322973}

\bibitem[{{Tinker} {et~al.}(2008){Tinker}, {Kravtsov}, {Klypin}, {Abazajian},
  {Warren}, {Yepes}, {Gottl{\"o}ber}, \& {Holz}}]{tinker08}
{Tinker}, J., {Kravtsov}, A.~V., {Klypin}, A., {et~al.} 2008, \apj, 688, 709,
  \dodoi{10.1086/591439}

\bibitem[{{Tinker} {et~al.}(2010){Tinker}, {Robertson}, {Kravtsov}, {Klypin},
  {Warren}, {Yepes}, \& {Gottl{\"o}ber}}]{tinker10}
{Tinker}, J.~L., {Robertson}, B.~E., {Kravtsov}, A.~V., {et~al.} 2010, \apj,
  724, 878, \dodoi{10.1088/0004-637X/724/2/878}

\bibitem[{{Vikhlinin} {et~al.}(2009){Vikhlinin}, {Burenin}, {Ebeling},
  {Forman}, {Hornstrup}, {Jones}, {Kravtsov}, {Murray}, {Nagai}, {Quintana}, \&
  {Voevodkin}}]{vikhlinin09}
{Vikhlinin}, A., {Burenin}, R.~A., {Ebeling}, H., {et~al.} 2009, \apj, 692,
  1033, \dodoi{10.1088/0004-637X/692/2/1033}

\bibitem[{{Voges} {et~al.}(1999){Voges}, {Aschenbach}, {Boller},
  {Br{\"a}uninger}, {Briel}, {Burkert}, {Dennerl}, {Englhauser}, {Gruber},
  {Haberl}, {Hartner}, {Hasinger}, {K{\"u}rster}, {Pfeffermann}, {Pietsch},
  {Predehl}, {Rosso}, {Schmitt}, {Tr{\"u}mper}, \& {Zimmermann}}]{voges99}
{Voges}, W., {Aschenbach}, B., {Boller}, T., {et~al.} 1999, \aap, 349, 389

\bibitem[{{Voit}(2005)}]{voit05}
{Voit}, G.~M. 2005, Reviews of Modern Physics, 77, 207,
  \dodoi{10.1103/RevModPhys.77.207}

\bibitem[{{von der Linden} {et~al.}(2014){von der Linden}, {Allen},
  {Applegate}, {Kelly}, {Allen}, {Ebeling}, {Burchat}, {Burke}, {Donovan},
  {Morris}, {Blandford}, {Erben}, \& {Mantz}}]{von_der_linden14}
{von der Linden}, A., {Allen}, M.~T., {Applegate}, D.~E., {et~al.} 2014,
  \mnras, 439, 2, \dodoi{10.1093/mnras/stt1945}

\bibitem[{{Wang} {et~al.}(2016){Wang}, {Mo}, {Yang}, {Zhang}, {Shi}, {Jing},
  {Liu}, {Li}, {Kang}, \& {Gao}}]{wang16}
{Wang}, H., {Mo}, H.~J., {Yang}, X., {et~al.} 2016, \apj, 831, 164,
  \dodoi{10.3847/0004-637X/831/2/164}

\bibitem[{{Wen} \& {Han}(2015)}]{wen15}
{Wen}, Z.~L., \& {Han}, J.~L. 2015, \apj, 807, 178,
  \dodoi{10.1088/0004-637X/807/2/178}

\bibitem[{{Wen} {et~al.}(2012){Wen}, {Han}, \& {Liu}}]{wen12}
{Wen}, Z.~L., {Han}, J.~L., \& {Liu}, F.~S. 2012, \apjs, 199, 34,
  \dodoi{10.1088/0067-0049/199/2/34}

\bibitem[{{Yan} {et~al.}(2020){Yan}, {Raza}, {Van Waerbeke}, {Mead},
  {McCarthy}, {Tr{\"o}ster}, \& {Hinshaw}}]{yan20}
{Yan}, Z., {Raza}, N., {Van Waerbeke}, L., {et~al.} 2020, \mnras, 493, 1120,
  \dodoi{10.1093/mnras/staa295}

\bibitem[{{Yang} {et~al.}(2005){Yang}, {Mo}, {van den Bosch}, \&
  {Jing}}]{yang05}
{Yang}, X., {Mo}, H.~J., {van den Bosch}, F.~C., \& {Jing}, Y.~P. 2005, \mnras,
  356, 1293, \dodoi{10.1111/j.1365-2966.2005.08560.x}

\bibitem[{{Yang} {et~al.}(2006){Yang}, {Mo}, {van den Bosch}, {Jing},
  {Weinmann}, \& {Meneghetti}}]{yang06}
{Yang}, X., {Mo}, H.~J., {van den Bosch}, F.~C., {et~al.} 2006, \mnras, 373,
  1159, \dodoi{10.1111/j.1365-2966.2006.11091.x}

\bibitem[{{Yang} {et~al.}(2007){Yang}, {Mo}, {van den Bosch}, {Pasquali}, {Li},
  \& {Barden}}]{yang07}
---. 2007, \apj, 671, 153, \dodoi{10.1086/522027}

\bibitem[{{Yang} {et~al.}(2021){Yang}, {Xu}, {He}, {Gu}, {Katsianis}, {Meng},
  {Shi}, {Zou}, {Zhang}, {Liu}, {Wang}, {Dong}, {Lu}, {Li}, {Chen}, {Wang},
  {Mo}, {Fu}, {Guo}, {Leauthaud}, {Luo}, {Zhang}, \& {Zu}}]{yang21}
{Yang}, X., {Xu}, H., {He}, M., {et~al.} 2021, \apj, 909, 143,
  \dodoi{10.3847/1538-4357/abddb2}

\end{thebibliography}
\bibliographystyle{aasjournal}

\end{document}